\documentclass[a4paper, 11pt]{article}

\usepackage{graphicx}%
\usepackage{multirow}%
\usepackage{amsmath,amssymb,amsfonts}%
\usepackage{amsthm}%
\usepackage{mathrsfs}%
\usepackage[title]{appendix}%
\usepackage{xcolor}%
\usepackage{textcomp}%
\usepackage{manyfoot}%
\usepackage{booktabs}%
\usepackage{algorithm}%
\usepackage{algorithmicx}%
\usepackage{algpseudocode}%
\usepackage{listings}%
\usepackage{natbib}
\usepackage{rotating}

\newcommand{\Autoref}[1]{%
	\begingroup%
	\def\chapterautorefname{Chapter}%
	\def\sectionautorefname{Section}%
	\def\subsectionautorefname{Subsection}%
	\autoref{#1}%
	\endgroup%
}

\usepackage{caption}
\usepackage[a4paper,top=3cm,bottom=3.5cm,left=2.5cm,right=2.5cm,marginparwidth=1.75cm]{geometry}

\usepackage[colorlinks=true, allcolors=blue]{hyperref}

\title{On prior smoothing with discrete spatial data in the context of disease mapping}
\author{Garazi Retegui$^{1,2*}$, Alan E. Gelfand$^{3}$, Jaione Etxeberria$^{1,2}$, Mar{\'i}a Dolores Ugarte$^{1,2}$\\
	\\
	\small $^1$ Department of Statistics, Computer Science and Mathematics, Public University of \\
	\small Navarre (UPNA), Arrosadia Campus, Pamplona, 31006, Navarra, Spain.\\
	\small $^2$ Institute for Advanced Materials and Mathematics (INAMAT2), Public University \\ 
	\small of Navarre (UPNA), Arrosadia Campus, Pamplona, 31006, Navarra, Spain.\\
	\small $^3$ Department of Statistical Science, Duke University, Old Chemistry Building,\\
	\small Durham, NC 27707, USA.\\
	\small $^{*}$ Corresponding author.\\
	\\
	\small garazi.retegui@unavarra.es\\ \\
	\date{}
}

\setcitestyle{authoryear,open={(},close={)}} 

\begin{document}
	\maketitle
	
	\begin{abstract}
		\noindent Disease mapping attempts to explain observed health event counts across areal units, typically using Markov random field models. These models rely on spatial priors to account for variation in raw relative risk or rate estimates. Spatial priors introduce some degree of smoothing, wherein, for any particular unit, empirical risk or incidence estimates are either adjusted towards a suitable mean or incorporate neighbor-based smoothing. While model explanation may be the primary focus, the literature lacks a comparison of the amount of smoothing introduced by different spatial priors. Additionally, there has been no investigation into how varying the parameters of these priors influences the resulting smoothing. This study examines seven commonly used spatial priors through both simulations and real data analyses. Using areal maps of peninsular Spain and England, we analyze smoothing effects using two datasets with associated populations at risk. We propose empirical metrics to quantify the smoothing achieved by each model and theoretical metrics to calibrate the expected extent of smoothing as a function of model parameters. We employ areal maps in order to quantitatively characterize the extent of smoothing within and across the models as well as to link the theoretical metrics to the empirical metrics.
	\end{abstract}

	\section{Introduction \label{S:intro}}
	Disease mapping is part of spatial epidemiology and refers to a collection of statistical techniques developed to provide accurate estimates of the geographical distribution of disease risk or rates in sets of areal units within a particular study region. Extensive literature exists on disease mapping, covering the specification, fitting, and interpretation of various models \citep[see for example][]{banerjee2015, martinez2019}. 
	Typically, neighbor-based modeling is employed. This approach provides the joint distribution for the areal units through neighbors, specifying a local conditional distribution such that the expected value of the spatial variable at a given unit is an average of its neighboring units. Such specifications are Gaussian Markov random fields \citep[GMRFs, ][]{rue2005} in the form of conditionally autoregressive (CAR) models \citep{besag1974, besag1991}.
	Working at the scale of areal units, these models attempt to explain the variation in risk of incidence or mortality from a disease (or diseases) over areal units in the region. 
	The local \emph{averaging}, implicit in all of these models, introduces some degree of smoothing to the inference. That is, for any particular unit, we immediately have an \emph{empirical} estimate of risk or incidence/mortality rate but our modeling smooths this empirical estimate either toward a suitable mean or incorporates some version of neighbor-based smoothing. Thus, customarily, with areal unit data, an inherent component is spatial smoothing \citep{banerjee2015}. However, the degree of smoothing must be carefully managed. Insufficient smoothing results in noisy maps dominated by random variation, while excessive smoothing can obscure the detection of both high- and low-risk areas. Since one of the main goals of disease mapping is to identify spatial patterns that inform public health decisions, it is crucial to understand and assess the extent of smoothing induced by the chosen model to ensure valid and actionable conclusions. Hence, while model assessment is often emphasized and may be the primary objective, the \emph{amount} of smoothing introduced by a particular model choice has not been studied in the literature. This paper aims to address this gap.

	Model behavior has historically been investigated with regard to goodness of fit, which measures the discrepancy between observed data and the values predicted by the model, while penalizing for model complexity. Examples of model selection criteria include the deviance information criteria \citep[DIC,][]{spiegelhalter2002} and the Watanabe-Akaike information criterion \citep[WAIC,][]{watanabe2010}, both widely used in the literature. However, as \cite{stern2000} point out, when the aim is smoothing, evaluating model performance becomes challenging. 
	In particular, they address the issue of model criticism in the identification of divergent regions whose disease rates are not consistent with a proposed model. They highlight initial considerations regarding the importance of understanding the smoothing effect in implementation of the models.  We emphasize that model performance with regard to choice of prior should not be viewed as a matter of goodness-of-fit since perfect fit can be achieved without implementing any smoothing.
	
	In this regard, in the simplest univariate spatial case, with no risk factors, the most smoothing would occur if every area had the same intercept. The least (in fact, none) smoothing would occur if each area was fitted independently with its local maximum likelihood estimate (MLE). The various spatial specifications provide something in between -- neighbor-based smoothing. 
	Therefore, we might seek to quantify the smoothing achieved by a given model specification.	
	One can imagine metrics to quantify the amount of smoothing associated with a particular choice. For example, \cite{white2017} propose the use of predictive mean square error or rank probability scores to measure the smoothness, while  \cite{quick2021} and \cite{song2021} offer thoughts on measuring the informativeness/performance of various CAR distributions. 
	In any event, smoothing does not seek to minimize a goodness of fit criterion \citep{duncan2020}.
	
	So, the contribution of this work is an investigation, both theoretical and empirical, into the extent of smoothing achieved by a given model specification. If we adopt a particular likelihood and introduce a choice of parametric spatial prior, we can ask (i) theoretically, how smoothing is expected to vary as we vary parameters in the prior, (ii) empirically, how much smoothing do we actually achieve? Further, we can ask how the smoothing varies across a collection of neighbor-based priors? Our contribution is developed within the context of disease mapping using areal data in a Bayesian framework. Specifically, this work focuses on models that smooth rates rather than risks. We specify different spatial priors, with an emphasis on CAR spatial priors, such as iCAR, LCAR or BYM, as well as a Gaussian process (GP) prior (references and details below). Moreover, we limit the theoretical analysis of smoothing behavior to the parameters of the spatial priors.  We do not consider the effects of population size (which, in practice, we could not control) or spatial structure (confining our story to first-order neighborhoods) on the amount of smoothing induced. We fix the adjacency matrix to a commonly used neighborhood matrix defined by \cite{besag1991}, considering two areas as neighbors if they share a common border. However, to investigate the behavior of the models as the number of areas increases within a spatial region, we vary the level of disaggregation, specifically within peninsular Spain. 
	In summary, our contribution offers novelty in illuminating theoretical smoothing behavior with regard to prior parameters across a collection of priors.  We see potential practical utility in terms of comparative expected smoothness across these priors after model fitting, anticipating practitioner decision-making between less smoothed maps which are noisier or more smoothed maps which obscure risk detection. We illustrate this issue in \autoref{S:illustration}.

	In this study, the Poisson-Gamma non-spatial model \citep{clayton1987} is helpful as a baseline to develop  theoretical insight into smoothing behavior since explicit calculation is available. However, to address the aforementioned questions for the models of interest, we employ both simulation studies and real data analyses. In particular, we conduct two simulation studies with different aims. The first simulation study focuses on formally quantifying the extent of smoothing and the factors influencing the level of expected smoothing for each choice of spatial prior under consideration. Additionally, it investigates whether model smoothing comparisons using theoretical metrics align with those using empirical metrics. Therefore, this simulation study performs an analysis \emph{within} each spatial prior. The second simulation study compares the extent of smoothing \emph{across} the proposed priors under different simulated scenarios and two different spatial structures with varying areal units and their associated neighbor structures. Then, we analyze two real datasets corresponding to lung cancer mortality counts and the corresponding population at risk for two different spatial regions, peninsular Spain and England. We define informative and non-informative prior distributions for the parameters of the spatial priors and investigate the amount of smoothing induced both across and within spatial priors.

	The format of the paper is as follows. In \autoref{S:methods}, we provide a description of the employed Bayesian model and spatial priors used to quantify and compare the extent of smoothing. Additionally, we derive the theoretical metric proposed for analyzing the factors influencing the extent of smoothing for each spatial prior, and we present the empirical smoothing metrics used throughout the study. \Autoref{S:Po-Gamma} presents the theoretical expected smoothing for the ``baseline'' Poisson-Gamma model.
	In \autoref{S:SS}, we present the two simulation studies: one within priors and one across priors. The real data analysis is presented in \autoref{S:illustration}. Lastly, in \autoref{S:discussion}, we summarize our contributions along with possible future investigation.

	\section{Modeling and Methods}\label{S:methods}
	In this section, the Poisson-logitNormal provides the likelihood for hierarchical modeling of disease rates. In \autoref{S:Po-logitN}, we provide details with the collection of priors we consider. We propose a general theoretical criterion, as a function of the prior parameters to calibrate  how much smoothing is expected to be induced. The empirical smoothing metrics are presented in \autoref{S:empirical}.

	\subsection{The Poisson-logitNormal model\label{S:Po-logitN}}
	
	We assume the study region $S$ is partitioned into $A$ non-overlapping areal units. Let $\mathbf{O} = \left(O_1, \dots, O_A\right)'$ denote the vector of observed counts of a disease across $A$ geographical regions or districts, and $\mathbf{n} = \left(n_1, \dots, n_A\right)'$ denote the population at risk within these units. We consider the extent of smoothing achieved by a given model specification, when working with observed counts and population at risk,  for inferring about rates. Hence, we introduce $\mathbf{r} = \left(r_1, \dots, r_A\right)'$ as a vector of the rates. 	
	Conditional on the rates, we model the observed counts as independent Poisson random variables, $
	O_i|r_i \sim Poisson\left(n_ir_i\right)$ for each $i = 1, \dots, A$, and we consider a logit-Normal distribution for the $r_i$. 
	Specifically, we model $\textrm{logit}\left(r_i\right)$ as
	\begin{equation}\label{Eq3}
		\textrm{logit} (r_i) = \alpha + \kappa_i
	\end{equation}
	where $\alpha$ is the overall rate and the vector $\boldsymbol{\kappa}=\left(\kappa_1,\dots, \kappa_A\right)'$  denotes the collection of spatial random effects. 
	Since we are focused on smoothing, we ignore covariates for our development. Recall that, as a spatial model, we interpret this model hierarchically, i.e., the $\kappa_{i}$'s are latent and the model specifies $O_{i}$ conditionally independent given $\kappa_{i}$ (and $\alpha$).

	With discrete spatial data, it is customary to model spatial dependence through a choice of neighbor weight matrix which provides the precision matrix, $\mathbf{Q}$, equivalently, through the conditional variances. $\mathbf{Q}$ is not nonsingular for some of the prior models; its inverse does not exist.  So, notationally, we  assume a normal distribution for the logit rates, that is, 
	$\textrm{logit} \left(\mathbf{r}\right)\sim N\left(\alpha \mathbf{1_A},\sigma^2\mathbf{Q}^-\right)$ with $\mathbf{r}$ the vector of $r_i$'s, $\mathbf{1_A}$ a column vector of ones of size $A$, $\mathbf{Q}^-$ the spatial correlation matrix and $^-$ denoting the Moore-Penrose generalized inverse. When $\mathbf{Q}$ is full rank, we replace $\mathbf{Q}^-$ with $\mathbf{Q}^{-1}$. 
	Therefore, we have three objects to look at here: (i) $\sigma^2 \mathbf{Q}^-$, the covariance matrix of the prior defined (if it exists), (ii) $\frac{1}{\sigma^2}\mathbf{Q}$, the precision matrix of the prior (which always exists), and (iii) var$(r_{i}|\{r_{j}, j \neq i\}) =\frac{\sigma^2}{Q_{ii}}$, the conditional variance for $r_i$ (or $\kappa_i$)  where $Q_{ii}$ denotes the $i$th diagonal element of $\mathbf{Q}$. The resulting conditional variances are smaller than the marginal variances, reflecting neighbor-based smoothing.

	Analyzing the smoothing induced by different spatial priors suggests investigation of the conditional variances associated with the various choices of  \( \boldsymbol{Q} \), i.e., the $Q_{ii}$ through var$(r_{i}|\{r_{j}, j \neq i\}) = \frac{\sigma^2}{Q_{ii}}$ as a function of model parameters.  Smaller conditional variance implies more smoothing. 
	If we sum the conditional variances across the set of areal units, we obtain a proposed metric for the overall amount of smoothing, referred to as the \emph{total conditional variance} (TCV):
	\begin{equation*}
		\text{TCV} = \sum_{i=1}^{n} \operatorname{var}(r_i \mid \{r_j,\, j \neq i\}) = \sum_{i=1}^{n} \frac{\sigma^2}{Q_{ii}}.
	\end{equation*}
	This metric serves as a scalar summary of the smoothing implied by a given prior. In the simulation study, for any case, we can directly calculate this metric, sometimes as an explicit parametric function. With real data, we can obtain the posterior distribution of this metric under a given prior. Further, smaller TCV implies more smoothing.  We see that the magnitude of $\sigma^2$ will always exercise control over the amount of expected smoothing.

	In the following lines, seven different priors are considered for $\boldsymbol{\kappa}$ and their smoothing effect are considered: an independent (non)spatial prior, five different CAR priors, and a Gaussian Process \citep[GP,][]{cressie2015} prior using a covariance function employing the distance between centroids. 
	
	For the non-spatial or independence prior case,
	suppose we place a logit normal prior on $r_{i}$, i.e., $\textrm{logit}\left(r_i\right) \sim N(\mu, \sigma^2)$. Following notation from \autoref{Eq3}, we view $\mu=\alpha$ and the $\kappa_i$ are i.i.d $N(0, \sigma^{2})$.  This situation is analogous to the Poisson-Gamma setting in \autoref{S:Po-Gamma} below.  The Bayesian smoothing is toward the mean and this is a baseline for comparison.  Further, the $\textrm{TCV}= \sigma^{2}A$; we have only $\sigma^2$ to control smoothing.
	
	Next, consider the CAR-type or neighbor-based priors. We examine five such priors commonly found in the spatial statistics literature for modeling discrete spatial autocorrelation: the intrinsic CAR prior \citep[iCAR,][]{besag1974}, the Besag, York, and Mollié prior specification \citep[BYM,][]{besag1991}, the proper CAR prior \citep[p-CAR,][]{jin2007}, the alternative developed by Leroux and Breslow \citep[LCAR,][]{leroux2000}, 
	and the BYM2 prior \citep{riebler2016}. Each specification is a special case of a Gaussian Markov random field (GMRF) and can be written in the general form $\boldsymbol{\kappa}\sim N\left(\boldsymbol{0},\sigma^2\mathbf{Q}^-\right)$.

	\begin{itemize}
		\item iCAR: $\mathbf{Q} = \mathbf{D} - \mathbf{W}$, where $\mathbf{W}$ is the spatial proximity matrix defined as $w_{ii} = 0$ and $w_{ij} = 1$ if the geographical units $i$ and $j$ are neighbors and 0 otherwise, $\mathbf{D} $ is a diagonal matrix whose elements are the number of neighbors of the $i$th area, denoted by $w_i^+$. Thus, $\mathbf{D} \equiv diag(w_1^+;\dots;w_A^+)$. Therefore, var$(r_{i}|\{r_{j}, j \neq i\}) =\frac{\sigma^2}{Q_{ii}} = \frac{\sigma^2}{w_i^+}$ and $\textrm{TCV} = \sum_{i} \frac{\sigma^2}{w_i^+}$. Again, $\sigma^2$ exerts control over expect smoothing. As noted above, a comparison which we do not explore here is say, between a specification with first order neighbors vs. one with first and second order neighbors. 
		
		\item BYM: This model is defined using the sum of two random effects, $\boldsymbol{\kappa} = \mathbf{u} + \mathbf{v}$ where $\mathbf{u}$ follows an iCAR prior and $\mathbf{v}$ follows an independent prior.  With an iCAR prior on $u_i$, there is no proper distribution for $\mathbf{u}$. However, adding $\mathbf{v}$, $\boldsymbol{\kappa}$ has a proper distribution, i.e., $\boldsymbol{\kappa} \sim N\left(\boldsymbol{0}, \sigma^2 (\mathbf{D} - \mathbf{W})^{-} + \tau^{2}\mathbf{I_A}\right) = N\left(\boldsymbol{0}, \sigma^2 \left((\mathbf{D} - \mathbf{W})^{-} + \frac{\tau^{2}}{\sigma^2}\mathbf{I_A}\right)\right) = N\left(\boldsymbol{0}, \sigma^2 \left((\mathbf{D} - \mathbf{W})^{-} + \nu \mathbf{I_A}\right)\right)$ where $\nu \equiv \frac{\tau^2}{\sigma^2}$.  Thus, the structure matrix is $((\mathbf{D} - \mathbf{W})^{-} + \nu \mathbf{I_A}))^{-1} $.  Therefore, the conditional variance is $\frac{\sigma^{2}}{((\mathbf{D} - \mathbf{W})^{-} + \nu \mathbf{I_A})_{ii}^{-1}}$.  Again, the smaller the $\sigma^2$, the more smoothing. Further, it seems that $\sigma^2$ and $\nu$ behave reciprocally.  Increasing $\nu$, i.e., $\tau^2$ relative to $\sigma^2$ will decrease smoothing, as intuition suggests. We have no closed form but $\textrm{TCV}=\sum_{i}\frac{\sigma^{2}}{((\mathbf{D} - \mathbf{W})^{-} + \nu \mathbf{I_A})_{ii}^{-1}}$ can be calculated in the model fitting. 
		
		\item p-CAR: $\mathbf{Q} = \mathbf{D} - \eta\mathbf{W}$, where $\eta$ is the spatial dependence parameter and $1/\epsilon_{min} < \eta < 1/\epsilon_{max}$ defines a proper distribution \citep[see][for more details]{jin2007}. $\epsilon_{min}$ and $\epsilon_{max}$ are the minimum and maximum eigenvalues of $\mathbf{D}^{-1/2}\mathbf{W}\mathbf{D}^{-1/2}$. For this prior, $\frac{\sigma^{2}}{Q_{ii}} = \frac{\sigma^{2}}{w_{i}^+}$, then $\textrm{TCV} = \sum_{i}\frac{\sigma^{2}}{w_{i}^+}$. As with iCAR, smoothing only depends upon $\sigma^2$; $\eta$ is not expected to have any impact. 
		
		\item LCAR: $\mathbf{Q} = \lambda \mathbf{R} + (1-\lambda)\mathbf{I_A}$, where $\mathbf{R} = \mathbf{D} - \mathbf{W}$ and $\lambda\in[0,1]$ is a spatial dependence parameter. $\lambda$ weights the structured and unstructured spatial components in $\mathbf{Q}$. This specification yields the independence case if $\lambda = 0$, and iCAR if $\lambda = 1$. For the LCAR prior, var$(r_{i}|\{ r_{j}, j \neq i\}) = \frac{\sigma^{2}}{Q_{ii}}= \frac{\sigma^{2}}{\lambda w_{i+} + (1-\lambda)} = \frac{\sigma^{2}}{\lambda(w_{i+} -1) +1}$. Therefore, $\textrm{TCV} = \sum_{i}\frac{\sigma^{2}}{\lambda(w_{i+} -1) +1}$. Again, we see that as $\sigma^2$ decreases we have more smoothing. Conversely, as $\lambda$ increases, we also observe more smoothing.  Further, we see directly the relationship between $\sigma^2$ and $\lambda$. To maintain similar expected smoothing, if one goes up, the other must go down.
		
		\item BYM2: The structure matrix is an adaptation of the reparameterization of the BYM model introduced by \cite{dean2001}. It is defined as $\mathbf{Q} = \left(\lambda\mathbf{R_*}^- + (1-\lambda)\mathbf{I_A}\right)^-$, where $\mathbf{R_*}$ denotes the spatial neighborhood matrix from the iCAR model, $\mathbf{R} = \mathbf{D} - \mathbf{W}$, scaled according to the geometric mean of the marginal variance \citep[see][for further details]{riebler2016} and $\lambda\in[0,1]$. The BYM2 spatial prior has a scaled spatially structured component and its variance matrix represents a weighted average of the variances of the structured and unstructured components, which facilitates the interpretation of the model.  In this case the conditional variance is $\frac{\sigma^{2}}{Q_{ii}}= \frac{\sigma^{2}}{(\lambda \mathbf{R_*}^- + (1- \lambda) \mathbf{I_A})_{ii}^{-1}}$ and $\textrm{TCV} = \sum_{i} \frac{\sigma^{2}}{(\lambda \mathbf{R_*}^- + (1- \lambda) \mathbf{I_A})_{ii}^{-1}}$.   We can not obtain the inverse explicitly but we can calculate it for any example.  The behavior in $\sigma^2$ is as above.
	\end{itemize}

	Finally, we consider a spatial prior based on a geostatistical model. A geostatistical model operates at point level and, in simplest form, specifies dependence between locations as a decreasing function of distance between them. The distributional specification is customarily through a Gaussian process (GP) with a suitable choice of covariance function. 
	Since we are analyzing discrete spatial data, i.e., a finite number of areal units, to use a GP, we supply a finite set of locations as the centroids of each of the areal units within the study area. Then, the joint probability of distribution of $\kappa_1,\dots, \kappa_A$ is multivariate Normal, i.e, $\boldsymbol{\kappa} \sim N(\mathbf{0}, \sigma^{2}\mathbf{R(\theta)})$ where each element of the covariance matrix $\mathbf{R(\theta)}$ is specified by a correlation function $\rho(s_i - s_j; \boldsymbol{\theta})$. 
	We illustrate with an exponential choice, $\rho(s_i - s_j; \boldsymbol{\theta}) = e^{-||s_i - s_j||/\psi}$,  
	where $\mid\mid s_i - s_j \mid\mid$ is the Euclidean distance between centroids $s_i$ and $s_j$ and $\psi$ is a range parameter. That is, for a given distance, the  larger the value of $\psi$ the stronger the spatial correlation.
	
	\begin{table}[b!]
		\centering
		\caption{Total conditional variance (TCV) for the seven spatial priors analyzed in this work.\label{tab5.0}}
		\begin{tabular}{l|ccc }
			\toprule
			\bf Prior & \bf Explicit expression& \bf TCV & \bf Parameters\\ 
			\midrule
			indep (iid) & \checkmark & $\sigma^2A$ & $\sigma^2$\\[0.25cm]
			GP & & $\sum_{i} \frac{\sigma^{2}}{Q_{ii}}$ &$\sigma^2$, $\psi$\\[0.25cm]
			iCAR &\checkmark& $\sum_{i} \frac{\sigma^2}{w_i^+}$ & $\sigma^2$\\[0.25cm]
			BYM & &$\sum_{i}\frac{\sigma^{2}}{((\mathbf{D} - \mathbf{W})^{-} + \nu \mathbf{I_A})_{ii}^{-1}}$ & $\sigma^2$, $\nu = \frac{\tau^2}{\sigma^2}$\\[0.25cm]
			pCAR &\checkmark &$\sum_{i} \frac{\sigma^2}{w_i^+}$  & $\sigma^2$\\[0.25cm]
			LCAR & \checkmark&$\sum_{i}\frac{\sigma^{2}}{\lambda(w_{i+} -1) +1}$& $\sigma^2$, $\lambda$\\[0.25cm]
			BYM2 & &$\sum_{i} \frac{\sigma^{2}}{(\lambda \mathbf{R_*}^{-} + (1- \lambda) \mathbf{I_A})_{ii}^{-1}}$& $\sigma^2$, $\lambda$\\
			\bottomrule
		\end{tabular}
	\end{table}	
	
	The conditional variances for the GP will be var$(r_{i}|\{r_{j}, j \neq i \}) = \frac{\sigma^{2}}{Q_{ii}}$ and $\textrm{TCV} = \sum_{i} \frac{\sigma^{2}}{Q_{ii}}$. Behavior in $\sigma^{2}$  will be as above. However, the conditional variance will depend on the set of intersite distances so is not available analytically (but, of course, these variances can be investigated in the model fitting).  If we consider the case of just $A=2$ sites with an exponential correlation function, the conditional variance is $\sigma^{2}(1- e^{-2 \mid\mid s_i - s_j\mid\mid/\psi})$. The suggestion is that as $\psi$ decreases, we get closer to independence, that is, we expect less smoothing.  So, in our study below we consider the effect of both $\sigma^2$ and $\psi$ with regard to smoothing. The TCVs for the different spatial priors defined in this section are summarized in \autoref{tab5.0}.

	\subsubsection{Model Fitting}
	Model fitting and inference is accomplished through the NIMBLE software \citep{deValpine2017}. NIMBLE is an algorithm library that provides MCMC in \texttt{R} \citep{deValpine2023}. Three MCMC chains were run for each model, each with 30000 iterations, discarding the first 5000 as burn-in. One out of every 75 iterations is saved leading a total of 999 iterations for inference. Convergence is assessed through the Gelman-Rubin method and the effective sample size which are implemented in \texttt{R} package \texttt{coda}. Graphical checks of chains and their autocorrelations were performed to assess convergence. 
	
	To complete model specification, the following hyperpriors were used. Usual choices for the intercept $\alpha$ and standard deviations $\sigma$ are flat priors and vague uniform priors, respectively. These choices are maintained for all priors defined above. The specific hyperparameter distributions defined for each prior are as follows: the spatial dependence parameter $\eta \sim U(-1,1) $ (following instructions of NIMBLE user manual \citep{deValpine2024}); the spatial dependence parameter $\lambda \sim U(0,1) $; and for the range parameter $\psi$ of the exponential correlation function, we employ uniform priors with support allowing ranges up to the maximum interpoint distance over the region \citep{wang2014}.

	\subsection{Empirical smoothing \label{S:empirical}}
	
	To examine whether model smoothing comparison using theoretical metrics aligns with that arising empirically and to empirically quantify the extent of smoothing achieved with each prior,
	we introduce several empirical measures of smoothness. With $\hat{r}_{i} = O_i/n_i$, we consider either a direct mean square smoothness (MSS) criterion, or relative mean square smoothness (RMSS) criterion, namely,
	\begin{eqnarray*}
		MSS &=& \sum_{i}(E(r_i|O_i) - \hat{r}_i)^{2}\\
		RMSS &=& \sum_{i} \frac{(E(r_i|O_i) - \hat{r}_i)^{2}}{E(r_i|O_i)}.
	\end{eqnarray*}
	
	To be more precise, in the simulation study, under a given prior $M$, suppose we draw $B$ replicate sets of observations, $\{O_{i}^{b}, i=1,2,...,A, b=1,2,...,B\}$.  For replicate $b$, we calculate the MSS criterion above, comparing posterior mean under the prior $M$ with $\hat{r}_{i}$. Call it $MSS^{M}_{b}$ and average over $b$ to obtain $E(MSS^{M})$ under prior model $M$. That is,
	\begin{eqnarray*}
		E(MSS^{M}) = \frac{1}{B} \sum_{b=1}^{B} MSS_b^M = \frac{1}{B} \sum_{b=1}^{B} \sum_{i=1}^{A}(r^{M,b}_i - \hat{r}^b_i)^{2},
	\end{eqnarray*}
	where $\hat{r}^b_i = O^b_i/n_i$ and $r^{M,b}_i$ is the posterior mean of the estimated rate under the prior model $M$ for area $i$ and replicate $b$.
	We do an analogous calculation to obtain $E(RMSS^{M})$. 
	\begin{eqnarray*}
		E(RMSS^{M}) = \frac{1}{B} \sum_{b=1}^{B} RMSS_b^M = \frac{1}{B} \sum_{b=1}^{B} \sum_{i = 1}^{A} \frac{\left(r^{M,b}_i - \hat{r}^b_i\right)^2}{r^{M,b}_i}.
	\end{eqnarray*}
	
	In addition, we obtain the resulting maximum empirical smoothing ($maxMSS$) and the maximum empirical relative smoothing ($maxRMSS$). Similar to the criteria above, we can calculate these quantities for replicate $b$ and prior $M$, and average over the replicates to obtain the expected maximum smoothing and relative smoothing associated with prior $M$.
	
	To further facilitate comparison of spatial priors, spatial structures, or disaggregation levels under a given spatial prior, we propose using proportions. As previously mentioned, we view the maximum smoothing as occurring when every area has the same intercept. Thus, the smoothing measure is \emph{conceptually} bounded as follows:
	\begin{equation*}
		0\leq \sum_{i}\left[E\left(r_i|O_i\right) - \hat{r}_i\right]^2 \leq \sum_{i}\left(\bar{r} - \hat{r}_i\right)^2
	\end{equation*}
	where the lower bound corresponds to no smoothing and the upper bound represents our conceptual maximum smoothing, with $\bar{r}$ being the mean of the rates. To quantify the degree of smoothing, we compute the following smoothing proportion (SP):
	\begin{equation*}
		SP = \frac{\sum_{i}\left[E\left(r_i|O_i\right) - \hat{r}_i\right]^2}{\sum_{i}\left(\bar{r} - \hat{r}_i\right)^2}.
	\end{equation*}
	
	\section{The Poisson-Gamma model\label{S:Po-Gamma}}
	
	The Poisson-Gamma model yields explicit expressions for the smoothing, allowing for theoretical assessment regarding how and how much smoothing is induced. Therefore, this model serves as a useful \emph{baseline} for clarification/illumination/connection of the TCV metric in \autoref{S:Po-logitN} as well for computing the empirical smoothing metrics in \autoref{S:empirical}. 
	
	Recall that the Poisson-Gamma model presumes $O_i |\eta_i \sim Poisson(E_i\eta_i)$ where $E_i$ is the expected number of cases, usually under internal standardization, and $\eta_i$ is the relative risk for areal unit $i$. For more discussion about expected number of cases and relative risks refer to \citet[Chapter 6.4,][]{banerjee2015}. Then, it presumes i.i.d. $\eta_i\sim Gamma(a_\eta, b_\eta)$ with scale parameter $a_\eta$ and shape parameter $b_\eta$, i.e., with mean $\mu_\eta = a_\eta/b_\eta$ and variance $\sigma^2_\eta = a_\eta/b_\eta^2$. 
	The posterior distribution of $\eta_i$ in the Poisson-Gamma emerges in closed form as $Gamma(a_\eta + O_i, b_\eta + E_i)$. Thus, considering that $a_\eta = \mu^2_\eta/\sigma^2_\eta $ and $b_\eta = \mu_\eta/\sigma^2_\eta $, the posterior mean for areal unit $i$ is
	\begin{eqnarray*}
		E\left(\eta_i \mid O_i\right) &=& \frac{a_\eta  + O_i}{b_\eta  + E_i} = \frac{\frac{\mu^2_\eta }{\sigma^2_\eta } + O_i}{\frac{\mu_\eta }{\sigma^2_\eta } + E_i} 
		= \frac{\frac{\mu_\eta }{\sigma^2_\eta }}{\frac{\mu_\eta }{\sigma^2_\eta } + E_i} \mu_\eta  + \frac{E_i}{\frac{\mu_\eta }{\sigma^2_\eta } + E_i}\frac{O_i}{E_i}\\ 
		& =&  (1-w_{\eta,i})\mu_\eta  +  w_{\eta,i}\frac{O_{i}}{E_{i}}
	\end{eqnarray*}
	where $w_{\eta,i} = \frac{E_i}{\frac{\mu_{\eta}}{\sigma^2_\eta}  + E_i}$. For this non-spatial case, the (Bayesian) smoothing is toward the mean.
	
	As we work with $O_i|r_i \sim Poisson(n_ir_i)$, we have to modify the relative risk smoothing to rate smoothing in the Poisson-Gamma model. Since $E_i\eta_i= n_ir_i$, we obtain $r_i = \frac{E_i}{n_i}\eta_i$. For convenience, internal standardizing sets $E_i = n_i\frac{\sum_{i} O_i}{\sum_{i}n_i}$, so $\frac{E_i}{n_i}=\frac{\sum_{i} O_i}{\sum_{i}n_i}$, that is, constant over $i$. Let's call it $\bar{r}$, an average rate for the map.  With $\eta_{i}$ having the Gamma distribution above, $E(r_i) = E(\bar{r}\eta_{i})= \bar{r}\mu_{\eta} \equiv \mu_{r}$ and $var(r_i) = var(\bar{r}\eta_{i}) = \bar{r}^{2} \sigma_{\eta}^{2} \equiv \sigma_{r}^{2}$.
	Next, rewriting $w_{\eta,i} = \frac{E_i}{\frac{\mu_\eta}{\sigma^2_\eta}  + E_i}$ in terms of $n_i$, we obtain $w_{\eta,i} = \frac{n_i}{\frac{\mu_r}{\sigma^2_r}  + n_i}$. Further, the posterior mean for areal unit $i$ in terms of rates is
	\begin{eqnarray*}
		E\left(r_i \mid O_i\right) = \bar{r}E\left(\eta_i \mid O_i\right) = \bar{r}\left( \left(1-w_{\eta,i}\right)\mu_\eta  +  w_{\eta,i}\frac{O_{i}}{E_{i}}\right) = \left(1-w_{\eta,i}\right)\mu_r  +  w_{\eta,i}\frac{O_{i}}{n_{i}}.
	\end{eqnarray*}
	As before, 
	the smoothing is Bayesian toward the mean. As a result, the smaller $w_{\eta,i}$, the less weight is placed on the MLE, the more smoothing is expected. Given that $w_{\eta,i} = \frac{n_i}{\frac{\mu_r}{\sigma^2_r}  + n_i}$, when $\mu_r$ is fixed, a smaller variance $\sigma^2_r$ implies more smoothing. Further, as expected, we see that the amount of smoothing depends on the population sizes $n_{i}$.  In particular, larger values of $n_{i}$ will yield less smoothing. As we noted above, we view the $\{n_i\}$ as fixed/given here so we only investigate smoothing with regard to $\sigma^2$ for a given $\mu$.
	
	For the no smoothing case, we have $\hat{r}_{i}= O_{i}/n_{i}$. The difference between the posterior mean smoothing for areal unit $i$ and $\hat{r}_{i}$ is
	\begin{equation}\label{Eq1}
		E\left(r_i \mid O_i\right)-\hat{r}_{i} = \left(\left(1-w_{\eta,i}\right)\mu_r  +  w_{\eta,i}\frac{O_{i}}{n_{i}}\right) - \hat{r}_{i} = \frac{\mu_r}{\sigma^2_r n_i + \mu_r}\left(\mu_r - \hat{r}_{i}\right).
	\end{equation}
	So, we can obtain explicitly the value of any of our proposed empirical metrics.  We can immediately see that, as $\sigma_{r}^{2} \rightarrow \infty$, no smoothing results, all of our metrics are equal to $0$.  Consequently, what is of interest for a given map, is the behavior of the smoothing as $\sigma_{r}^{2} \rightarrow \infty$ (given $\mu_{r}$). Here, at $\sigma_{r}^{2}=0$, \autoref{Eq1}  becomes $\mu_{r} - \hat{r}_{i}$, providing the maximum smoothing.  With a set of $\{O_i\}$, we can quantify the empirical smoothing associated with the map. 
	
	
	
	In this regard, we can offer a simple simulation study to illustrate the quantification of the empirical smoothing associated with a given map across fixed $\mu_{\eta}$ and $\sigma_{\eta}^{2}$. Taking the expected number of cases and population data from the real dataset of Spain examined in \autoref{S:illustration}, we can generate replicate $B$ sets, $\{O_{ib}, b=1,2,...B\}$, and use them to calculate the proposed empirical metrics, using  \autoref{Eq1}.  We note that while \autoref{Eq1} is expressed in terms of  $\mu_{r}$ and $\sigma_{r}^{2}$ , the Poisson-Gamma model is defined for risks. Consequently, in the simulation study we fixed  $\mu_{\eta}$ and $\sigma_{\eta}^{2}$. To align with this, \autoref{Eq1} can be re-expressed as
	\begin{eqnarray*}
		E\left(r_i \mid O_i\right)-\hat{r}_{i} =  \frac{\mu_r}{\sigma^2_r n_i + \mu_r}\left(\mu_r - \hat{r}_{i}\right) = \frac{\mu_\eta}{\bar{r}\sigma^2_\eta n_i + \mu_\eta}\left(\bar{r}\mu_\eta - \hat{r}_{i}\right).
	\end{eqnarray*}
	This study is presented in \autoref{s:A} and corroborates our theoretical discussion. Specifically, it is seen that as the value of $\sigma^2_\eta$ increases, the amount of smoothing decreases. This decay in smoothing is similar across the values of $\mu_\eta$.  The orange dotted lines show the theoretical criteria values at $\sigma^2_\eta=0$.

	\section{Simulation studies\label{S:SS}}
	Here, we present the results of two simulation studies to investigate the extent of smoothing for the priors in \autoref{S:Po-logitN}. The first simulation study, outlined in \autoref{S:SS1}, seeks to evaluate the theoretical extent of smoothing and the factors that influence the expected level of smoothing for each prior through the TCV. Additionally, it evaluates the empirical smoothing achieved by each prior. Thus, this simulation study focuses on comparing the smoothing extent within priors. In \autoref{S:SS2} we compare the extent of smoothing across priors. This simulation study mirrors real-world situations where we encounter actual data frames and we consider choice of smoothing model to explain the variation in rates.
	
	\subsection{Within prior simulation study \label{S:SS1}}
	
	Here, we consider the seven priors proposed in \autoref{S:Po-logitN} individually: an independent (non)spatial prior, the five different CAR priors, and the GP prior using a covariance function employing the distance between centroids. More precisely, we compare theoretical smoothing with empirical smoothing to assess whether the model's smoothing, as evaluated using the TCV metric aligns with that arising through the empirical metrics. 
	To accomplish our objective, we employed Scenario~2, as described in \autoref{S:SS2} focusing on the spatial structure of peninsular Spain with varying levels of disaggregation. Specifically, we present results for a disaggregation level of 
	$A=47$, corresponding to the provinces of peninsular Spain, and then increase the number of areas to $A=100$ and $A=300$. It is worth noting that similar studies were conducted in other scenarios, yielding consistent conclusions (but omitted here). It is also worth noting, and is intuitive, that increased disaggregation results in less smooth maps, hence increased prior smoothing.
	
	For each prior, we vary the values of the parameters affecting the smoothing, as discussed in \autoref{S:Po-logitN}. The results for iCAR and LCAR priors are presented in this section, while the results for the GP and BYM priors are detailed in \autoref{s:B}. Results for the remaining priors are omitted, as their conclusions align with those drawn from the spatial priors discussed.

	We start with the iCAR prior. The amount of smoothing induced by this prior is influenced only by the value of $\sigma^2$, with smoothing decreasing as $\sigma^2$ increases (refer to \autoref{S:Po-logitN}). Five different options are considered for $\sigma^2$, ranging from $10^{-4}$ to $0.25$. \autoref{tab1} presents the empirical smoothing criteria values considered in this work (average of MSS and RMSS, $maxMSS$, $maxRMSS$ and the SP), alongside the theoretical smoothing criteria (TCV). Results indicate that theoretical smoothing aligns with empirical metrics, with smoothing decreasing as $\sigma^2$ increases. Additionally, empirical smoothing remains nearly constant for very small values of $\sigma^2$, while the theoretical smoothing criterion approaches zero. This suggests we have reached maximum smoothing, similar to the baseline case of the Poisson-Gamma model (see \autoref{S:Po-Gamma}).
	Conversely, as $\sigma^2$ values increase, smoothing continues to decrease, but the pace of decrease slows. In contrast, the theoretical measure consistently increases. 
	Across different numbers of areas, higher empirical values (MSS, RMSS, and maximum values) observed with $A=300$ result in lower smoothing and lower SP values compared to 
	$A=47$, complicating comparisons among spatial structures. Additionally, for a TCV value of zero, the SP is lower for $A=300$ than for $A=47$ indicating that the maximum smoothing decreases as the number of areas increases. Moreover, a faster increase in TCV is observed with a greater number of areas, suggesting that smoothing approaches an asymptote more quickly as the spatial disaggregation increases.
	
	\begin{table}[b!]
		\centering
		\caption{Empirical smoothing criteria, the average values of MSS, RMSS,the maximum values, $maxMSS$ and $maxRMSS$, and smoothing proportion together with the theoretical smoothing metric, TCV, for the different parameter values of the iCAR priors.\label{tab1}} 
			\begin{tabular}{c|rrrrrr}
				\hline
				parameters & TCV & SP & MSS & RMSS & $maxMSS$ & $maxRMSS$ \\ 
				\hline
				$\sigma^2$&&&&&&\\
				\hline
				\multicolumn{7}{l}{\bf A = 47}\\
				$10^{-4}$ & 0.001 & 0.885 & 1077.924 & 10.278 & 8630.649 & 81.235 \\ 
				0.0025 & 0.029  & 0.294  & 357.542 & 3.292 & 3775.424 & 27.436\\ 
				0.0081 & 0.093 & 0.101 & 122.749 & 1.146 & 1128.705 & 7.349  \\
				0.04 & 0.461 &  0.021 &  25.117 & 0.253 & 149.187 & 1.812  \\ 
				0.25 &  2.883 & 0.013 &  15.605 & 0.160 & 50.161 & 0.690 \\ 
				\multicolumn{7}{l}{ }\\
				\multicolumn{7}{l}{\bf A = 300}\\
				$10^{-4}$ & 0.006  & 0.122  & 1390.674 & 13.277 & 8654.513 & 79.023 \\ 
				0.0025  & 0.149 & 0.056  & 639.943 & 6.372 & 6868.727 & 63.323\\ 
				0.0081 & 0.483 & 0.039  & 445.332 & 4.571 & 5532.243 & 52.831\\ 
				0.04 & 2.387 & 0.026 & 299.177 & 3.320 & 3756.458 & 41.699\\ 
				0.25  & 14.920  & 0.021 & 235.699 & 2.806 & 2432.384 & 32.641 \\ 
				\hline
			\end{tabular} 
	\end{table}

	\begin{table}[!t]
		\centering
		\caption{Empirical smoothing criteria, the average values of MSS, RMSS, the maximum values, $maxMSS$ and $maxRMSS$, and the smoothing proportion (SP), together with the theoretical smoothing metric (TCV) for the different parameter values of the LCAR priors.\label{tab2}} 
		\resizebox{0.83\textwidth}{!}{
			\begin{tabular}{cc|rrrrrr}
				\hline
				\multicolumn{2}{c|}{parameters}  & TCV & SP & MSS & RMSS & $maxMSS$ & $maxRMSS$ \\ 
				\hline
				$\sigma^2$ & $\lambda$ &&&&&&\\
				\hline
				\multicolumn{8}{l }{\bf A=47}\\ 
				$10^{-4}$  &  0.1  & 0.003  & 0.835 &1016.718 & 9.722 & 7935.193 & 72.096 \\ 
				&0.5  & 0.002 & 0.870 & 1058.832 & 10.114 & 8314.968 & 77.020 \\ 
				&0.9 & 0.001 & 0.883  & 1075.035 & 10.255 & 8550.978 & 80.161 \\ 
				\multicolumn{8}{c }{ }\\ 
				0.0025  &  0.1 & 0.087 & 0.142  & 173.387 & 1.664 & 1247.626 & 9.168  \\ 
				& 0.5 & 0.044 & 0.219 & 266.848 & 2.477 & 2561.572 & 17.316  \\ 
				&0.9 &  0.030 & 0.280 & 340.966 & 3.142 & 3560.948 & 25.552 \\
				\multicolumn{8}{c }{ }\\ 
				0.0081  &  0.1  & 0.281 & 0.039 & 47.908 & 0.490 & 342.899 & 4.010 \\ 
				&0.5 & 0.144 & 0.065 & 79.681 & 0.769 & 609.584 & 5.033  \\ 
				&0.9 & 0.098 & 0.094 & 114.048 & 1.070 & 1020.058 & 6.781 \\
				\multicolumn{8}{c }{ }\\ 
				0.04  &  0.1  & 1.387 & 0.014  & 17.037 & 0.178 & 83.085 & 1.141\\ 
				&0.5 & 0.711 & 0.016 & 19.620 & 0.201 & 105.762 & 1.370  \\ 
				&0.9 & 0.486 & 0.019 & 23.676 & 0.240 & 139.651 & 1.726\\ 
				\multicolumn{8}{c }{ }\\ 
				0.25  &  0.1 & 8.671 & 0.014   & 16.969 & 0.271 & 118.981 & 5.775 \\ 
				&0.5 & 4.446 & 0.013 & 16.139 & 0.201 & 89.375 & 2.612 \\ 
				&0.9 & 3.038 & 0.013 & 15.715 & 0.177 & 73.380 & 1.595  \\ 
				\multicolumn{8}{c }{ }\\ 
				\multicolumn{8}{l }{\bf A=300 }\\ 
				$10^{-4}$  &  0.1  & 0.021 & 0.136 & 1552.462 & 14.609 & 8816.281 & 82.139 \\ 
				&0.5 & 0.010 & 0.135 & 1533.572 & 14.438 & 8756.314 & 81.193 \\ 
				&0.9 &  0.006 & 0.130 & 1483.746 & 14.034 & 8682.165 & 80.145\\ 
				\multicolumn{8}{c }{ }\\ 
				0.0025  &  0.1 & 0.524 & 0.080  &  911.212 & 9.200 & 8173.991 & 76.906 \\ 
				& 0.5 & 0.245 & 0.070  &  804.111 & 8.103 & 7578.713 & 70.156 \\ 
				&0.9  & 0.162 & 0.061 & 696.449 & 6.964 & 7105.241 & 65.484 \\ 
				\multicolumn{8}{c }{ }\\ 
				0.0081  &  0.1  & 1.697 & 0.057  & 644.881 & 6.486 & 7003.704 & 64.969\\
				&0.5  & 0.794 & 0.048  & 550.793 & 5.556 & 6280.349 & 58.458\\ 
				&0.9 & 0.524 & 0.042  & 479.849 & 4.864 & 5759.609 & 54.421 \\ 
				\multicolumn{8}{c }{ }\\ 
				0.04  &  0.1 & 8.380 & 0.033  & 378.600 & 3.874 & 4167.480 & 42.132 \\ 
				&0.5 & 3.920 & 0.030  & 337.976 & 3.561 & 4047.001 & 42.015 \\ 
				&0.9 & 2.588 & 0.028  & 315.952 & 3.412 & 3922.069 & 41.976 \\ 
				\multicolumn{8}{c }{ }\\ 
				0.25  &  0.1 & 52.374 & 0.027 &305.599 & 3.285 & 2896.865 & 33.383  \\ 
				&0.5  & 24.497 & 0.024&  270.367 & 2.998 & 2721.260 & 33.064 \\ 
				&0.9 & 16.176 & 0.023 & 258.903 & 2.937 & 2734.291 & 34.104  \\ 
				\hline
		\end{tabular} }
	\end{table}
	
	For the LCAR prior, the level of smoothing induced is influenced by the values of $\sigma^2$ and the spatial dependence parameter $\lambda$. Therefore, in this simulation study we vary the $\sigma^2$ and $\lambda$ values. Specifically, we considered three values for $\lambda$: $0.1$, $0.5$, and $0.9$ with a full range of $\sigma^2$ values, from very small to very large. We present results for $\sigma^2 = 10^{-4}, 0.0025, 0.0081, 0.04, 0.25$, to assess whether theoretical smoothing aligns with that determined through empirical metrics. Recall that, theoretically, as $\sigma^2$ increases, the expected smoothing decreases, whereas as  $\lambda$ increases, the smoothing increases. \autoref{tab2} shows the empirical smoothing criteria, the average values of MSS and RMSS, the maximum values, $maxMSS$ and $maxRMSS$, and the smoothing proportion, together with the TCV for the different parameter values. 
	The results show a strong alignment between theoretical and empirical metrics across both levels of spatial disaggregation. For $\sigma^2 = 10^{-4}$, empirical smoothing remains nearly constant regardless of changes in the $\lambda$ parameter, while the theoretical smoothing metric approaches zero, indicating that maximum smoothing has been achieved. In contrast, as $\sigma^2$ increases, the $\lambda$ parameter still affects the amount of smoothing, but the effect is almost negligible, whereas the TCV increases. The smoothing tends to an asymptote.
	Similar to the iCAR prior, the results indicate that the maximum smoothing decreases as the number of areas increases, and that smoothing reaches its asymptote more quickly as spatial disaggregation increases. Furthermore, higher TCV values are associated with lower SP values as the number of areas increases.
	
	The priors discussed here, iCAR and LCAR, are neighbor-based priors that provide explicit expressions for the TCV. To assess the robustness of the theoretical metrics across different spatial priors, we also include results for the GP and BYM priors. The GP prior is not neighbor-based, while the BYM prior lacks an explicit expression for the TCV. Detailed results are provided in \autoref{s:B}. The findings for the GP and BYM priors consistently demonstrate that theoretical smoothing metrics align closely with empirical observations. As the $\sigma^2$ parameter decreases, smoothing increases, eventually reaching a maximum, and asymptotically approaches its lower limit as $\sigma^2$ continues to decline.  Furthermore, the effect of increasing the number of areas aligns with the patterns observed for the iCAR and LCAR spatial priors.
	
	\subsection{Across priors simulation study \label{S:SS2}}
	
	To compare the amount of smoothing across priors, we delineate various scenarios by varying spatial correlation and rate variation. The definition and results of the scenarios varying the rate variation are shown in \autoref{s:C}. To compare the smoothing across priors we use empirical smoothing metrics. 
	
\begin{figure}[t!]
		\centering
		\includegraphics[page = 2, width = 6.5cm]{Fig/Figure1.pdf}
		\includegraphics[page = 1, width = 6.75 cm]{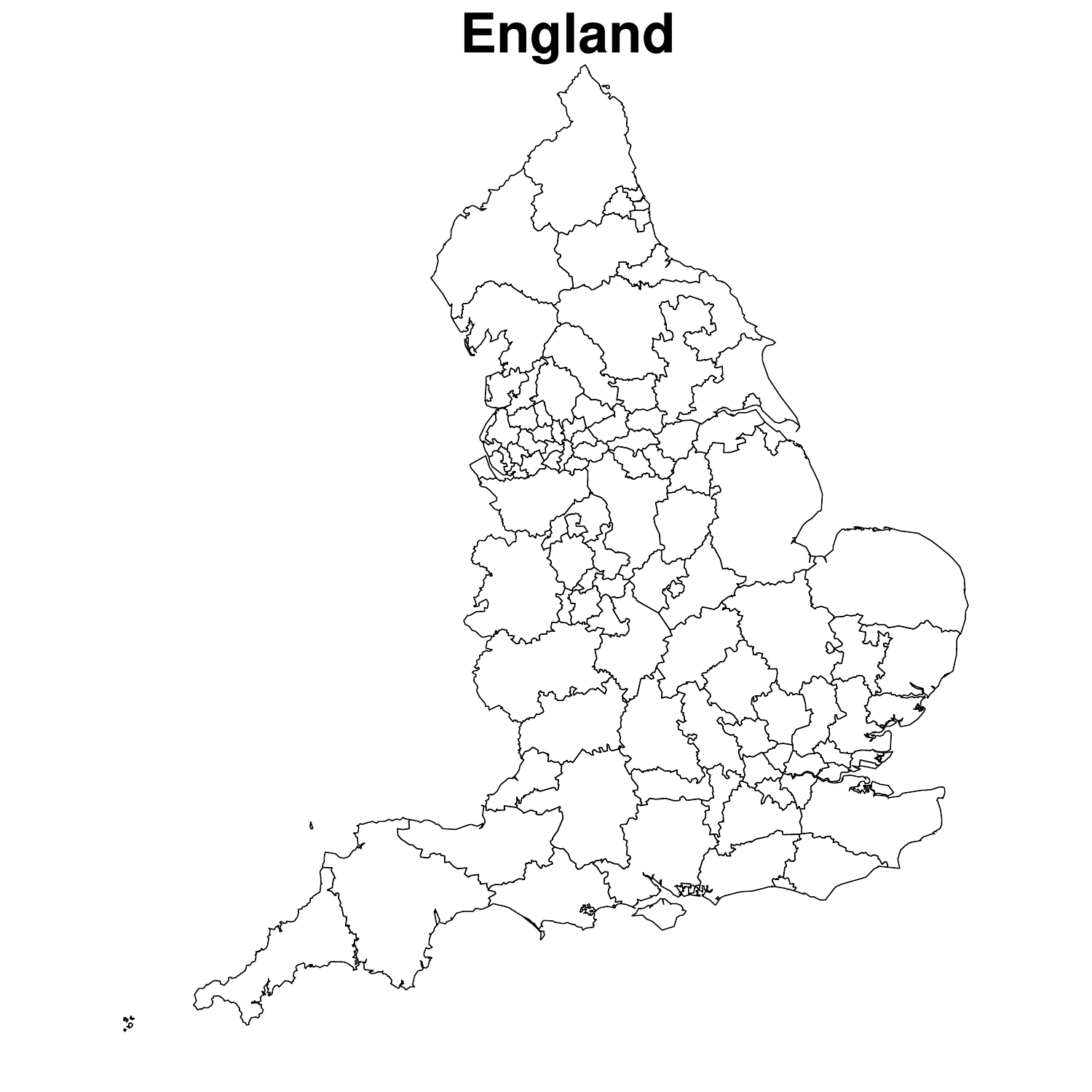}
		\caption{Spatial structures corresponding to the 47 provinces of peninsular Spain (left) and the 106 clinical commissioning groups of England (right).}
		\label{Fig1}
	\end{figure}

	For the simulation study across priors, we analyze two spatial regions and vary the level of spatial disaggregation in one of them. Both regions correspond to the areal units and their associated neighbor structures found in the real data examined in the following section. Specifically, the first spatial region represents peninsular Spain, with three levels of spatial disaggregation: 47 provinces, and 100, and 300 areas. To define the spatial structures for the 100 and 300 areas, we preserve the proportion of municipalities per province as observed in the real case of peninsular Spain. The second spatial structure corresponds to the 106 clinical commissioning groups of England (see \autoref{Fig1}).
	To simulate the true rates based on these sets of areal units, we initially establish a high-resolution grid for each spatial region. Then, we define the rate surface $r(s)$ on the logit scale, since is the link function of the Poisson-logitNormal models (see \autoref{Eq3}), following a GP prior. That is, $logit(r(s)) = \phi(s)$ where $\boldsymbol{\phi} \sim N\left(\boldsymbol{\mu}(\boldsymbol{\theta}), \mathbf{C}(\boldsymbol{\theta}) \right)$. More specifically, to induce spatial variation we consider two or more fixed locations, assuming that the mean $\boldsymbol{\mu}(\boldsymbol{\theta})$ depends on the distance from these locations, declining through an exponential function\footnote{These locations can be imagined as exposure sites.}. The results will vary depending on the selection of the fixed locations; therefore, we propose three different scenarios for the surface. In Scenario 1 we aim to create a south-east to north-west pattern by fixing two nearby locations. In Scenarios 2 and 3, we increase the number of selected locations according to the spatial structure to induce more fluctuation in the surface. 
	
	To smooth the surfaces, we adopt the spatial covariance matrix, $\mathbf{C}(\boldsymbol{\theta}) = \sigma_C^2\mathbf{R}(\boldsymbol{\varphi})$, with $\mathbf{R}(\boldsymbol{\varphi})$ a correlation matrix whose size corresponds to the number of grid points. The  $(i,j)$ elements of $\mathbf{R}(\boldsymbol{\varphi})$ are computed using a Matérn correlation function,
	\begin{eqnarray*}
		\rho(s_i,s_j;\varphi) = \frac{1}{2^{v-1}\Gamma(v)}\left(\mid \mid s_i-s_j\mid \mid \varphi\right)^v K_v\left(\mid \mid s_i-s_j \mid \mid \varphi\right).
	\end{eqnarray*}		
	For scenarios 1 and 2 we set $\sigma_C = 0.1$, the smoothness parameter $v = 2$, and the decay parameter $\varphi = 2$. For Scenario 3, we set $v = 1.25$ to achieve a less smooth surface\footnote{We note that the 
		$\phi(s)$ surface is specified to create random relative risks associated with areal units.  It has no connection with the GP prior in \autoref{S:Po-logitN}.}.
	
	Once we obtain the rate surfaces on the logit scale, we compute the rates for each areal unit of the study region, i.e. $r_i$ for each $i=1,\dots, A$. Specifically, we compute	\\$r_i = \frac{1}{|A_i|}\int_{s \in A_i} \frac{1}{(1+\exp(-w(s)))} ds \approx \frac{1}{H}\sum_{s \in A_i} \frac{1}{(1+\exp(-w(s)))}$ with $|A_i|$ the area of areal unit $A_i$ and $H$ the number of $s$ points in $A_i$.

	The defined scenarios exhibit varying degrees of spatial correlation. Scenario~1 shows high spatial correlation, while Scenario~2 introduces more fluctuation, reducing the spatial correlation. Scenario~3 offers the lowest spatial correlation. The spatial variability of all scenarios is similar, as the rates vary around a similar range. For the spatial distribution of the scenarios, please refer to \autoref{s:C}. 
	Three additional scenarios were established by adjusting the variability of the rates but are not displayed here due to their close resemblance to the results obtained from the aforementioned scenarios. Further details regarding these scenarios, along with the results from Scenario~2, can be found in \autoref{s:C}.
	
	Finally, to generate the observed counts $O_i$ from a Poisson distribution, it is necessary to establish the population. Specifically, the population data used in the simulation study have been taken from the real datasets examined in the following section, for each spatial structure. 
	We generate $B = 1000$ datasets for each scenario defined. 
	
	To compare the extent of smoothing across priors, we fit 
	the Poisson-logitNormal model with independent (non)spatial priors, referred to as iid, along with GP prior and the five CAR priors iCAR, BYM, pCAR, LCAR, and BYM2 as defined in \autoref{S:Po-logitN}, to each scenario. To compare the smoothness of the priors, we compute empirical smoothing metrics — the average values of MSS and RMSS, maximum MSS, maximum RMSS and SP (as detailed in \autoref{S:empirical}) — for each prior model and scenario.

	\begin{table}[b!]
		\caption{Average values of MSS, RMSS, maximum MSS, maximum RMSS and SP values reached by each spatial prior model in Scenario~1 and Scenario~3, across the spatial regions of Spain and England. For Spain, results are presented for spatial disaggregation levels of $A=47$ and $A=100$. \label{tab3}} 
		\resizebox{\textwidth}{!}{
			\begin{tabular}{lrrrrrcrrrrr}
				\hline
				&\multicolumn{5}{c}{Scenario 1}&&\multicolumn{5}{c}{Scenario 3}\\
				\cline{2-6}\cline{8-12}
				& MSS & RMSS & $maxMSS$ & $maxRMSS$ & SP && MSS & RMSS & $maxMSS$ & $maxRMSS$ & SP\\ 
				\hline
				\multicolumn{12}{l}{\textbf{Spain}}\\
				\multicolumn{12}{l}{\hspace*{0.5cm}\textbf{A = 47}}\\
				iid & 11.087 & 0.165 & 46.465 & 0.755 & 0.005 && 14.110 & 0.164 & 67.434 & 0.741 & 0.024 \\ 
				GP & 10.391 & 0.154 & 42.878 & 0.663 & 0.004 && 13.988 & 0.163 & 63.203 & 0.691 & 0.024 \\ 
				iCAR & 11.079 & 0.158 & 40.906 & 0.629 & 0.005 && 14.223 & 0.165 & 61.407 & 0.696 & 0.025 \\ 
				BYM & 11.251 & 0.165 & 45.586 & 0.742 & 0.005 && 14.183 & 0.164 & 68.380 & 0.747 & 0.025 \\ 
				pCAR & 10.452 & 0.154 & 41.479 & 0.646 & 0.004 && 14.015 & 0.163 & 63.008 & 0.693 & 0.024  \\ 
				LCAR & 10.643 & 0.185 & 44.878 & 1.820 & 0.005 && 13.883 & 0.162 & 61.801 & 0.715 & 0.024\\ 
				BYM2 & 11.183 & 0.163 & 44.346 & 0.711 & 0.005 && 14.210 & 0.164 & 67.026 & 0.726 & 0.025\\ 
				\multicolumn{12}{l}{\hspace*{0.5cm}\textbf{A = 100}}\\
				iid & 62.897 & 1.059 & 628.254 & 8.914 & 0.034 && 91.564 & 1.078 & 777.931 & 8.725 & 0.111  \\ 
				GP & 88.166 & 2.253 & 1653.805 & 58.756 & 0.048 && 138.821 & 2.368 & 2357.412 & 57.551 & 0.168  \\  
				iCAR & 55.832 & 0.974 & 798.806 & 13.354 & 0.030 && 93.497 & 1.102 & 1087.225 & 12.155 & 0.113 \\  
				BYM & 63.856 & 1.152 & 703.635 & 10.659 & 0.035 && 93.912 & 1.123 & 817.680 & 10.308 & 0.114 \\  
				pCAR & 55.218 & 0.970 & 790.579 & 13.176 & 0.030 && 92.138 & 1.089 & 1016.933 & 11.514 & 0.111 \\  
				LCAR & 109.278 & 3.263 & 2610.374 & 103.431 & 0.059 && 161.364 & 3.415 & 3497.335 & 116.578 & 0.195 \\  
				BYM2 & 60.038 & 1.041 & 693.155 & 10.243 & 0.033 && 93.924 & 1.102 & 885.888 & 10.150 & 0.114 \\ 
				\multicolumn{12}{l}{ }\\
				\multicolumn{12}{l}{\textbf{England}}\\
				\multicolumn{12}{l}{\hspace*{0.5cm}\textbf{A = 106}}\\
				iid & 110.923 & 1.655 & 705.137 & 5.358 & 0.038 && 140.171 & 1.810 & 522.930 & 11.352 & 0.111 \\ 
				GP & 93.230 & 1.502 & 835.847 & 14.164 & 0.032 && 142.254 & 1.892 & 912.298 & 13.544 & 0.113\\  
				iCAR & 95.771 & 1.480 & 614.975 & 9.880 & 0.033 && 133.768 & 1.690 & 559.085 & 7.151 & 0.106\\  
				BYM & 95.419 & 1.475 & 614.040 & 9.631 & 0.032 && 133.018 & 1.684 & 543.892 & 6.884 & 0.105\\ 
				pCAR & 94.420 & 1.471 & 612.910 & 9.738 & 0.032 && 132.974 & 1.680 & 536.851 & 6.821 & 0.105 \\  
				LCAR & 90.276 & 1.417 & 587.808 & 10.382 & 0.031 &&  124.653 & 1.586 & 523.366 & 7.218 & 0.099 \\ 
				BYM2 & 95.316 & 1.476 & 612.615 & 9.672 & 0.032 && 133.209 & 1.689 & 543.877 & 6.932 & 0.105 \\  
				\hline
		\end{tabular} }
	\end{table}

	\autoref{tab3} presents the empirical smoothing metrics for each prior model, scenario and spatial structure. The results include the spatial structures of 47 provinces and 100 areas in Spain, as well as the spatial structure of England. Metrics for additional spatial structures are provided in \autoref{s:C}.
	Notable, for all spatial structures, Scenario~3 exhibits greater empirical smoothing criteria values. This scenario has greater spatial variability, compared with Scenario~1 which has a clear spatial pattern. 
	Moreover, in comparing the smoothing metric values between spatial structures,  an increase in smoothness is observed as the spatial region is divided into a greater number of areas. Specifically, for the two spatial structures of Spain, the smoothness increases approximately fivefold when the number of areas rises from 47 to 100, with a similar trend observed in both scenarios. This pattern continues as the number of areas increases to 300, with the smoothing increasing threefold from 100 to 300 areas 
	(refer to \autoref{s:C}). Additionally, when comparing the spatial structures of Spain and England, both with approximately 100 areas, England generally exhibits higher MSS and RMSS values, but lower maximum smoothing values and a smaller proportion of smoothing.
	
	When comparing the results for each spatial structure, distinct findings emerge. In the case of the spatial distribution in Spain with 47 areas, all spatial priors exhibit very similar empirical criteria values. Largest disparity is observed for the LCAR prior in Scenario 1, showing similar MSS and $maxMSS$ compared to other priors but higher RMSS and $maxRMSS$, showing the strongest smoothing compared to other priors. 
	In contrast, increasing the number of areas in Spain to 100 leads to greater variability in the criteria values across different spatial priors. In terms of mean values (MSS and RMSS), the iCAR and pCAR priors produce the lowest values, whereas the GP and LCAR priors achieve the highest. For maximum values, the iid prior records the lowest values, while the LCAR prior achieves the highest. Similarly, the smoothing proportion calculations show comparable trends, with the iCAR and pCAR priors having the lowest values and the GP and LCAR priors exhibiting the highest. 
	
	When analyzing the spatial structure in England, the CAR spatial priors exhibit similar empirical criteria values. For mean values (MSS and RMSS), these priors yield the lowest values, with the LCAR prior showing a slight improvement in smoothing. Conversely, the iid prior achieves the highest mean values. Regarding the maximum values, the iid prior records the lowest values, while the GP prior achieves the highest. In terms of proportions, all priors show similar values in Scenario 1, whereas Scenario 3 reveals disparities, with the LCAR prior exhibiting less smoothing. It is worth noting that consistent conclusions arise across all of the analyzed scenarios (some of which are omitted here for brevity).  Additionally, the results show that decreasing the variability of the rates leads to higher smoothing criteria values.

	\section{Data illustration \label{S:illustration}}
	
	We now turn to the analysis of real datasets. Specifically, we investigate the smoothing effects induced by the seven spatial priors proposed in \autoref{S:Po-logitN} to clarify the theoretical and simulation findings for these datasets. As observed in \autoref{S:SS1}, the $\sigma^2$ parameter significantly affects the amount of smoothing induced by the spatial priors and is a common parameter across all priors. Therefore, we first assume different informative prior distributions for the $\sigma^2$ parameter of the proposed spatial priors. We define three informative prior distributions: small-size $Unif (0,0.01)$, medium-size $Unif (0.01,0.16)$, and large-size $Unif (0.16,100)$ prior distributions. Additionally, we include a fourth prior distribution for the $\sigma^2$ parameter, a uniform prior distribution on $(0, 1000)$. 
	
	We use two datasets representing two different spatial regions, as noted in the previous section. The first spatial region corresponds to peninsular Spain, divided into $A=47$ (the real provinces), $A=100$ and $A=300$. The second spatial region relates to 106 clinical commissioning groups of England. Both datasets include counts for lung cancer and the corresponding population at risk, focusing on aggregate data for females in Spain from 2019 to 2021 and for males in England in 2017. The Spanish Statistical Office (INE) provided the dataset for Spain. The cancer mortality data for England has been supplied by the National Cancer Registration and Analysis Service (NCRAS) and population data by the Office for National Statistics (ONS).
	Results for peninsular Spain with $A=47$ and $A=300$ are presented in this section, while results for Spain with $A=100$, along with the England spatial structure, are detailed in \autoref{s:C}.

	\subsection{Lung cancer deaths in Spain}
	
	\begin{figure}[t!]
		\begin{center}
			\includegraphics[width = 6.5cm, page=1]{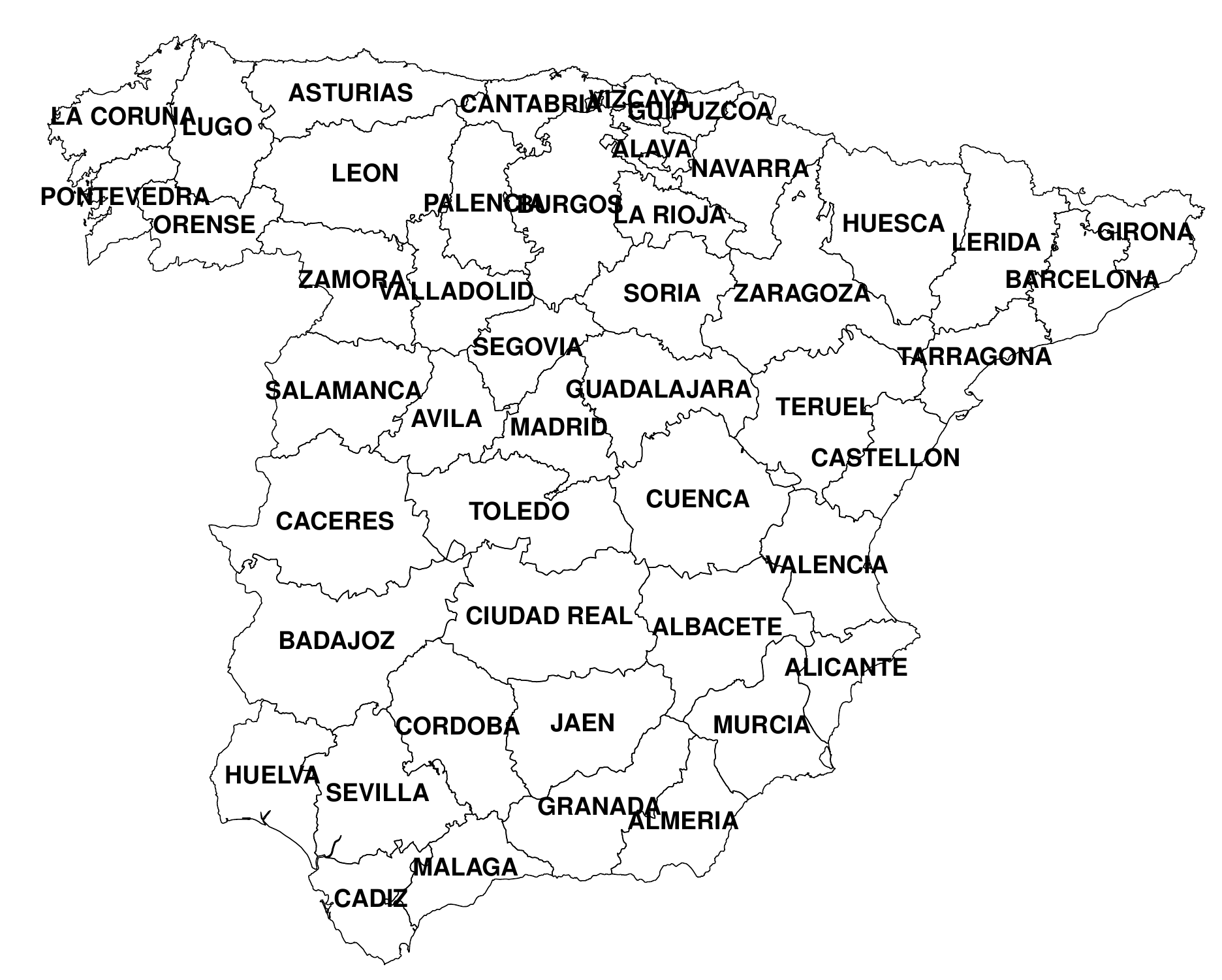} \hspace*{1cm}
			\includegraphics[width = 6.5cm, page=2]{./Fig/Figure2.pdf}
		\end{center}
		\caption{\label{Fig4} Administrative division of peninsular Spain into 47 provinces (left) and $A=300$ (right).}
	\end{figure}   
	
	The data set presents counts for lung cancer in females and the corresponding population at risk, with data aggregated from 2019 to 2021.
	A total of 14,650 deaths from lung cancer in Spain were recorded. The administrative division used for this study ranges from 131,606 to about 10,533,728 inhabitants per unit if we consider the spatial disaggregation of the provinces, and from 1,729 to 6,634,263 inhabitants per unit if we consider $A=300$. The spatial distribution of peninsular Spain for each case is provided in \autoref{Fig4}.

\begin{figure}[b!]
		\begin{center}
			\includegraphics[width = 6.5cm, page = 1]{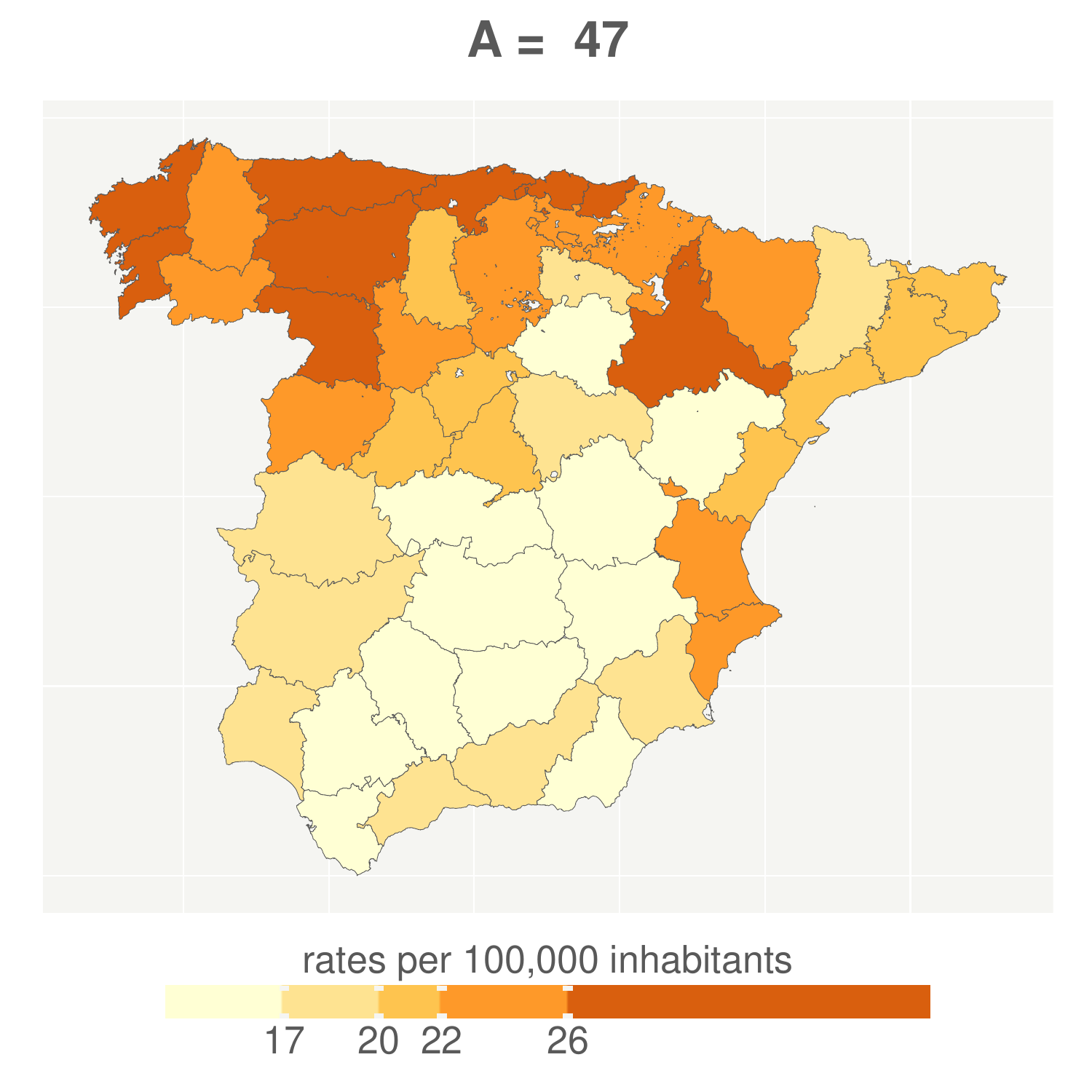} \hspace*{1cm}
			\includegraphics[width = 6.5cm, page = 2]{./Fig/Figure3.pdf}
		\end{center}
		\caption{\label{Fig5} Crude rates per 100,000 inhabitants in peninsular Spain for the administrative division of 47 provinces (left) and $A=300$ (right).}
	\end{figure}

	To provide an initial overview of the disease distribution, crude mortality rates per 100,000 inhabitants across different areas of Spain have been calculated and are displayed in left panel for $A=47$ and right panel for $A=300$ in \autoref{Fig5}. The crude rates range from 13 to 38 deaths per 100,000 inhabitants for $A=47$ and from 0 to 72 for $A=300$. The highest crude mortality rates are predominantly found in the north-western regions of Spain, as well as along the Mediterranean coast (including Tarragona, Castellón, Valencia, and Alicante) and in Zaragoza. In contrast, the central-southern regions, particularly areas south of Madrid, report the lowest mortality rates. This indicates variability in the spatial distribution of mortality rates across Spain.
	
	\begin{table}[t!]
		\centering
		\caption{\label{tab4} Empirical and theoretical expected smoothing for lung cancer dataset in the 47 provinces of Spain.} 
		\resizebox{\textwidth}{!}{
			\begin{tabular}{l|rrrrrrrrrr}
				\hline
				& MSS & RMSS & $maxMSS$ & $maxRMSS$ & TCV & $\sigma^2$ & $\tau^2$ & $\lambda$ & $\psi$ & SP \\ 
				\hline
				\multicolumn{11}{l}{\textbf{Small-size informative prior distribution}}\\
				iid & 5.180 & 0.238 & 23.608 & 1.090 & 0.460 & 0.010 & && & 0.196 \\ 
				GP & 4.853 & 0.213 & 32.867 & 1.486 & 0.002 & 0.010 &&& 2.043 & 0.183 \\ 
				iCAR & 6.278 & 0.276 & 46.994 & 2.042 & 0.109 & 0.010 &&&  & 0.237 \\ 
				BYM & 3.733 & 0.168 & 29.114 & 1.311 & 0.005 & 0.009 & 0.009 &&& 0.141 \\ 
				pCAR & 6.290 & 0.277 & 46.503 & 2.023 & 0.109 & 0.010 &  &&& 0.238 \\ 
				LCAR & 6.212 & 0.274 & 46.028 & 2.005 & 0.113 & 0.010 && 0.958 && 0.235 \\ 
				BYM2 & 4.720 & 0.210 & 38.144 & 1.690 & 0.003 & 0.010 && 0.816 && 0.178 \\ 
				\multicolumn{11}{l}{ }\\
				\multicolumn{11}{l}{\textbf{Large-size informative prior distribution}}\\
				iid & 3.232 & 0.153 & 11.229 & 0.607 & 8.068 & 0.172 & && & 0.122 \\ 
				GP & 3.087 & 0.138 & 15.129 & 0.713 & 0.176 & 0.192 & &&8.602 & 0.117 \\ 
				iCAR & 3.128 & 0.141 & 15.121 & 0.704 & 1.998 & 0.177 & && & 0.118 \\ 
				BYM & 3.292 & 0.156 & 11.375 & 0.628 & 1.980 & 0.207 & 0.172 &&& 0.124 \\ 
				pCAR & 3.126 & 0.142 & 15.548 & 0.720 & 2.043 & 0.181 &&&  & 0.118 \\ 
				LCAR & 3.100 & 0.141 & 14.608 & 0.681 & 2.222 & 0.176 & &0.859 && 0.117 \\ 
				BYM2 & 3.128 & 0.143 & 12.099 & 0.584 & 0.689 & 0.172 && 0.941 && 0.118 \\  
				\hline
			\end{tabular}
		}
	\end{table}
	
	\begin{table}[b!]
		\centering
		\caption{\label{tab5} Empirical and theoretical expected smoothing for lung cancer dataset in Spain $A=300$.} 
		\resizebox{\textwidth}{!}{
			\begin{tabular}{l|rrrrrrrrrr}
				\hline
				& MSS & RMSS & $maxMSS$ & $maxRMSS$ & TCV & $\sigma^2$ & $\tau^2$ & $\lambda$ & $\psi$ & SP \\ 
				\hline
				\multicolumn{11}{l}{\textbf{Small-size informative prior distribution}}\\
				iid & 102.392 & 4.887 & 2496.917 & 117.336 & 2.956 & 0.01 & && & 0.901 \\ 
				GP & 98.174 & 4.336 & 1952.709 & 71.743 & 0.001 & 0.01 &&& 1.752 & 0.864 \\ 
				iCAR & 99.310 & 4.384 & 2030.084 & 76.968 & 0.584 & 0.01 &&&  & 0.874 \\ 
				BYM & 94.050 & 4.317 & 2110.307 & 84.202 & 0.032 & 0.01 & 0.009 &&& 0.828 \\ 
				pCAR & 99.381 & 4.392 & 2040.590 & 77.702 & 0.584 & 0.01 & && & 0.875 \\ 
				LCAR & 99.522 & 4.397 & 2062.446 & 79.240 & 0.587 & 0.01 && 0.989 && 0.876 \\ 
				BYM2 & 94.907 & 4.298 & 2014.807 & 76.605 & 0.017 & 0.01 && 0.801 && 0.835 \\ 
				\multicolumn{11}{l}{ }\\
				\multicolumn{11}{l}{\textbf{Large-size informative prior distribution}}\\
				iid & 78.972 & 3.875 & 1657.737 & 67.081 & 49.163 & 0.164 &&&  & 0.695 \\ 
				GP & 89.695 & 4.145 & 1902.582 & 71.371 & 0.085 & 0.199 &&& 8.638 & 0.789 \\ 
				iCAR & 84.663 & 4.058 & 1813.075 & 76.190 & 10.135 & 0.170 &&&  & 0.745 \\ 
				BYM & 74.308 & 3.655 & 1193.745 & 50.983 & 10.363 & 0.172 & 0.163 &&& 0.654 \\ 
				pCAR & 85.103 & 4.080 & 1894.393 & 73.766 & 10.318 & 0.173 & && & 0.749 \\ 
				LCAR & 82.301 & 3.918 & 1801.324 & 76.173 & 10.455 & 0.166 & &0.932& & 0.724 \\ 
				BYM2 & 78.407 & 3.853 & 1533.736 & 69.441 & 3.522 & 0.164 && 0.948 && 0.690 \\ 
				\hline
			\end{tabular}
		}
	\end{table}

	\autoref{tab4} and \autoref{tab5}  display the empirical smoothing criteria, namely MSS and RMSS, along with their respective maximum values and the SP for $A=47$ and $A=300$, respectively, corresponding to small- and large-size prior distributions. It also includes the TCV measure and posterior mean of the estimated hyperparameters for each spatial prior and prior distribution for the variance of the proposed spatial priors for the Spain dataset divided by 47 and 300 areas. Results for medium-size and uniform spatial priors are available in \autoref{s:D}. In general, as the values specified for the informative prior increase, the variance increases, which leads to an increase in the theoretical measure and a decrease in the empirical smoothing criteria. This behavior aligns with what we have observed for the TCV metric in the simulation study. Moreover, as the number of areas increase, the smoothing increase as seen in the simulation study.
	For $A=47$, only the small-size prior differentiates between the spatial priors, with the lowest smoothing achieved by the BYM spatial prior and the highest by iCAR, pCAR and LCAR. As the values of the informative prior distribution for the variance increase, the spatial priors become indistinguishable in terms of smoothing. Results for the uniform and medium-size informative prior are essentially identical to those for the large-size prior. This suggest that the spatial priors achieve nearly the minimum average smoothing, as the MSS and RMSS values remain nearly constant across medium- and large-size informative prior while the theoretical metric continues to increase at the same rate. However, the maximum MSS and RMSS values still decrease for most spatial priors when transitioning from a medium-size to a large-size informative prior distribution.
	In contrast, for $A=300$, a smoothing decrease is observed as the values of the informative priors increase; however, this decrease is much smaller compared to that observed for $A=47$. Across spatial priors, disparities are more pronounced for large-size informative prior distribution. For small-size informative prior, the lowest smoothing is obtained by BYM2, while the highest smoothing is achieved by iid. Conversely, for large-size informative prior, BYM achieves the lowest smoothing, and the GP prior exhibits the highest smoothing. Results for the medium-size informative prior are consistent with those for the large-size informative prior, indicating that the GP spatial prior tends to induce more smoothing than the CAR (neighbor-based) priors as the prior on $\sigma^2$ becomes weaker. Moreover, the uniform prior yields values nearly identical to those of medium-size informative prior.
	
	When analyzing the posterior mean rates per 100,000 inhabitants obtained for each spatial prior, see \autoref{FigA7} and \autoref{FigA8} for $A=47$ and \autoref{FigA9} and \autoref{FigA10} for $A=300$ in \autoref{s:D},  a significant degree of smoothing is evident when using a small-size prior distribution across all spatial priors. Higher rates are concentrated in north-western Spain, gradually decreasing as we move outward from this hotspot. However, the lower rates initially observed in the central-southern provinces in the crude rates disappear, resulting in a smoother surface overall. This smoothing effect becomes more pronounced as the number of areas increase. For $A=47$, a similar smoothing effect is observed across all spatial priors for each variance prior distribution. When the number of areas in peninsular Spain is increased to $A=300$, the surfaces obtained with the different variance prior distributions become noticeably smoother compared to the crude rates, though differences among the spatial priors begin to emerge.
	With a small-size informative prior distribution, greater smoothing is observed, particularly with the iid prior, which produces similar rate estimates for most areas. The other spatial priors exhibit comparable rate distributions, with higher rates concentrated in Asturias and Cantabria that gradually decrease outward from this hotspot. Notably, with the GP prior, some disparities emerge, as higher rates are maintained in areas such as Zaragoza and along the Mediterranean coast. When assuming medium-size informative prior distributions, the disparities among the spatial priors diminish. However, for large-size informative prior distributions, the BYM prior produces a slightly less smooth surface.
	As mentioned earlier, the uniform prior presents very similar smoothing to that seen with the medium-size informative prior distribution for both spatial disaggregations.

	In conclusion, the behavior observed in the simulation studies is consistent when applied to real data sets, confirming that the choice of spatial prior influences the level of smoothing in the estimates. By analyzing the Spain and England datasets, each with distinct spatial structures, we also assess the robustness of our findings across different geographical contexts. Two main conclusions can be reached from the real data analyses.  First, as the number of spatial units increases, the effect of smoothing becomes more pronounced, underscoring the need for caution when interpreting disease rates in large-area settings, where oversmoothing may obscure meaningful spatial variation.
	Second, the choice of prior distribution for the hyperparameters of the spatial model influences the level of smoothing, with different specifications potentially masking important spatial patterns.
	These findings highlight the practical importance of understanding the smoothing effects introduced by spatial priors, which is essential for accurately interpreting and drawing conclusions about the spatial distribution of a disease.

	\section{Discussion \label{S:discussion}}
	
	Little discussion has appeared in the literature with regard to quantification and comparison of how and how much various neighbor-based priors introduce spatial smoothing in the context of disease mapping modeling.  This becomes of interest because model performance is \emph{not} a question of goodness-of-fit.  
	In this work, we address this gap through both simulation studies and real data applications, proposing metrics to evaluate spatial smoothing from both theoretical and empirical perspectives.
	We introduce the Total Conditional Variance (TCV), defined as the sum of the conditional variances across all areal units, to evaluate the expected level of smoothing associated with a given spatial prior, i.e., this metric allows for a comparison of the smoothness induced by different hyperprior distributions within that spatial prior.
	It also provides insight into the theoretical behavior of smoothing in the Poisson-logitNormal model. Notably, we observe that the smoothing effect reaches a maximum and then, when $\sigma^2$ increases, it approaches zero—behavior that mirrors the properties of the Poisson-Gamma model. On the empirical side, we propose additional metrics to quantify the actual smoothing achieved by different priors in practice. These empirical measures allow us to confirm that model behavior aligns with theoretical expectations. Furthermore, they facilitate meaningful comparisons across different spatial priors in terms of their smoothing effects.
	
	The within prior simulation study demonstrates that theoretical expectations are consistent with empirical performance, suggesting that $\sigma^2$ consistently exercises control over the expected level of smoothing across all spatial priors, even when additional parameters are present. The across prior simulation study further reveals that the choice of spatial prior plays an important role in the amount of smoothing achieved within a data-set. This effect becomes more pronounced as the number of areas increases. Furthermore, higher levels of smoothing are often observed when increasing the number of areas. These conclusions are further supported by the findings from the real data analyses. 
	These analyses also demonstrate that, beyond the spatial prior defined, the selection of hyperprior distributions is also important, leading to varying degrees of smoothing within the priors.
	
	When comparing the level of smoothing across spatial priors, it is not possible to draw a general conclusion, as the results vary depending on the spatial structure of the regions. This variability is evident in both the simulation studies and the real data analyses. However, across both spatial regions used in this work, the GP prior generally produces the highest levels of smoothing. In the case of England, a slight reduction in smoothing is observed when using the LCAR prior, consistent in both the real data and the simulations. For Spain, differences among the spatial priors become more pronounced as the number of areas increases, with the BYM prior exhibiting lower levels of smoothing in such settings.
	
	In conclusion, the utility of the analysis developed here is to better enable the practitioner working with areal disease data to appreciate the smoothing consequences of choice of disease mapping prior and specification of hyperpriors on the parameters of the prior.  Maps can provide visual comparison of resultant smoothing but explicit quantification may offer a useful supplement.  Further, since our methodology is generally applicable, such smoothing assessment under fitting of a particular disease mapping model specification could be useful in interpreting the results/implications of a disease mapping analysis.
	
	Future work could explore the dynamic disease mapping setting where some autoregressive specification is introduced into the modeling \citep{knorr2000}.  Then, we might ask about the effect of temporal dependence on smoothing over time.  Also, there is now a rich literature on multivariate disease mapping to capture interaction/association in disease risk and rates \citep[][Chapter 8]{martinez2019}.  When within unit dependence is introduced into the joint modeling, how can we formally assess resultant smoothing and how does such modeling affect smoothing compared with marginal modeling of the diseases?
	
	\section*{Acknowledgements}
	The authors thank Miguel Ángel Beltrán and Miguel Ángel Martínez-Beneito for useful conversation in the model fitting. The authors also	thank the reviewers and the associate editor for their helpul comments.
	
	\section*{Funding statement}
	The work was supported by Project PID2020-113125RB-I00/MCIN/AEI/10.13039/501100011033, 
	Project PID2024-155382OB-I00 funded by MICIU/AEI/10.13039/501100011033 and FEDER, UE and BIOSTATNET - PROYECTOS REDES DE INVESTIGACIÓN 2024 - RED2024-153680-T/MICIU/AEI/.
	Garazi Retegui is supported by PhD student scholarship from the Public University of Navarra together with Banco Santander (Ayudas Predoctorales Santander UPNA 2021-2022). Jaione Etxeberria and María Dolores Ugarte would like to acknowledge support from Project UNEDPAM/PI/PR24/05A.
	
	\section*{Declaration of conflicting interests}
	The authors declared no conflicts of interest.

	\section*{Data Availability}
	The data and code supporting the findings of this study are available at \url{https://github.com/spatialstatisticsupna/Prior_Smoothing}.

	\bibliography{BIB}
	
	\begin{appendices}
	
	\section{The Poisson-Gamma model \label{s:A}}
	A simulation study to illustrate the quantification of the empirical smoothing associated with a given map across given $\mu_{\eta}$ and $\sigma_{\eta}^{2}$ is presented in this section.  We can generate replicate $B$ sets, $\{O_{ib}, b=1,2,...B\}$ and use them to calculate the proposed empirical metrics, using the difference between the posterior mean smoothing for areal unit $i$ and $\hat{r}_{i}$, i.e., \begin{equation}\label{Eq1*}
		E\left(r_i \mid O_i\right)-\hat{r}_{i} = \left(\left(1-w_{\eta,i}\right)\mu_r  +  w_{\eta,i}\frac{O_{i}}{n_{i}}\right) - \hat{r}_{i} = \frac{\mu_r}{\sigma^2_r n_i + \mu_r}\left(\mu_r - \hat{r}_{i}\right).
	\end{equation}
	Specifically, we draw $\eta_i$'s (the relative risks) from $Gamma(a_\eta, b_\eta)$, fixing $\mu_\eta$ and considering a full range of $\sigma_\eta^2$'s from very small to very large. Then, for each $\sigma_\eta^2$, we generate $B = 1000$ replicates of the set $\{O_{ib}, b=1,2,...B\}$ from the $Poisson(E_i\eta_i)$. Therefore, in this simulation, we have to bring in the expected number of cases $E_i$. We have taken the expected number of cases from the real dataset of Spain examined in Section~5 of the main text. This model is not hierarchical; $\mu_\eta$ and $\sigma_\eta^2$ are fixed. Hence, we can compute a discrepancy $d = E\left(r_i \mid O_i\right)-\hat{r}_{i}$ explicitly (no model fitting), see \autoref{Eq1*}. With $B$ independent replicates, we obtain $B$ discrepancies and therefore the proposed empirical metrics.

	\begin{figure}[b!]
		\begin{center}
			\includegraphics[page=1, width=7.5cm]{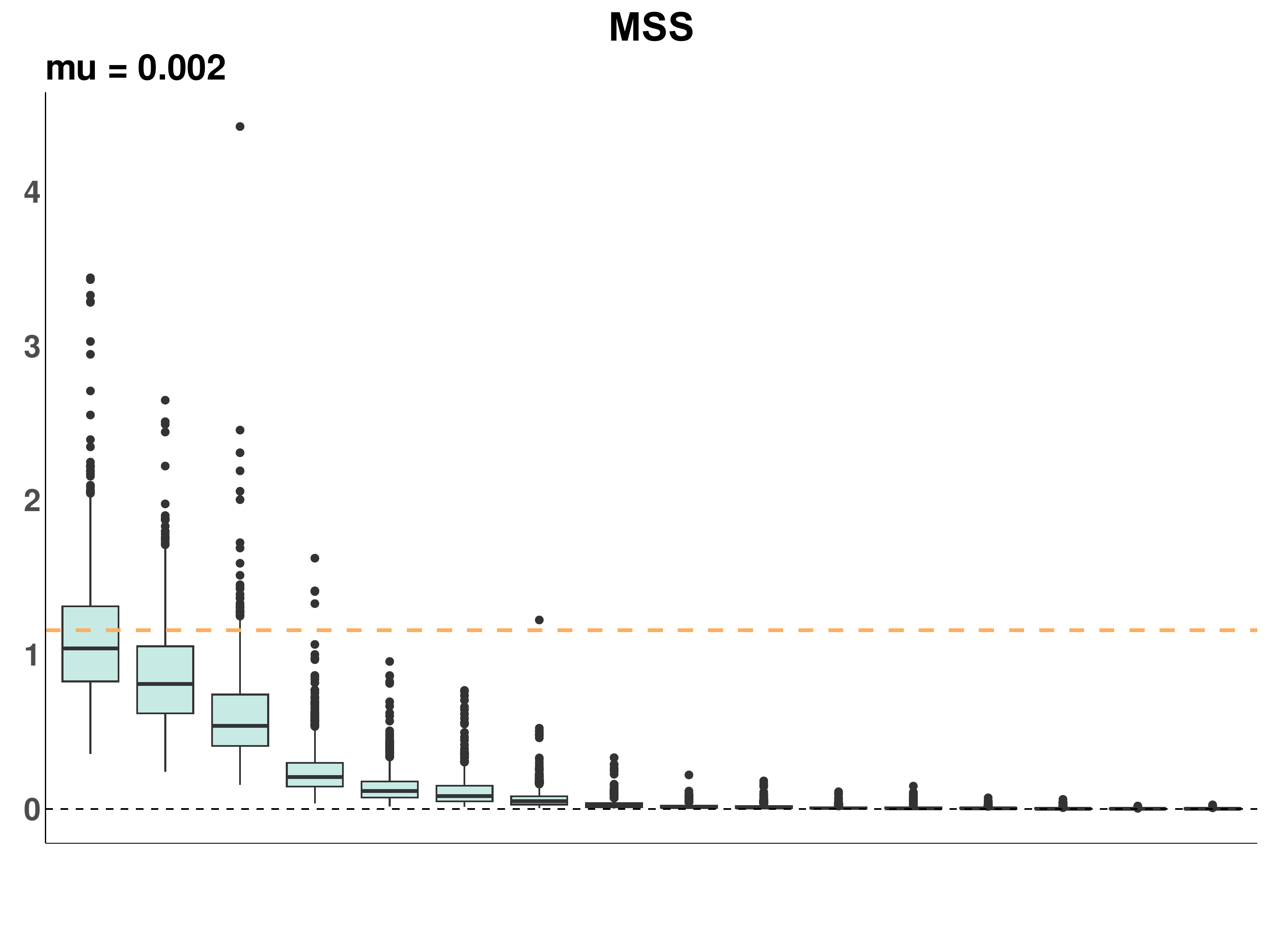}
			\includegraphics[page=2, width=7.5cm]{Fig_Supp/FigureA1.pdf}\\
			
			\includegraphics[page=3, width=7.5cm]{Fig_Supp/FigureA1.pdf}
			\includegraphics[page=4, width=7.5cm]{Fig_Supp/FigureA1.pdf}\\
			
			\includegraphics[page=5, width=7.5cm]{Fig_Supp/FigureA1.pdf}
			\includegraphics[page=6, width=7.5cm]{Fig_Supp/FigureA1.pdf}
		\end{center}
		\caption{Boxplots of the empirical smoothing criteria MSS (left) and MRSS (right) for the $B = 1000$ replicates and the theoretical maximum criteria value reached at $\sigma^2 = 0$ (orange dotted line).\label{FigA1}}
	\end{figure}
	
	We consider three different values for $\mu_{\eta}$ to see what quantitative limits we get with an actual map and how they depend on $\mu_{\eta}$. Specifically, we consider $\mu_\eta = 0.002, 0.02, 0.2$. \autoref{FigA1} shows the results obtained for the empirical smoothing metrics MSS and RMSS for the $B=1000$ replicates in boxplots and the orange dotted lines represent the theoretical criteria values obtained at $\sigma_\eta^2 = 0$ for each metric. As the value of $\sigma^2_\eta$ increases, both the MSS and the RMSS decrease, indicating a reduction in the amount of smoothing induced. When we increase $\mu_\eta$, the MSS criteria values increase, while the RMSS criteria values decrease. Interestingly, we find that the empirical criteria values approach zero for similar $\sigma^2_\eta$ values, regardless of different $\mu_{\eta}$ values. This suggests that while $\mu_{\eta}$ affects the variability in the criteria values, it does not significantly affect the overall decay to minimal smoothing.
	
	\begin{figure}[b!]
		\begin{center}
			\includegraphics[page=1, width=7.5cm]{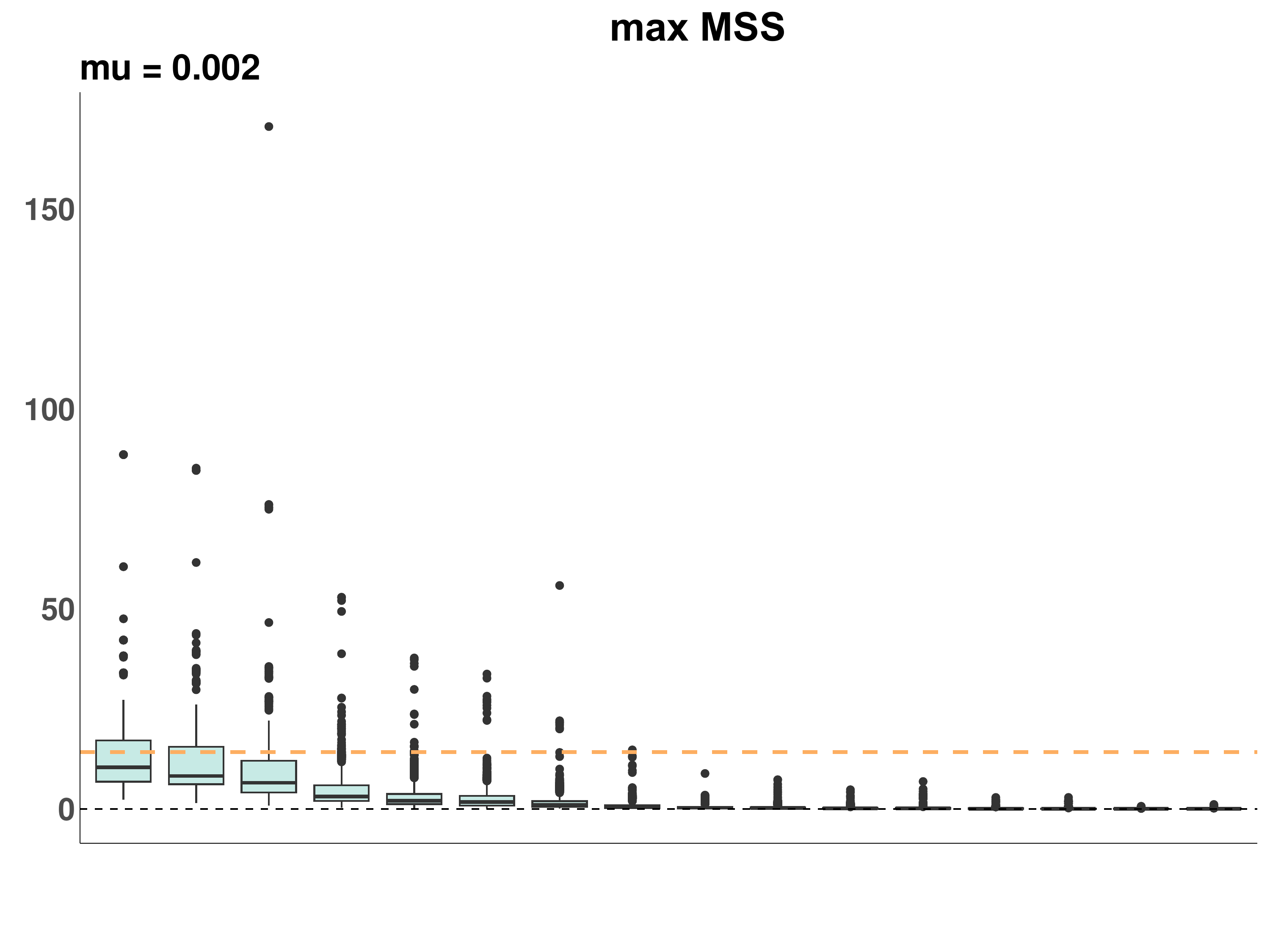}
			\includegraphics[page=2, width=7.5cm]{Fig_Supp/FigureA2.pdf}\\
			
			\includegraphics[page=3, width=7.5cm]{Fig_Supp/FigureA2.pdf}
			\includegraphics[page=4, width=7.5cm]{Fig_Supp/FigureA2.pdf}\\
			
			\includegraphics[page=5, width=7.5cm]{Fig_Supp/FigureA2.pdf}
			\includegraphics[page=6, width=7.5cm]{Fig_Supp/FigureA2.pdf}
		\end{center}
		\caption{Boxplots of the maximum empirical smoothing criteria maxMSS (left) and maxMRSS (right) for the $B = 1000$ replicates and the theoretical maximum criteria value reached at $\sigma^2 = 0$ (orange dotted line).\label{FigA2}}
	\end{figure}
	
	\autoref{FigA2} illustrates the empirical smoothing maximum metrics, with the orange dotted lines representing the theoretical criteria values at $\sigma_\eta^2 = 0$. The trends observed in the figure align with those seen in the MSS and RMSS analyses, as $\sigma_\eta^2$ increases, the empirical metrics decrease, demonstrating reduced smoothing. These observations confirm that the behavior of the empirical smoothing metrics is consistent with the theoretical expectations.

	\clearpage
	\section{Within prior simulation study \label{s:B}}
	In this section, we present the results of the within simulation study from Section~4.1 of the main paper, including detailed tables for the GP and BYM priors, which were commented on but not shown in the main text. Results for other priors, such as iid, pCAR, and BYM2, are not included here, as their conclusions align closely with those of the priors presented. Specifically, the findings consistently demonstrate that theoretical smoothing aligns with empirical metrics. As the $\sigma^2$ parameter decreases, smoothing increases, eventually reaching a maximum, and asymptotically approaches its lower limit as $\sigma^2$ continues to decline. These patterns reaffirm the robustness of the theoretical metrics across different spatial priors.

	\begin{table}[t!]
		\centering
		\caption{Empirical smoothing criteria, MSS, RMSS, the maximum values, $maxMSS$ and $maxRMSS$, and the smoothing proportion (SP), together with the theoretical smoothing metric (TCV) for the different parameter values of the GP priors.\label{tabA1}} \vspace*{0.2cm}
		\resizebox{0.8\textwidth}{!}{
			\begin{tabular}{cc|rrrrrr}
				\hline
				\multicolumn{2}{c|}{parameters} & TCV & SP & MSS & RMSS & $maxMSS$ & $maxRMSS$  \\ 
				\hline
				$\sigma^2$ & $\psi$ &&&&&&\\
				\hline
				\multicolumn{8}{l }{\bf A=47}\\ 
				$10^{-4}$  &  1  & 0.003 & 0.757  & 921.786 & 8.762 & 7277.863 & 63.933\\ 
				&5 & 0.001 & 0.870 & 1059.791 & 10.006 & 8098.247 & 74.204  \\ 
				&9  & 0.000 & 0.914  & 1113.283 & 10.505 & 8311.845 & 77.008\\ 
				\multicolumn{8}{c }{ }\\ 
				0.0025 &  1 & 0.081 & 0.090  & 109.356 & 1.076 & 737.043 & 6.246  \\ 
				& 5 & 0.020 & 0.265& 322.760 & 3.001 & 2486.355 & 16.808  \\ 
				& 9 & 0.011 & 0.393 & 478.732 & 4.414 & 3912.762 & 28.660  \\ 
				\multicolumn{8}{c }{ }\\ 
				0.0081  &  1 & 0.262 & 0.026  & 31.872 & 0.332 & 211.619 & 2.512  \\ 
				& 5 & 0.066 & 0.088  & 106.877 & 1.044 & 723.586 & 6.038 \\ 
				& 9 & 0.037 & 0.156 & 189.685 & 1.798 & 1288.293 & 9.354  \\ 
				\multicolumn{8}{c }{ }\\ 
				0.04  &  1& 1.292 & 0.013 & 15.747 & 0.165 & 63.082 & 0.888  \\ 
				&5  & 0.325 & 0.020& 23.908 & 0.250 & 144.691 & 1.745 \\ 
				& 9 & 0.183 & 0.031  & 37.646 & 0.388 & 254.721 & 2.726 \\ 
				\multicolumn{8}{c }{ }\\ 
				0.25  &  1  & 8.075 & 0.014   & 16.863 & 0.265 & 117.760 & 5.549\\ 
				&5 & 2.033 & 0.013 & 15.413 & 0.166 & 62.117 & 1.068 \\ 
				&9  & 1.144 & 0.013 & 15.653 & 0.163 & 60.147 & 0.819 \\ 
				\multicolumn{8}{c }{ }\\ 
				\multicolumn{8}{l }{\bf A=300 }\\ 
				$10^{-4}$  &  1  & 0.009 & 0.119  & 1360.494 & 13.052 & 8426.696 & 77.642\\ 
				&5 & 0.002 & 0.137  & 1564.212 & 14.626 & 8713.158 & 80.502 \\ 
				&9 & 0.001 & 0.141 & 1609.232 & 14.956 & 8786.160 & 81.190 \\ 
				\multicolumn{8}{c }{ }\\ 
				0.0025  &  1 & 0.218 & 0.042 & 476.860 & 4.954 & 5405.560 & 52.206  \\ 
				&5& 0.045 & 0.066 & 755.692 & 7.547 & 6880.340 & 63.334  \\ 
				&9& 0.025 & 0.082  & 934.401 & 9.261 & 7537.859 & 69.000 \\ 
				\multicolumn{8}{c }{ }\\ 
				0.0081 &  1  & 0.706 & 0.029  & 336.289 & 3.691 & 4170.625 & 45.016\\ 
				&5 & 0.147 & 0.042& 480.327 & 4.912 & 5384.742 & 52.169  \\ 
				& 9  & 0.082 & 0.052 & 597.341 & 6.005 & 6122.131 & 57.469 \\ 
				\multicolumn{8}{c }{ }\\ 
				0.04  &  1 & 3.484 & 0.024 & 269.913 & 3.145 & 3260.246 & 40.704  \\ 
				& 5& 0.725 & 0.028  & 313.641 & 3.511 & 3882.101 & 44.189 \\ 
				&9 & 0.403 & 0.031& 353.254 & 3.806 & 4298.695 & 45.686  \\ 
				\multicolumn{8}{c }{ }\\ 
				0.25  &  1 & 21.775 & 0.023 & 258.861 & 2.988 & 2578.379 & 33.602  \\ 
				& 5 & 4.530 & 0.023 & 263.842 & 3.101 & 3120.471 & 40.036  \\ 
				& 9 & 2.521 & 0.024 & 273.652 & 3.202 & 3340.384 & 41.919  \\ 
				\hline
		\end{tabular} }
	\end{table}
	
	For the GP prior, as discussed in Section~2.1 of the main paper, the level of smoothing induced by the prior is influenced by the values of $\sigma^2$ and the range parameter $\psi$. Therefore, in this simulation study we vary the $\sigma^2$ and $\psi$ values. Specifically, we considered five values ranging from $10^{-4}$ to $0.25$ for $\sigma^2$ and three different options for $\psi$: $1$, $5$, and $9$. \autoref{tabA1} shows the empirical smoothing criteria, MSS, RMSS and the maximum values,  $maxMSS$ and $maxRMSS$, and the smoothing proportion (SP), together with the theoretical smoothing (TCV) for the different parameter values. Theoretical smoothing aligns with that determined through empirical metrics. Remind that theoretical measures demonstrate that as the $\sigma$ value increases, the expected smoothing decreases, while conversely, as the $\psi$ value increases, the smoothing increases.  However, for the extreme values of $\sigma^2$, we appreciate different behavior for the $\psi$ parameter. Specifically, for $\sigma^2 = 10^{-4}$, changes in $\psi$ parameter slightly affect the amount of smoothing. Similar to the other priors analyzed, this suggests that the smoothing for the GP prior reaches a maximum value. Something similar happens for the empirical smoothing metrics with $\sigma^2= 0.25$, while the theoretical measure consistently increases. Therefore, we observe that smoothing tends to an asymptote, as seen with the iCAR and LCAR priors in the main text. Across different numbers of areas, higher empirical values (MSS, RMSS, and maximum values) observed with $A=300$ result in lower smoothing and lower SP values compared to $A=47$. Additionally, for a TCV value of zero, the SP is lower for $A=300$ than for $A=47$ indicating that the maximum smoothing decreases as the number of areas increases. Moreover, a faster increase in TCV is observed with a greater number of areas, suggesting that smoothing approaches an asymptote more quickly as the spatial disaggregation increases.
	
	Similar results for BYM prior are presented in \autoref{tabA2}. For this prior, the parameters affecting the induced smoothing are $\sigma^2$ and $\nu = \tau^2/\sigma^2$, with the former corresponding to the variance parameter of the iCAR and the $\tau^2$ to the iid component of the BYM. Both parameters influence the smoothing induced by the BYM prior similarly; as the values of these parameters increase, the smoothing decreases. We have considered eight different variance values for all priors, however, since the BYM prior has two variance parameters, only a subset of the results is provided. \autoref{tabA2} shows the results for $\sigma^2 = 10^{-4}, 0.0025, 0.04$ and $\nu = 0.25, 1, 4$.  Similar to other spatial priors, the theoretical smoothing aligns with the empirical metrics. As the $\sigma^2$ parameter decreases, smoothing increases, eventually reaching a maximum, and asymptotically approaches its lower limit as $\sigma^2$ continues to decline. he influence of $\sigma^2$ on smoothing is more pronounced than that of changes in $\nu$ (or equivalently, $\tau^2$). For instance, at $\sigma^2 = 0.0025$, where neither the maximum nor the asymptote has been reached, increasing $\nu$ from 0.25 to 4 (a 16-fold increase) results in a 7-fold decrease in smoothing. In contrast, increasing $\sigma^2$ 16-fold, from 0.0025 to 0.04, with $\nu = 0.25$, leads to an 11-fold decrease in smoothing.
	
	\begin{table}[t!]
		\centering
		\caption{Empirical smoothing criteria, MSS, RMSS and the maximum values, $maxMSS$ and $maxRMSS$, together with the theoretical smoothing metric (TCV) for the different parameter values of the BYM priors with small $\sigma^2$ values.\label{tabA2} } 
		\begin{tabular}{cc|rrrrrr}
			\hline
			\multicolumn{2}{c|}{parameters} & TCV & SP & MSS & RMSS & $maxMSS$ & $maxRMSS$  \\ 
			\hline
			$\sigma^2$ & $\nu$ &&&&&&\\
			\hline
			\multicolumn{8}{l }{\bf A=47}\\ 
			$10^{-4}$&  0.25 & 0.002 & 0.841  & 1025.096 & 9.791 & 8382.667 & 77.842 \\ 
			&1 & 0.006 & 0.736 & 896.442 & 8.562 & 7572.311 & 67.477 \\ 
			&4&0.021 & 0.487 & 593.096 & 5.605 & 5009.017 & 39.009  \\ 
			\multicolumn{7}{c }{ }\\
			0.0025 & 0.25 & 0.060 & 0.179 & 218.214 & 2.039 & 1993.345 & 13.046 \\ 
			& 1  & 0.154 & 0.078  & 94.690 & 0.933 & 644.274 & 6.183\\ 
			& 4  & 0.514 & 0.025  & 30.550 & 0.320 & 223.763 & 2.775\\ 
			\multicolumn{7}{c }{ }\\
			0.04 & 0.25 & 0.954 & 0.015  & 18.185 & 0.188 & 91.670 & 1.203 \\ 
			& 1 & 2.460 & 0.013 & 15.955 & 0.165 & 60.792 & 0.813  \\ 
			& 4 & 8.221 & 0.013  & 15.545 & 0.162 & 51.527 & 0.719 \\ 
			\multicolumn{8}{c }{ }\\ 
			\multicolumn{8}{l }{\bf A=300 }\\ 
			$10^{-4}$&  0.25 & 0.014 & 0.973 & 1747.379 & 16.240 & 21406.162 & 196.635 \\ 
			& 1  & 0.038 & 0.948 & 1703.569 & 15.917 & 21534.963 & 198.934\\ 
			& 4 & 0.130 &  0.881 & 1583.856 & 14.957 & 21751.433 & 203.050 \\
			\multicolumn{7}{c }{ }\\
			0.0025 & 0.25  & 0.359 & 0.536  &963.597 & 9.345 & 19225.167 & 173.260\\ 
			& 1 & 0.955 & 0.493  & 886.736 & 8.713 & 19360.504 & 177.645 \\ 
			& 4 & 3.262  & 0.405  & 729.027 & 7.317 & 18115.810 & 172.701 \\ 
			\multicolumn{7}{c }{ }\\
			0.04 & 0.25 & 5.749 & 0.237 &427.349 & 4.649 & 15424.682 & 153.284  \\ 
			& 1 & 15.288 & 0.218  & 392.745 & 4.292 & 13915.657 & 139.127 \\ 
			& 4  & 52.198 & 0.200  & 359.935 & 3.943 & 10422.177 & 105.438\\ 
			\hline
		\end{tabular}
	\end{table}

	\clearpage
	\section{Across priors simulation study \label{s:C}}
	
\begin{figure}[b!]
		\begin{center}
			\includegraphics[page = 1, width = 15.5cm]{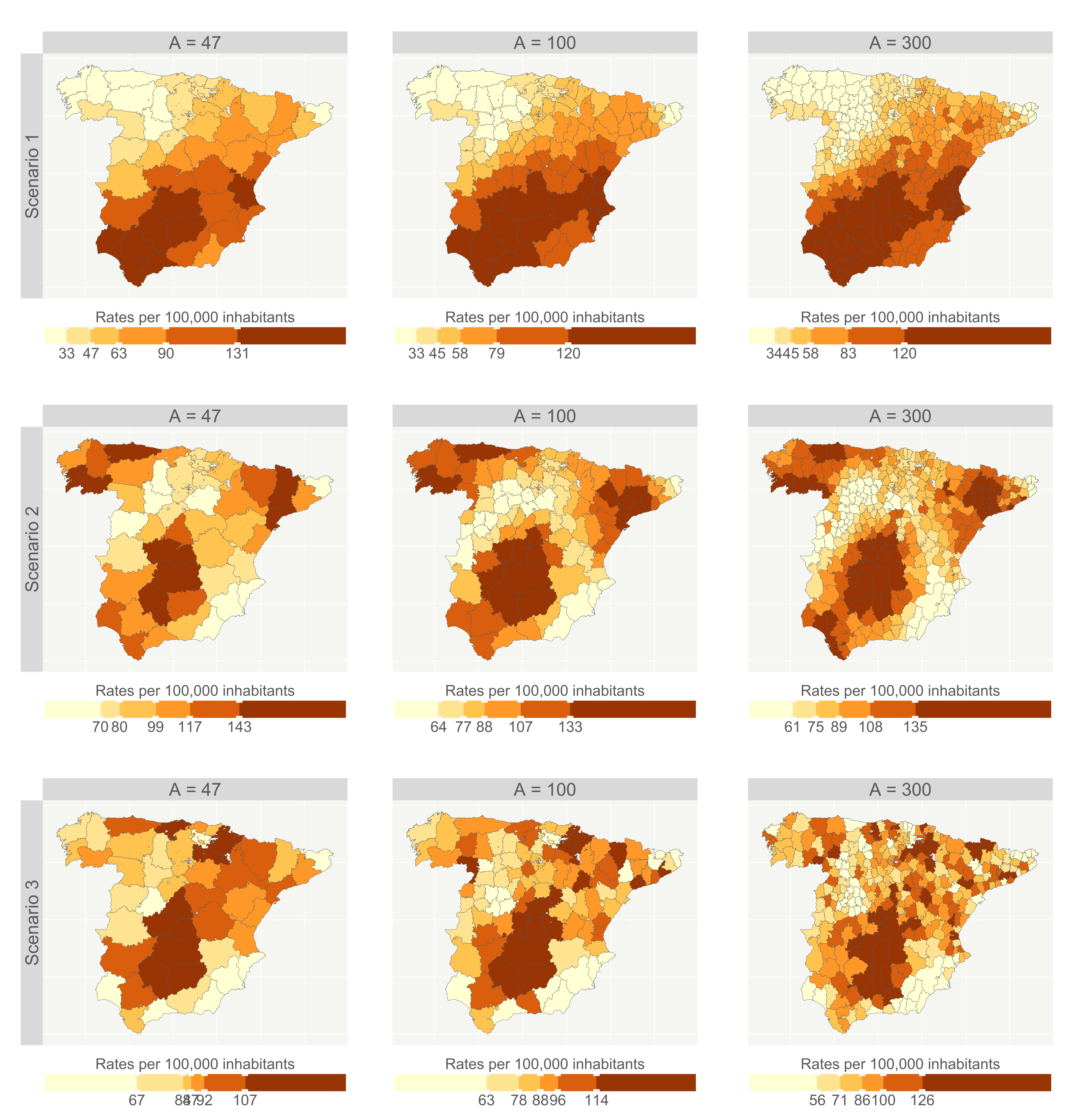}
		\end{center}\vspace{-0.8cm}
		\caption{\label{FigA3} Spatial distribution of the simulated rates for Scenarios 1, 2 and 3, organized by columns for each spatial disaggregation within Spain and by rows for each scenario. Please note that the scales are independent and vary for each scenario and spatial structure. }
	\end{figure}

\begin{figure}[t!]
		\begin{center}
			\includegraphics[page = 1, width = 16cm]{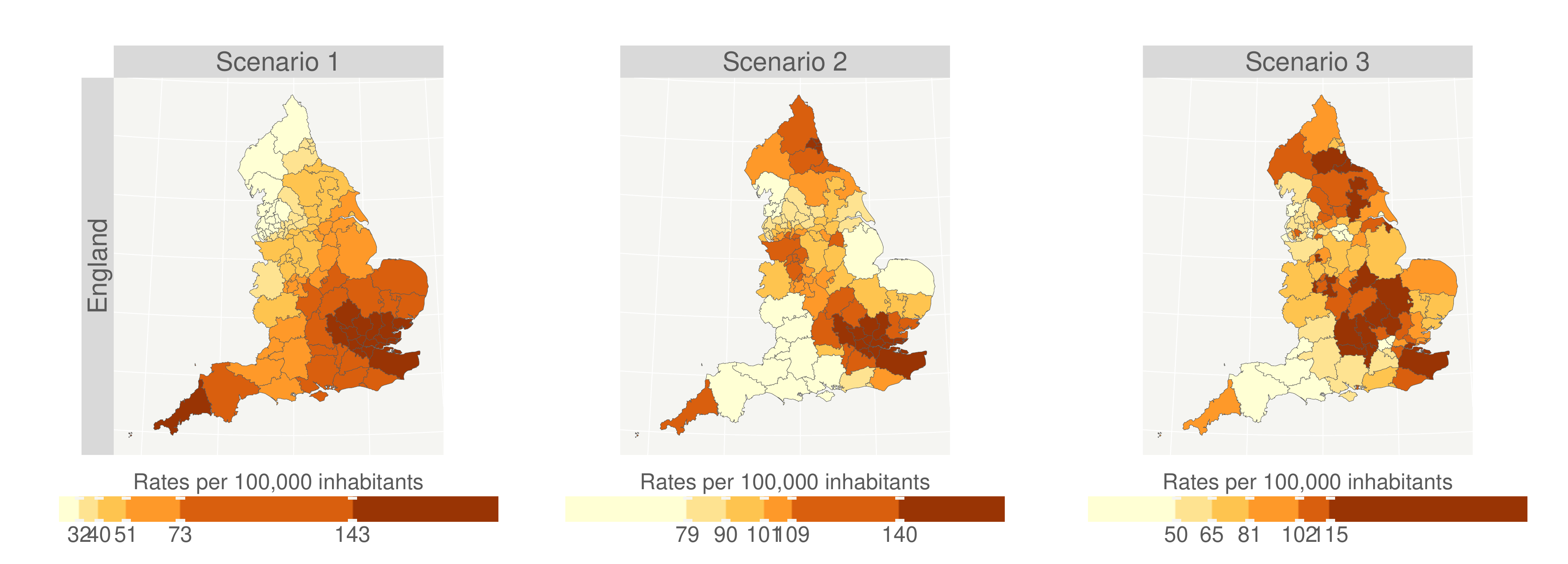}
		\end{center}\vspace{-0.8cm}
		\caption{\label{FigA4} Spatial distribution of the simulated rates for Scenarios 1, 2 and 3 for England. Please note that the scales are independent and vary for each scenario and spatial structure. }
	\end{figure}

	In this section, the definition of the simulation scenarios proposed in Section~4.2 of the main text and the scenarios not presented there together with the results achieved are shown. Specifically, the spatial distribution of Scenarios 1, 2 and 3 presented in the main text are illustrated in \autoref{FigA3} for the different spatial disaggregation of Spain and in \autoref{FigA4} for England. Note that in Scenario 1 we aim to create a south-east to north-west pattern and in Scenarios 2 and 3, we try to induce more fluctuation in the spatial distribution of the rates.
	
	\begin{table}[b!]
		\caption{MSS, RMSS, maximum MSS, maximum RMSS and SP values reached by each spatial prior model in Scenario~1 and Scenario~3, across the spatial region of Spain for a spatial disaggregation level of $A=300$. \label{TabA3}} 
		\resizebox{\textwidth}{!}{
			\begin{tabular}{lrrrrrcrrrrr}
				\hline
				&\multicolumn{5}{c}{Scenario 1}&&\multicolumn{5}{c}{Scenario 3}\\
				\cline{2-6}\cline{8-12}
				& MSS & RMSS & $maxMSS$ & $maxRMSS$ & SP && MSS & RMSS & $maxMSS$ & $maxRMSS$ & SP\\ 
				\hline
				\multicolumn{12}{l}{\textbf{Spain}}\\
				\multicolumn{12}{l}{\hspace*{0.5cm}\textbf{A = 300}}\\
				iid & 271.273 & 4.187 & 7391.257 & 95.765 & 0.124 && 400.208 & 4.519 & 14216.642 & 134.995 & 0.238\\ 
				GP & 360.850 & 6.626 & 13454.975 & 278.851 & 0.164 && 540.376 & 7.281 & 21741.857 & 332.041 & 0.321 \\ 
				iCAR & 253.244 & 4.131 & 11010.573 & 139.223 & 0.115 && 408.508 & 4.475 & 15075.607 & 128.434 & 0.242 \\  
				BYM & 253.637 & 4.181 & 10127.206 & 123.187 & 0.115 && 401.615 & 4.399 & 14033.807 & 117.283 & 0.239\\ 
				pCAR & 252.575 & 4.122 & 10993.998 & 138.858 & 0.115 && 406.254 & 4.457 & 15134.750 & 130.278 & 0.241  \\ 
				LCAR & 541.164 & 11.126 & 20058.857 & 628.809 & 0.246 && 797.425 & 13.398 & 29197.637 & 708.722 & 0.475 \\ 
				BYM2 & 253.374 & 4.126 & 10909.737 & 137.247 & 0.115 && 406.582 & 4.446 & 14789.707 & 125.227 & 0.241 \\  
				\hline
		\end{tabular} }
	\end{table}
	
	\autoref{TabA3} summarizes the empirical smoothing metrics for each prior model and scenario, based on the spatial structure of Spain divided into $A=300$ areas. Notably, Scenario~3 shows higher empirical smoothing criteria values compared to Scenario~1. An increase in smoothness is observed compared to the results presented in the main paper with $A=47$ and $A=100$. Specifically, as the number of areas increases from 100 to 300, smoothing metrics increase approximately threefold. Comparing results across different spatial priors, variability in the criteria values is observed. The LCAR prior exhibits the highest smoothing empirical criteria values, followed by the GP prior. In contrast, the iCAR, BYM, pCAR and BYM2 priors present the lowest smoothing metrics in terms of mean values (MSS and RMSS) and smoothing proportion, with an improvement in proportion when using the iid prior in Scenario~3. Among these, the iid prior records the lowest maximum smoothing values in Scenario~1, followed by the BYM prior. In Scenario~3 the BYM prior achieves the lowest maximum smoothing value.
	
	Additionally, Scenario~2 has been introduced in the main text, but its results have not been presented. \autoref{TabA3.2} displays the empirical criteria values across the spatial regions of Spain and England. The conclusions drawn are similar to those in Scenarios 1 and 3. For Spain, as the number of areas increases, the smoothing criteria values also increase—approximately fivefold from $A=47$ to $A=100$and threefold from $A=100$ to $A=300$. The MSS and RMSS values are very similar to those in Scenario~3; however, Scenario~2 yields lower maximum and SP values compared to Scenario~3.  Moreover, for $A=47$, all spatial priors produce similar smoothing criteria values. As the number of areas increases, disparities emerge, with the LCAR and GP priors yielding higher values. For England, Scenario~2 results in higher SP values than Scenarios 1 and 3. However, the overall conclusions remain consistent with those observed in Scenario~3, with the LCAR spatial prior producing the lowest criteria values.

	\begin{table}[t!]
		\centering
		\caption{MSS, RMSS, maximum MSS, maximum RMSS and SP values reached by each spatial prior model in Scenario~2, across the spatial regions of Spain and England. \label{TabA3.2}} 
		\resizebox{0.63\textwidth}{!}{
			\begin{tabular}{lrrrrr}
				\hline
				&\multicolumn{5}{c}{Scenario 2}\\
				\cline{2-6}
				&  MSS & RMSS & $maxMSS$ & $maxRMSS$ & SP\\ 
				\hline
				\multicolumn{6}{l}{\textbf{Spain}}\\
				\multicolumn{6}{l}{\hspace*{0.5cm}\textbf{A = 47}}\\
				iid & 15.637 & 0.164 & 53.515 & 0.751 & 0.013 \\ 
				GP & 15.063 & 0.157 & 50.914 & 0.694 & 0.012 \\ 
				iCAR & 15.560 & 0.160 & 49.340 & 0.684 & 0.013  \\ 
				BYM & 15.651 & 0.165 & 52.070 & 0.749 & 0.013 \\ 
				pCAR & 15.341 & 0.159 & 51.446 & 0.712 & 0.013  \\ 
				LCAR & 15.329 & 0.165 & 55.327 & 0.978 & 0.013 \\ 
				BYM2 & 15.686 & 0.164 & 51.782 & 0.734 & 0.013 \\ 
				\multicolumn{6}{l}{\hspace*{0.5cm}\textbf{A = 100}}\\
				iid & 92.459 & 1.045 & 756.422 & 9.238 & 0.071 \\ 
				GP &  134.526 & 2.204 & 2461.891 & 52.016 & 0.103 \\  
				iCAR &  84.542 & 0.984 & 968.526 & 12.469 & 0.065 \\  
				BYM &91.343 & 1.115 & 686.256 & 10.320 & 0.070 \\  
				pCAR & 84.383 & 0.983 & 958.465 & 12.351 & 0.065 \\  
				LCAR &165.474 & 3.601 & 3983.208 & 138.790 & 0.127 \\  
				BYM2 &  89.109 & 1.048 & 756.678 & 10.155 & 0.068 \\ 
				\multicolumn{6}{l}{\hspace*{0.5cm}\textbf{A = 300}}\\
				iid &397.705 & 4.195 & 10062.791 & 99.976 & 0.222 \\ 
				GP &  466.972 & 5.712 & 16137.290 & 210.630 & 0.261 \\ 
				iCAR & 354.039 & 4.029 & 13240.278 & 136.449 & 0.197 \\  
				BYM &  354.091 & 4.045 & 12322.173 & 124.119 & 0.198 \\ 
				pCAR & 353.466 & 4.020 & 13193.608 & 135.804 & 0.197 \\ 
				LCAR &  841.082 & 11.433 & 31179.719 & 566.844 & 0.471 \\ 
				BYM2 & 355.065 & 4.033 & 13048.514 & 133.553 & 0.198 \\  
				\multicolumn{6}{l}{ }\\
				\multicolumn{6}{l}{\textbf{England}}\\
				\multicolumn{6}{l}{\hspace*{0.5cm}\textbf{A = 106}}\\
				iid &  171.005 & 1.642 & 745.233 & 6.634 & 0.175\\ 
				GP & 151.728 & 1.483 & 1295.708 & 12.876 & 0.155\\  
				iCAR &  151.548 & 1.481 & 823.279 & 8.481 & 0.155\\  
				BYM &  150.847 & 1.475 & 801.485 & 8.239 & 0.154\\ 
				pCAR & 151.317 & 1.478 & 799.713 & 8.224 & 0.155 \\  
				LCAR & 143.892 & 1.407 & 799.239 & 8.268 & 0.147 \\ 
				BYM2 &  150.992 & 1.476 & 799.063 & 8.214 & 0.154\\  
				\hline
			\end{tabular} 
		}
	\end{table}

\begin{figure}[t!]
		\begin{center}
			\includegraphics[page = 1, width = 15.5cm]{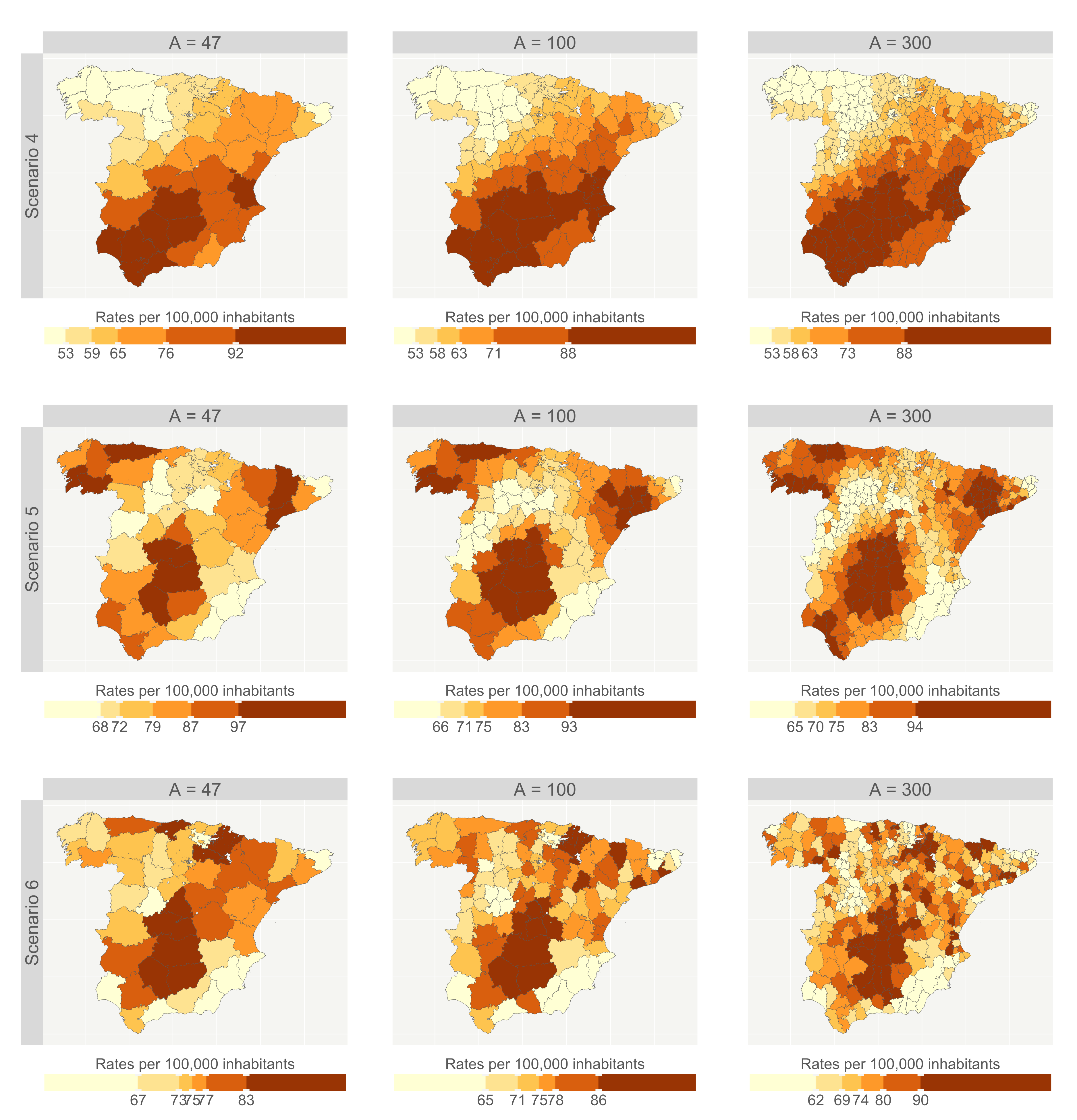} 
		\end{center}\vspace{-0.8cm}
		\caption{\label{FigA5} Spatial distribution of the simulated rates for Scenarios 3, 4 and 5, organized by columns for each scenario and by rows for each spatial structure. Please note that the scales are independent and vary for each scenario and spatial structure. }
	\end{figure}

\begin{figure}[t!]
		\begin{center}
			\includegraphics[page = 1, width = 16cm]{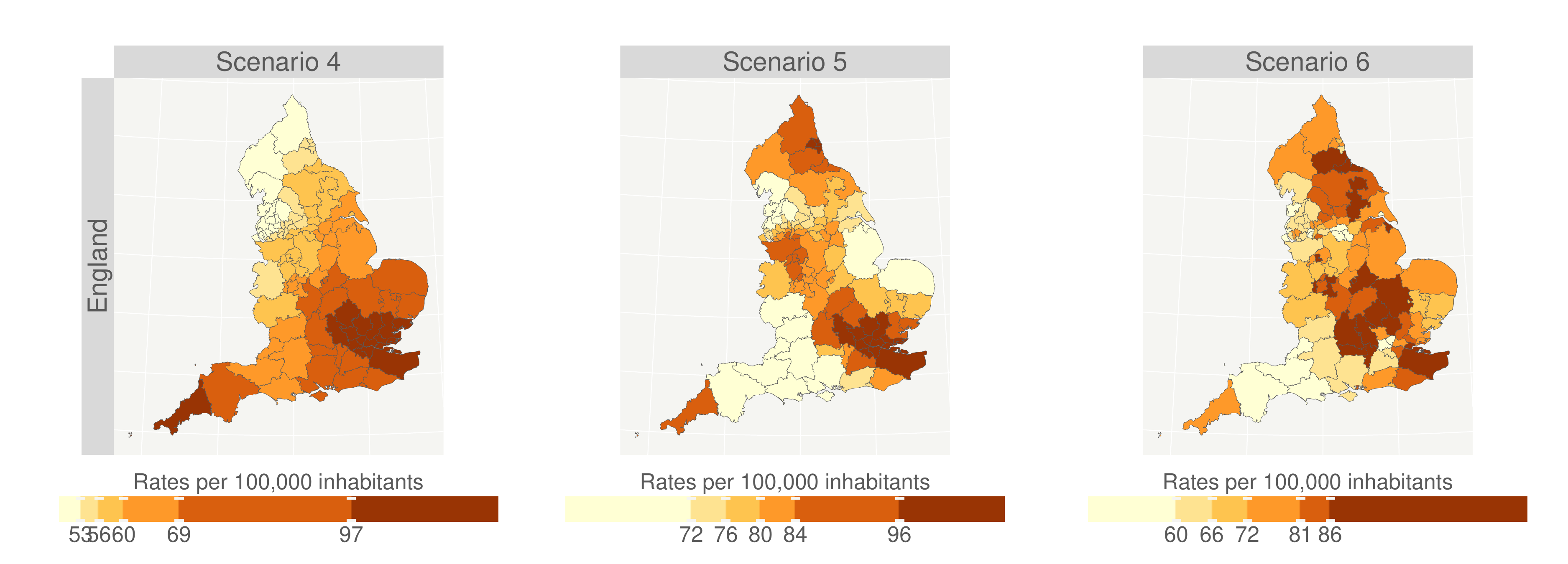}
		\end{center}\vspace{-0.8cm}
		\caption{\label{FigA6} Spatial distribution of the simulated rates for Scenarios 4, 5 and 6 for England. Please note that the scales are independent and vary for each scenario and spatial structure. }
	\end{figure}

	On the other hand, not presented in the main text, three additional scenarios have been established by adjusting the variability of the rates. \autoref{FigA5} and \autoref{FigA6} show the spatial distribution of these scenarios, named Scenarios 4, 5 and 6. The spatial distribution is the same as in Scenarios 1, 2 and 3, respectively, however, the variability of the rate values is lower. As in the previous scenarios, to generate the observed counts $O_i$ from a Poisson distribution, we used population data from the real data-sets discussed in the main text for each spatial structure. For each defined scenario, we generate $B = 1000$ data-sets.
	
	\begin{table}[t!]
		\caption{MSS, RMSS, maximum MSS, maximum RMSS and SP values reached by each spatial prior model in Scenario~4 and Scenario~5, across the spatial regions of Spain and England. For Spain, results are presented for spatial disaggregation levels of $A=47$, $A=100$ and $A=300$. \label{tabA4}} 
		\resizebox{\textwidth}{!}{
			\begin{tabular}{lrrrrrcrrrrr}
				\hline
				&\multicolumn{5}{c}{Scenario 4}&&\multicolumn{5}{c}{Scenario 5}\\
				\cline{2-6}\cline{8-12}
				& MSS & RMSS & $maxMSS$ & $maxRMSS$ & SP && MSS & RMSS & $maxMSS$ & $maxRMSS$ & SP\\ 
				\hline
				\multicolumn{12}{l}{\textbf{Spain}}\\
				\multicolumn{12}{l}{\hspace*{0.5cm}\textbf{A = 47}}\\
				iid & 10.929 & 0.162 & 45.702 & 0.704 & 0.028 && 12.765 & 0.161 & 53.720 & 0.748 & 0.063  \\  
				GP &  9.642 & 0.142 & 48.931 & 0.765 & 0.025 && 11.847 & 0.149 & 48.603 & 0.682 & 0.058 \\  
				iCAR & 9.817 & 0.144 & 44.443 & 0.696 & 0.025  && 12.287 & 0.154 & 47.169 & 0.661 & 0.060\\ 
				BYM & 10.815 & 0.161 & 45.624 & 0.725 & 0.028 && 12.762 & 0.163 & 49.745 & 0.741 & 0.063 \\ 
				pCAR & 9.694 & 0.143 & 43.885 & 0.687 & 0.025 && 12.287 & 0.154 & 49.686 & 0.692 & 0.060\\ 
				LCAR & 9.641 & 0.143 & 44.395 & 0.742 & 0.025 && 12.141 & 0.154 & 49.063 & 0.788 & 0.060 \\ 
				BYM2 & 10.553 & 0.156 & 43.898 & 0.691 & 0.027 && 12.668 & 0.161 & 49.574 & 0.718 & 0.062 \\ 
				\multicolumn{12}{l}{\hspace*{0.5cm}\textbf{A = 100}}\\
				iid & 65.658 & 0.986 & 653.623 & 9.847 & 0.187 && 78.384 & 1.008 & 823.618 & 11.151 & 0.287  \\ 
				GP & 78.811 & 1.631 & 1379.312 & 37.751 & 0.223 && 94.256 & 1.662 & 1537.365 & 40.012 & 0.344 \\
				iCAR & 61.106 & 0.958 & 1084.312 & 16.496 & 0.173 && 72.563 & 0.965 & 1104.517 & 15.027 & 0.264\\ 
				BYM & 67.878 & 1.100 & 684.153 & 10.374 & 0.193 && 78.963 & 1.091 & 678.563 & 10.151 & 0.290\\ 
				pCAR & 60.837 & 0.953 & 1074.152 & 16.319 & 0.172 && 72.583 & 0.962 & 1091.752 & 14.817 & 0.264  \\ 
				LCAR & 88.104 & 2.264 & 1716.193 & 68.534 & 0.250 && 105.789 & 2.298 & 2006.156 & 73.045 & 0.387 \\ 
				BYM2 & 64.580 & 1.014 & 843.261 & 12.481 & 0.183 && 76.503 & 1.024 & 830.749 & 11.457 & 0.280 \\ 
				\multicolumn{12}{l}{\hspace*{0.5cm}\textbf{A = 300}}\\
				iid & 275.192 & 3.982 & 8495.797 & 116.054 & 0.460 && 333.614 & 4.161 & 10036.928 & 121.375 & 0.597  \\ 
				GP & 307.244 & 4.990 & 10944.034 & 176.044 & 0.513 && 363.394 & 5.116 & 12179.068 & 177.102 & 0.649 \\
				iCAR & 273.193 & 4.176 & 10217.453 & 144.385 & 0.456 && 317.633 & 4.170 & 11184.733 & 140.868 & 0.567 \\ 
				BYM & 272.074 & 4.173 & 9860.089 & 137.727 & 0.454 && 316.304 & 4.168 & 10699.736 & 133.129 & 0.565 \\ 
				pCAR & 272.897 & 4.170 & 10207.870 & 144.162 & 0.455 && 317.615 & 4.162 & 11146.578 & 140.105 & 0.567 \\ 
				LCAR & 431.726 & 7.815 & 14343.542 & 319.852 & 0.725 && 529.146 & 8.241 & 17656.398 & 319.131 & 0.953 \\ 
				BYM2 & 273.164 & 4.171 & 10156.053 & 143.041 & 0.456 && 318.033 & 4.170 & 11093.555 & 139.104 & 0.568\\ 
				\multicolumn{12}{l}{ }\\
				\multicolumn{12}{l}{\textbf{England}}\\
				\multicolumn{12}{l}{\hspace*{0.5cm}\textbf{A = 106}}\\
				iid & 113.992 & 1.632 & 605.900 & 8.137 & 0.202 && 130.526 & 1.591 & 937.371 & 11.344 & 0.493 \\  
				GP & 99.639 & 1.533 & 965.608 & 15.464 & 0.176 && 127.607 & 1.572 & 1273.908 & 15.685 & 0.481 \\  
				iCAR & 94.705 & 1.451 & 698.478 & 11.848 & 0.168 &&  120.930 & 1.494 & 980.092 & 12.255 & 0.457 \\ 
				BYM & 94.074 & 1.441 & 682.796 & 11.594 & 0.167 && 119.871 & 1.481 & 950.124 & 11.893 & 0.453 \\ 
				pCAR & 94.433 & 1.444 & 683.870 & 11.592 & 0.167 && 120.479 & 1.484 & 946.637 & 11.794 & 0.455 \\  
				LCAR & 91.907 & 1.407 & 697.710 & 11.802 & 0.163 &&  118.397 & 1.460 & 970.533 & 12.091 & 0.447 \\ 
				BYM2 & 94.076 & 1.439 & 681.890 & 11.560 & 0.167 && 120.097 & 1.483 & 949.285 & 11.870 & 0.454 \\ 
				\hline
		\end{tabular} }
	\end{table}
	
	\begin{table}[t!]
		\centering
		\caption{MSS, RMSS, maximum MSS, maximum RMSS and SP values reached by each spatial prior model in Scenario~6, across the spatial regions of Spain and England. \label{tabA4.2}} 
		\resizebox{0.63\textwidth}{!}{
			\begin{tabular}{lrrrrr}
				\hline
				&\multicolumn{5}{c}{Scenario 6}\\
				\cline{2-6}
				&  MSS & RMSS & $maxMSS$ & $maxRMSS$ & SP\\ 
				\hline
				\multicolumn{6}{l}{\textbf{Spain}}\\
				\multicolumn{6}{l}{\hspace*{0.5cm}\textbf{A = 47}}\\
				iid & 11.778 & 0.158 & 49.944 & 0.665 & 0.116 \\ 
				GP & 12.081 & 0.161 & 68.497 & 0.964 & 0.119 \\ 
				iCAR & 12.044 & 0.160 & 67.935 & 0.964 & 0.119 \\
				BYM & 12.177 & 0.163 & 56.251 & 0.736 & 0.120 \\ 
				pCAR & 11.737 & 0.157 & 59.534 & 0.846 & 0.116 \\ 
				LCAR & 11.643 & 0.155 & 59.900 & 0.852 & 0.115 \\ 
				BYM2 & 12.080 & 0.161 & 53.256 & 0.709 & 0.119 \\ 
				\multicolumn{6}{l}{\hspace*{0.5cm}\textbf{A = 100}}\\
				iid & 78.411 & 1.042 & 969.126 & 12.686 & 0.398 \\ 
				GP & 98.167 & 1.726 & 1583.281 & 37.600 & 0.498 \\ 
				iCAR & 79.809 & 1.063 & 1282.483 & 16.504 & 0.402 \\ 
				BYM & 80.741 & 1.107 & 737.266 & 10.327 & 0.414 \\
				pCAR & 78.634 & 1.047 & 1230.045 & 15.963 & 0.397 \\ 
				LCAR & 106.679 & 2.248 & 1921.919 & 67.285 & 0.542 \\ 
				BYM2 & 80.199 & 1.072 & 943.110 & 12.265 & 0.408 \\ 
				\multicolumn{6}{l}{\hspace*{0.5cm}\textbf{A = 300}}\\
				iid & 337.693 & 4.459 & 12504.208 & 155.379 & 0.614 \\ 
				GP & 382.223 & 5.637 & 13602.569 & 230.489 & 0.696\\
				iCAR & 346.015 & 4.502 & 12652.754 & 149.399 & 0.629 \\ 
				BYM & 341.638 & 4.449 & 12190.376 & 142.424 & 0.621\\
				pCAR & 344.175 & 4.488 & 12713.951 & 151.691 & 0.625 \\ 
				LCAR & 503.004 & 8.421 & 16821.484 & 344.613 & 0.921 \\ 
				BYM2 & 344.047 & 4.470 & 12514.029 & 147.053 & 0.625 \\ 
				\multicolumn{6}{l}{ }\\
				\multicolumn{6}{l}{\textbf{England}}\\
				\multicolumn{6}{l}{\hspace*{0.5cm}\textbf{A = 106}}\\
				iid & 123.024 & 1.692 & 780.262 & 11.410 & 0.410 \\ 
				GP & 130.181 & 1.831 & 1203.402 & 17.244 & 0.433 \\ 
				iCAR & 117.370 & 1.634 & 815.434 & 11.380 & 0.391 \\ 
				BYM & 115.262 & 1.604 & 754.206 & 10.564 & 0.384 \\ 
				pCAR & 116.628 & 1.611 & 754.073 & 10.501 & 0.389 \\ 
				LCAR & 113.044 & 1.568 & 771.292 & 10.812 & 0.377 \\ 
				BYM2 & 115.618 & 1.609 & 753.311 & 10.562 & 0.386 \\ 
				
				\hline
			\end{tabular} 
		}
	\end{table}
	
	\autoref{tabA4} and \autoref{tabA4.2} present the empirical smoothing metrics (MSS, RMSS, $maxMSS$, $maxRMSS$ and SP) for each prior model, scenario and spatial structure. The results include the spatial structures of 47 provinces, 100 and 300 areas in Spain, as well as the spatial structure of England. 
	The SP values indicate that Scenario~4, 5 and 6 exhibit more smoothing than Scenario~1, 2 and 3, respectively. This suggests that as the variability of the rates decreases, an increase in smoothing is appreciated across all spatial structures considered.
	However, the MSS and RMSS values are generally similar for Scenario~4 and Scenario~1, except for the spatial structure of Spain with $A=47$ and the LCAR and GP priors with $A=100$ and $A=300$, where Scenario~4 shows lower values. In Scenario~5 and 6, lower MSS values are observed compared to Scenario~2 and 3, respectively, across all spatial structures. Overall, both Scenario~4, 5 and 6 exhibit higher relative maximum smoothing values than Scenario~1, 2 and 3, respectively. Despite these differences, the conclusions are consistent with those for Scenario~1, 2 and 3. Notably, Scenario~5 and 6 demonstrates higher empirical smoothing criteria values compared to Scenario~4.
	Additionally, smoothing increases as the spatial region is divided into a greater number of areas. Comparing the spatial structures of Spain and England, both with approximately 100 areas, England exhibits higher smoothing metric values, indicating a greater degree of smoothness. For the spatial structure of Spain, as the number of areas increases, larger disparities emerge among spatial priors, with greater smoothing observed for the LCAR and GP priors. In contrast, for the England spatial structure, all CAR priors show similar smoothing, with better performance in terms of smoothing than the GP or iid spatial priors. Disparities among CAR priors arise only in Scenario~5 and 6, where the LCAR prior exhibits lower smoothing.
	
	\clearpage
	\section{Data illustration  \label{s:D}}
	
	This section presents the results of the Data Illustration, Section~5, of the main paper. Specifically, we investigate the smoothing effects induced by the seven spatial priors proposed to determine whether the theoretical and simulation findings hold for real datasets. To achieve this, we consider four prior distributions for the $\sigma^2$ parameter of the spatial priors: the small-size, medium-size and large-size informative prior distribution, as well as the uniform prior distribution. For more information refer to Section~5 of the main paper. We examine two distinct datasets, each reflecting specific spatial structures. The main paper focuses on the spatial structure of peninsular Spain with  $A=47$ and $A=300$. Here, we present the results for the medium-size and uniform priors, including the posterior mean rates for each spatial prior per 100,000 inhabitants with $A=47$ and $A=300$, which are discussed but not shown in the main text. Additionally, we extend the analysis by providing results for $A=100$, along with results for the spatial distribution of England.

	\subsection{Lung cancer deaths in Spain}
	
	The results for peninsular Spain with $A=47$ and $A=300$, which are discussed but not shown in the main paper, are presented in \autoref{SubSectionSpain1}. Additionally, \autoref{SubSectionSpain2} provides the results for peninsular Spain with $A=100$.
	
	\subsubsection{Peninsular Spain with $A=47$ and $A=300$}\label{SubSectionSpain1}
	\begin{table}[b!]
		\centering
		\caption{\label{tabA5} Empirical and theoretical expected smoothing for lung cancer dataset in the 47 provinces of Spain.} 
		\resizebox{\textwidth}{!}{
			\begin{tabular}{l|rrrrrrrrrr}
				\hline
				& MSS & RMSS & $maxMSS$ & $maxRMSS$ & TCV & $\sigma^2$ & $\tau^2$ & $\lambda$ & $\psi$ & SP \\ 
				\hline
				\multicolumn{11}{l}{\textbf{Medium-size informative prior distribution}}\\
				iid & 3.243 & 0.151 & 13.62 & 0.646 & 2.555 & 0.054 &&&  & 0.122 \\ 
				GP & 3.333 & 0.148 & 18.641 & 0.875 & 0.055 & 0.095 &&& 6.692 & 0.126 \\ 
				iCAR & 3.475 & 0.155 & 23.903 & 1.088 & 0.844 & 0.075 & && & 0.131 \\ 
				BYM & 3.192 & 0.144 & 19.688 & 0.898 & 0.040 & 0.054 & 0.014&& & 0.121 \\ 
				pCAR & 3.385 & 0.152 & 22.274 & 1.016 & 0.959 & 0.085 & && & 0.128 \\ 
				LCAR & 3.365 & 0.151 & 21.821 & 0.999 & 0.936 & 0.072 && 0.836 && 0.127 \\ 
				BYM2 & 3.374 & 0.151 & 23.119 & 1.051 & 0.039 & 0.037 && 0.847 && 0.127 \\ 
				\multicolumn{11}{l}{ }\\
				\multicolumn{11}{l}{\textbf{Uniform prior distribution}}\\
				iid & 3.234 & 0.150 & 13.508 & 0.638 & 2.554 & 0.054 & && & 0.122 \\ 
				GP & 3.315 & 0.147 & 18.270 & 0.857 & 0.067 & 0.107 &&& 6.992 & 0.125 \\ 
				iCAR & 3.455 & 0.154 & 23.480 & 1.071 & 0.849 & 0.075 &  &&& 0.130 \\ 
				BYM & 3.313 & 0.148 & 21.781 & 0.995 & 0.029 & 0.063 & 0.006 &&& 0.125 \\ 
				pCAR & 3.372 & 0.151 & 21.907 & 1.003 & 0.975 & 0.087 &&&  & 0.127 \\ 
				LCAR & 3.367 & 0.151 & 22.496 & 1.028 & 0.933 & 0.072 && 0.836 && 0.127 \\ 
				BYM2 & 3.374 & 0.151 & 22.890 & 1.043 & 0.039 & 0.037 && 0.847 && 0.127 \\ 
				\hline
			\end{tabular}
		}
	\end{table}
	
	\begin{table}[t!]
		\centering
		\caption{\label{tabA6} Empirical and theoretical expected smoothing for lung cancer dataset in Spain $A=300$.} 
		\resizebox{\textwidth}{!}{
			\begin{tabular}{l|rrrrrrrrrr}
				\hline
				& MSS & RMSS & $maxMSS$ & $maxRMSS$ & TCV & $\sigma^2$ & $\tau^2$ & $\lambda$ & $\psi$ & SP \\ 
				\hline
				\multicolumn{11}{l}{\textbf{Medium-size informative prior distribution}}\\
				iid & 88.054 & 4.282 & 2181.796 & 96.491 & 18.380 & 0.061 & && & 0.775 \\ 
				GP & 94.101 & 4.224 & 1872.191 & 67.208 & 0.030 & 0.100 & &&6.214 & 0.828 \\ 
				iCAR & 87.930 & 4.146 & 1900.501 & 76.246 & 5.951 & 0.100 & && & 0.774 \\ 
				BYM & 86.758 & 4.087 & 1883.046 & 73.130 & 0.318 & 0.064 & 0.015 &&& 0.764 \\ 
				pCAR & 87.784 & 4.154 & 1940.299 & 74.936 & 6.500 & 0.109 &&&  & 0.773 \\ 
				LCAR & 90.790 & 4.206 & 1943.723 & 72.983 & 3.866 & 0.063 & &0.957 && 0.799 \\ 
				BYM2 & 87.592 & 4.124 & 1901.570 & 75.470 & 0.227 & 0.036 && 0.758 && 0.771 \\ 
				\multicolumn{11}{l}{ }\\
				\multicolumn{11}{l}{\textbf{Uniform prior distribution}}\\
				iid & 88.099 & 4.285 & 2186.229 & 97.113 & 18.355 & 0.061 & && & 0.775 \\ 
				GP & 92.163 & 4.198 & 1927.707 & 70.914 & 0.044 & 0.123 &&& 6.370 & 0.811 \\ 
				iCAR & 87.930 & 4.147 & 1896.422 & 76.657 & 5.970 & 0.100 & && & 0.774 \\ 
				BYM & 87.248 & 4.112 & 1894.586 & 74.960 & 0.338 & 0.074 & 0.009 &&& 0.768 \\ 
				pCAR & 87.715 & 4.151 & 1943.074 & 75.170 & 6.629 & 0.111 &  &&& 0.772 \\ 
				LCAR & 90.033 & 4.228 & 1920.578 & 76.643 & 3.955 & 0.064 && 0.957 && 0.792 \\ 
				BYM2 & 87.553 & 4.120 & 1895.213 & 75.129 & 0.227 & 0.036 && 0.758 && 0.771 \\ 
				\hline
			\end{tabular}
		}
	\end{table}
	
	\autoref{tabA5} and \autoref{tabA6} present the empirical smoothing criteria, namely MSS and RMSS, along with their respective maximum values and the SP for $A=47$ and $A=300$, respectively, under the medium-size informative and uniform prior distributions. These tables also include the TCV measure and the posterior mean of the estimated hyperparameters for each spatial prior and variance prior distribution, for the Spain dataset divided into 47 and 300 areas. While commented on in the main paper, these results are illustrated here for completeness. The uniform prior yields values nearly identical to those of medium-size informative prior in both cases. For $A=47$, the spatial priors become indistinguishable in terms of smoothing. 
	In contrast, for $A=300$, the BYM prior achieves the lowest smoothing, whereas the GP prior exhibits the highest smoothing. As noted in the main paper, results for the medium-size informative prior align with those for the large-size informative prior, confirming that the GP spatial prior induces more smoothing than the CAR (neighbor-based) priors as the prior on $\sigma^2$ becomes weaker. 
	
\begin{figure}[h!]
		\begin{center}
			\includegraphics[width = 13.5cm, page = 1]{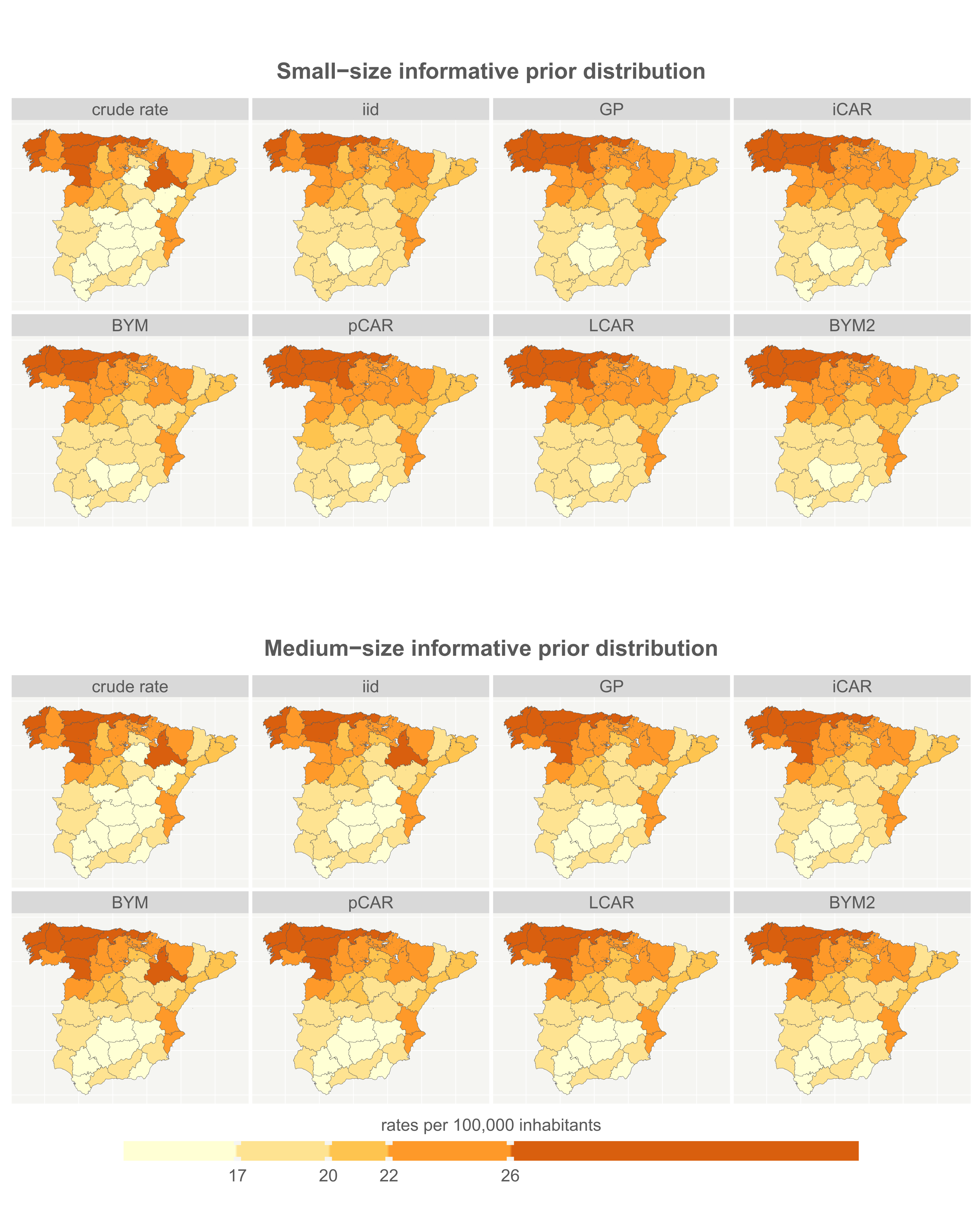}
		\end{center}\vspace{-0.8cm}
		\caption{\label{FigA7} Crude rates and posterior mean rates obtained by each spatial prior per 100,000 inhabitants for Peninsular Spain with $A=47$.}
	\end{figure}

\begin{figure}[h!]
		\begin{center}
			\includegraphics[width = 13.5cm]{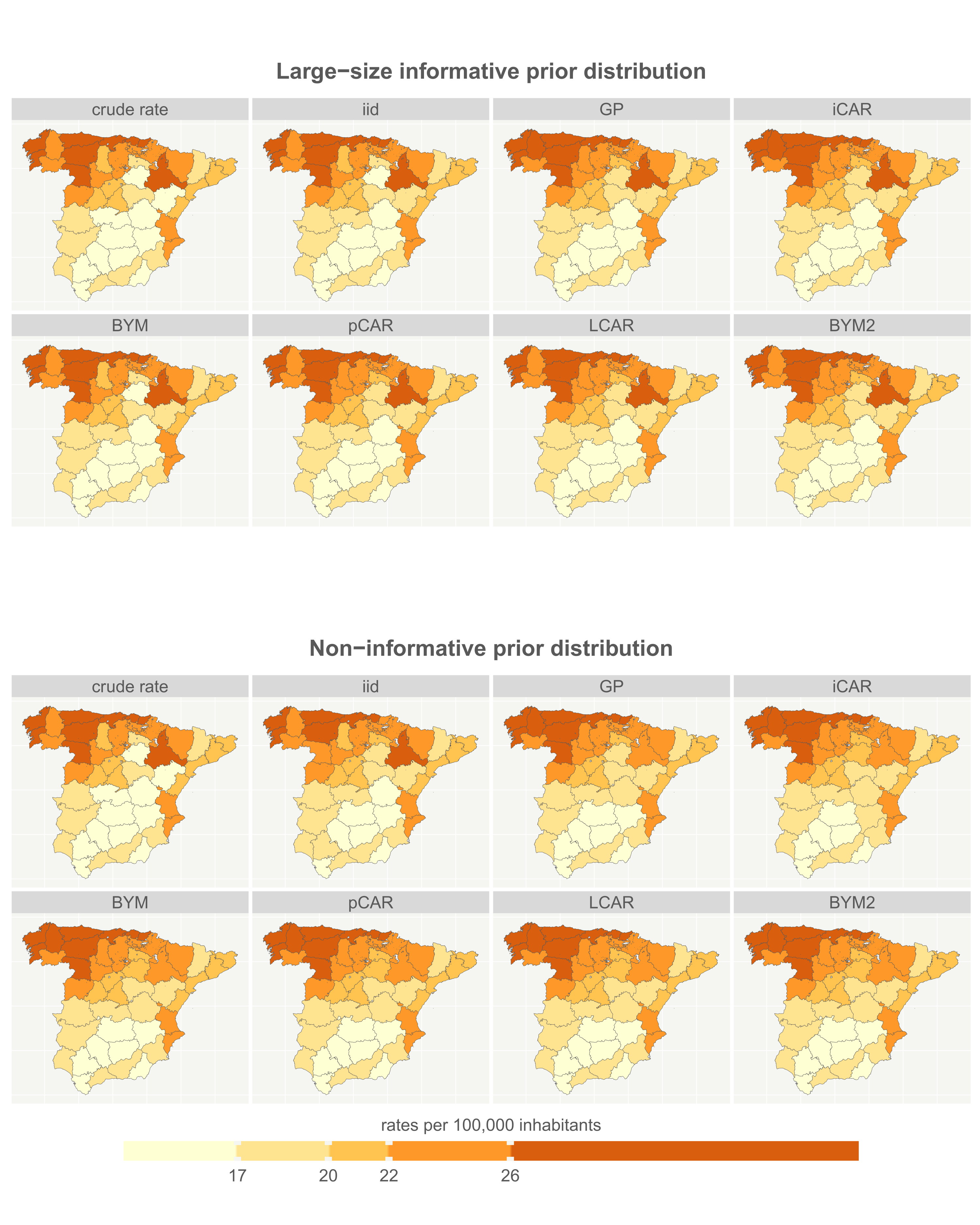}
		\end{center}\vspace{-0.8cm}
		\caption{\label{FigA8} Crude rates and posterior mean rates obtained by each spatial prior per 100,000 inhabitants for Peninsular Spain with $A=300$.}
	\end{figure}

\begin{figure}[h!]
		\begin{center}
			\includegraphics[width = 13.5cm]{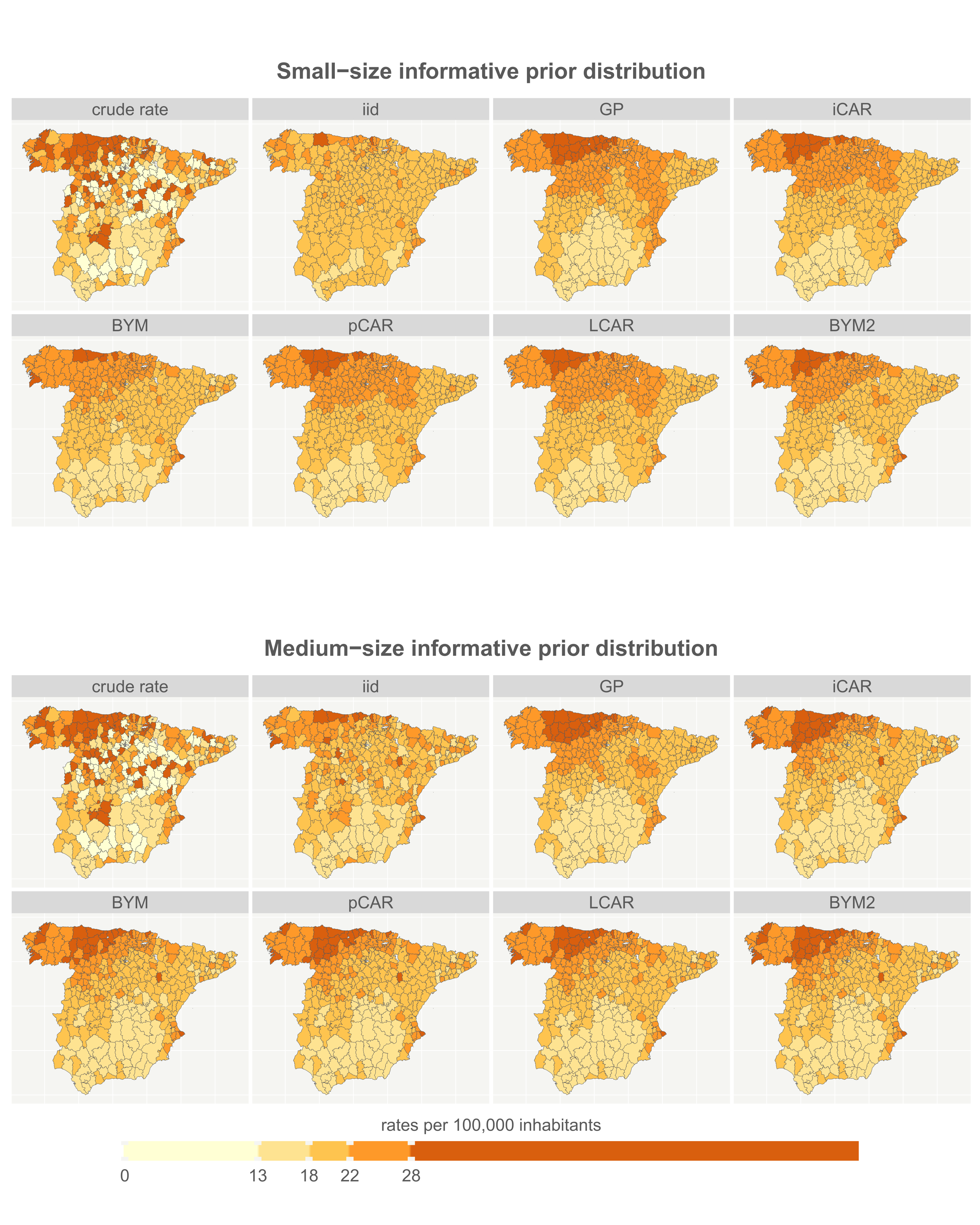}
		\end{center}\vspace{-0.8cm}
		\caption{\label{FigA9} Crude rates and posterior mean rates obtained by each spatial prior per 100,000 inhabitants for Peninsular Spain with $A=300$.}
	\end{figure}

\begin{figure}[h!]
		\begin{center}
			\includegraphics[width = 13.5cm]{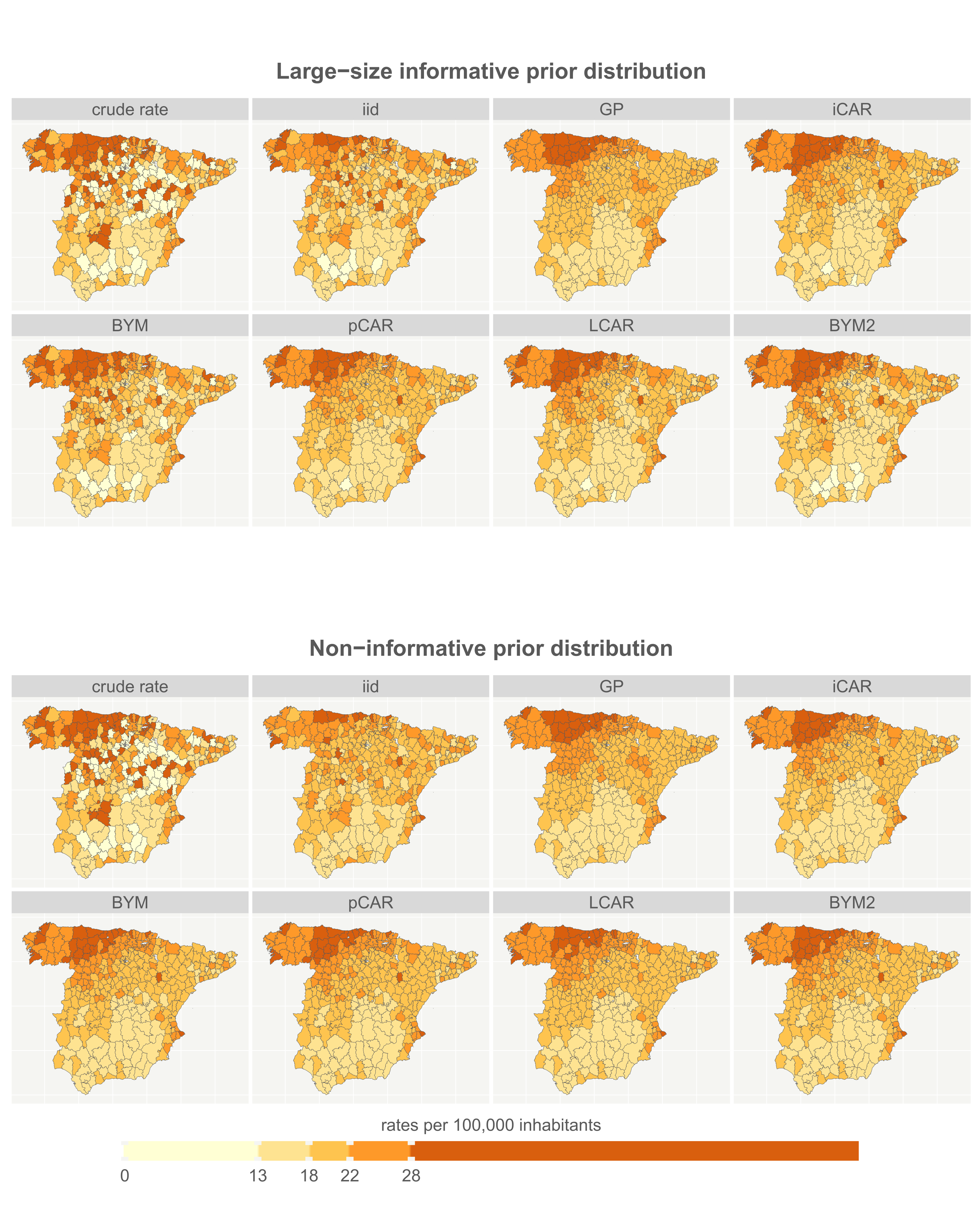}
		\end{center}\vspace{-0.8cm}
		\caption{\label{FigA10} Crude rates and posterior mean rates obtained by each spatial prior per 100,000 inhabitants for Peninsular Spain with $A=300$.}
	\end{figure}

	\autoref{FigA7} and \autoref{FigA8} present the crude rates and the posterior mean rates obtained for each spatial prior per 100,000 inhabitants with $A=47$. Similar figures for $A=300$ are presented in \autoref{FigA9} and \autoref{FigA10}. Relevant comments are provided in the main paper.
	
	\clearpage
	\subsubsection{Peninsular Spain with $A=100$}\label{SubSectionSpain2}
	
	\begin{table}[b!]
		\centering
		\caption{\label{tabA7} Empirical and theoretical expected 
			smoothing for lung cancer data-set in Spain.} 
		\resizebox{\textwidth}{!}{
			\begin{tabular}{l|rrrrrrrrrr}
				\hline
				& MSS & RMSS & $maxMSS$ & $maxRMSS$ & TCV & $\sigma^2$ & $\tau^2$ & $\lambda$ & $\psi$ &SP \\ 
				\hline
				\multicolumn{10}{l}{\textbf{Informative with small values prior distribution}}\\
				iid & 24.577 & 1.149 & 402.573 & 17.652 & 0.979 & 0.010 & && & 0.638 \\ 
				GP & 23.615 & 1.029 & 340.736 & 12.308 & 0.002 & 0.010 &&& 1.894 & 0.613 \\ 
				iCAR & 26.106 & 1.120 & 314.849 & 11.986 & 0.210 & 0.010 & && & 0.677 \\ 
				BYM & 22.085 & 0.992 & 257.681 & 10.551 & 0.0110 & 0.009 & 0.009 & && 0.573 \\ 
				pCAR & 26.101 & 1.121 & 311.472 & 11.987 & 0.210 & 0.010 & && & 0.677 \\ 
				LCAR & 25.930 & 1.116 & 311.172 & 11.933 & 0.215 & 0.010 & & 0.969 && 0.673 \\ 
				BYM2 & 23.796 & 1.039 & 294.808 & 11.188 & 0.006 & 0.010 && 0.845 && 0.617 \\ 
				\multicolumn{10}{l}{ }\\
				\multicolumn{10}{l}{\textbf{Informative with medium values prior distribution}}\\
				iid & 18.666 & 0.889 & 221.171 & 8.566 & 4.948 & 0.049 & && & 0.484 \\ 
				GP & 20.757 & 0.948 & 318.560 & 11.698 & 0.059 & 0.103 & &&6.565 & 0.538 \\ 
				iCAR & 20.965 & 0.947 & 277.920 & 10.599 & 1.707 & 0.079 & && & 0.544 \\ 
				BYM & 19.750 & 0.904 & 221.156 & 8.949 & 0.089 & 0.057 & 0.014&& & 0.512 \\ 
				pCAR & 20.631 & 0.938 & 261.058 & 10.140 & 1.903 & 0.088 &&& & 0.535 \\ 
				LCAR & 20.583 & 0.934 & 263.017 & 10.200 & 1.812 & 0.074 && 0.855 && 0.534 \\ 
				BYM2 & 20.654 & 0.935 & 259.259 & 10.08 & 0.070 & 0.035 & &0.863 && 0.536 \\ 
				\multicolumn{10}{l}{ }\\
				\multicolumn{10}{l}{\textbf{Informative with large values prior distribution}}\\
				iid & 17.481 & 0.839 & 102.561 & 5.067 & 16.676 & 0.167 & && & 0.453 \\ 
				GP & 20.355 & 0.943 & 315.828 & 11.595 & 0.155 & 0.191 &&& 8.685 & 0.528 \\ 
				iCAR & 19.598 & 0.912 & 237.064 & 9.352 & 3.730 & 0.173 & && & 0.508 \\ 
				BYM & 18.105 & 0.868 & 109.543 & 6.591 & 3.791 & 0.186 & 0.167 &&& 0.470 \\ 
				pCAR & 19.364 & 0.904 & 216.573 & 8.773 & 3.808 & 0.176 & && & 0.502 \\ 
				LCAR & 19.153 & 0.897 & 222.317 & 8.937 & 4.091 & 0.172 && 0.879 && 0.497 \\ 
				BYM2 & 18.593 & 0.887 & 174.806 & 8.626 & 1.351 & 0.167 && 0.950 && 0.482 \\ 
				\multicolumn{10}{l}{ }\\
				\multicolumn{10}{l}{\textbf{Uniform prior distribution}}\\
				iid & 18.741 & 0.893 & 224.493 & 8.719 & 4.946 & 0.049 & && & 0.486 \\ 
				GP & 20.864 & 0.951 & 322.217 & 11.790 & 0.069 & 0.115 & &&7.013 & 0.541 \\ 
				iCAR & 20.966 & 0.947 & 278.583 & 10.617 & 1.715 & 0.079 &&&  & 0.544 \\ 
				BYM & 20.415 & 0.927 & 252.109 & 9.873 & 0.071 & 0.068 & 0.005 &&& 0.530 \\ 
				pCAR & 20.635 & 0.939 & 260.669 & 10.125 & 1.934 & 0.090 &&& & 0.535 \\ 
				LCAR & 20.789 & 0.941 & 273.100 & 10.470 & 1.771 & 0.073 && 0.858 && 0.539 \\ 
				BYM2 & 20.681 & 0.936 & 260.321 & 10.104 & 0.070 & 0.035 && 0.867 && 0.536 \\ 
				\hline
			\end{tabular}
		}
	\end{table}

\begin{figure}[b!]
		\begin{center}
			\includegraphics[width = 13.5cm]{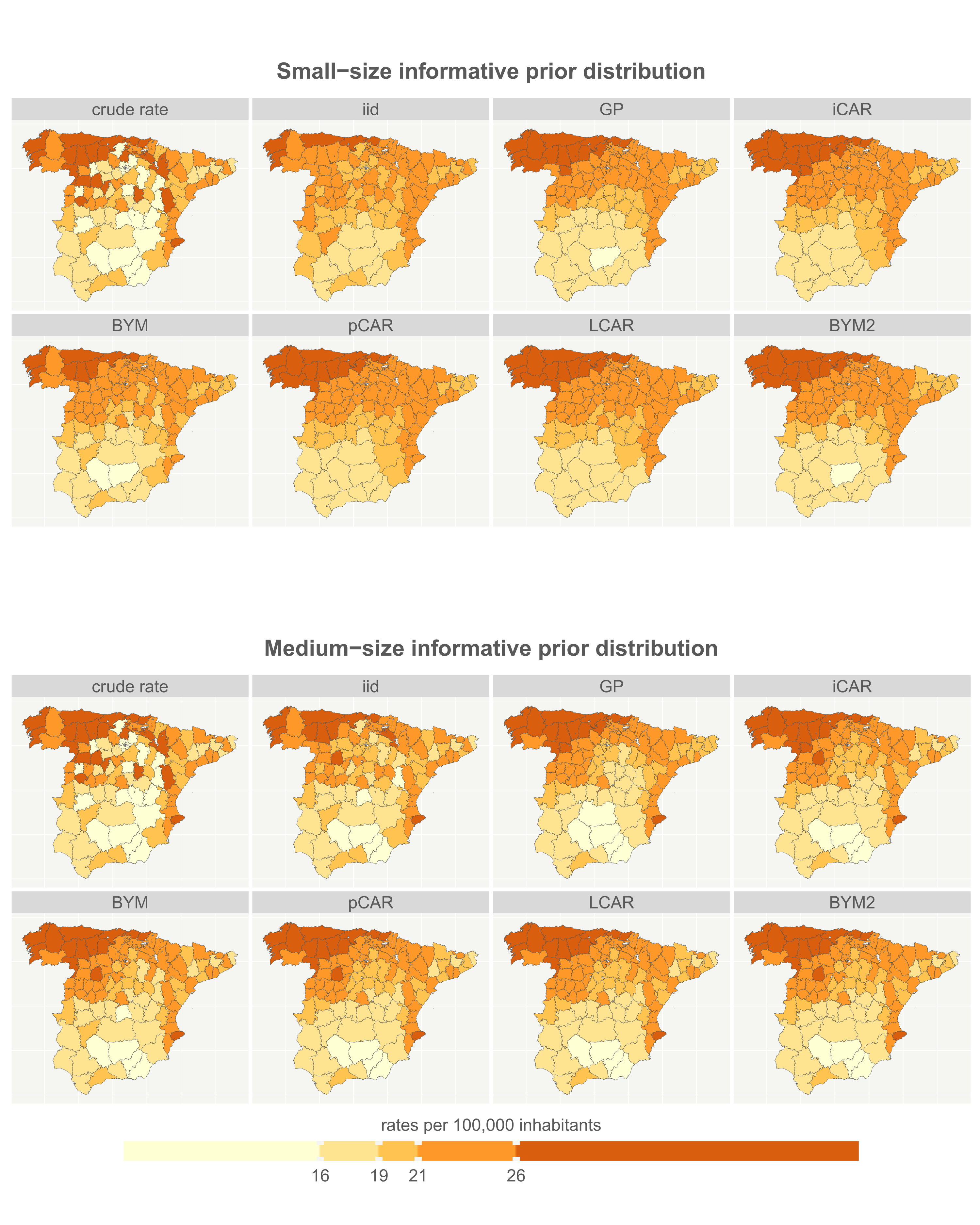}
		\end{center}
		\caption{\label{FigA11} Crude rates and posterior mean rates obtained by each spatial prior per 100,000 inhabitants when using a small-size and large-size informative prior distribution.}
	\end{figure}

	The data set presents counts for lung cancer in females and the corresponding population at risk, with data aggregated from 2019 to 2021. To provide an initial overview of the disease distribution, crude mortality rates per 100,000 inhabitants across different areas of Spain have been calculated and are displayed in the first grid of \autoref{FigA11} for $A=100$. The crude rates range between 6 and 44 deaths per 100,000 inhabitants. Similar to the spatial patterns observed for peninsular Spain with $A=47$ and $A=300$, the highest crude mortality rates are concentrated in the northwestern regions of Spain, along the Mediterranean coast (including Tarragona, Castellón, Valencia, and Alicante), and in Zaragoza. In contrast, the central-southern regions, particularly those south of Madrid, exhibit the lowest mortality rates.
	
	\begin{figure}[b!]
		\begin{center}
			\includegraphics[width = 13.5cm]{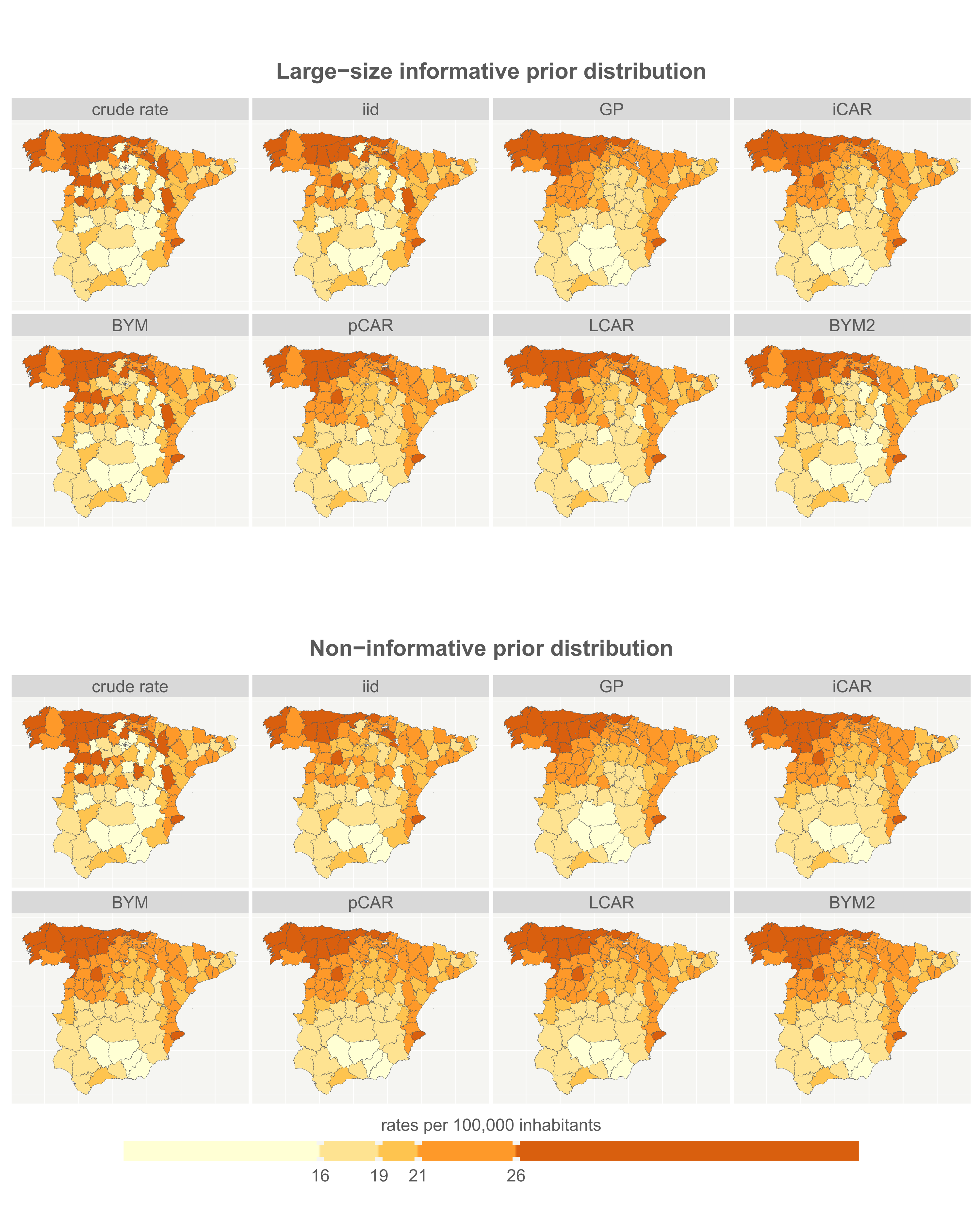}
		\end{center}
		\caption{\label{FigA12} Crude rates and posterior mean rates obtained by each spatial prior per 100,000 inhabitants when using a small-size and large-size informative prior distribution.}
	\end{figure}  
	
	In this study, we assume different prior distributions for $\sigma^2$ parameter of the proposed spatial priors. 
	\autoref{tabA7} displays the empirical smoothing criteria, namely MSS and RMSS, along with their respective maximum values and smoothing proportion (SP). It also includes the theoretical measure (TCV) and posterior mean of the estimated hyperparameters for each spatial prior and prior distribution for the variance of the proposed spatial priors. Let's start by analyzing the smoothing induced by the spatial priors under informative prior distributions. In general, as the values specified for the informative prior increase, the variance increases, which leads to an increase in the theoretical measure and a decrease in the empirical smoothing criteria. This behavior aligns with what we have observed in the theoretical metric and the simulation study. Furthermore, it is evident that for some spatial priors we achieve nearly the minimum average smoothing when a medium-size informative prior is assumed, as the MSS and RMSS values remain nearly constant while the TCV continues to increase at the same pace. On another note, the maximum MSS and RMSS values still decrease for the spatial priors when we transition from a medium-size to a large-size informative prior distribution. Additionally, larger disparities across spatial priors are observed with a small-size informative prior distribution compared to other informative prior distributions. Specifically, with a small-size informative prior distribution, the iCAR and pCAR spatial prior presents the highest empirical smoothing criteria values.
	When comparing the smoothing obtained with a uniform prior distribution to that with informative prior distributions, we see that the values are almost identical to those obtained with medium-size informative prior distributions.

	\autoref{FigA11} and \autoref{FigA12} illustrate the crude mortality rates and posterior mean rates per 100,000 inhabitants for each spatial prior for the informative and uniform prior distributions. Notably, a significant degree of smoothing is evident when using a small-size prior distribution across all spatial priors. Among these, the BYM prior demonstrates the least pronounced smoothing effect, preserving some areas with lower rates and preventing full convergence toward the mean. As the size of the informative prior distribution increases, the degree of smoothing gradually decreases. Even with the medium-size informative prior distribution, smoothing remains noticeable. However, with the large-size informative prior distribution, most divergent areas are preserved. However, some degree of smoothing is appreciated in regions significantly diverging from their neighbors, such as areas of the provinces of Teruel, Cuenca, and Albacete, which exhibit notably lower rates compared to surrounding areas. As seen in \autoref{tabA7}, the uniform prior presents very similar smoothing to that seen with the medium-size informative prior distribution.

	\subsection{Lung cancer deaths in England}
	
	\begin{figure}[b!]
		\begin{center}
			\includegraphics[width = 8cm]{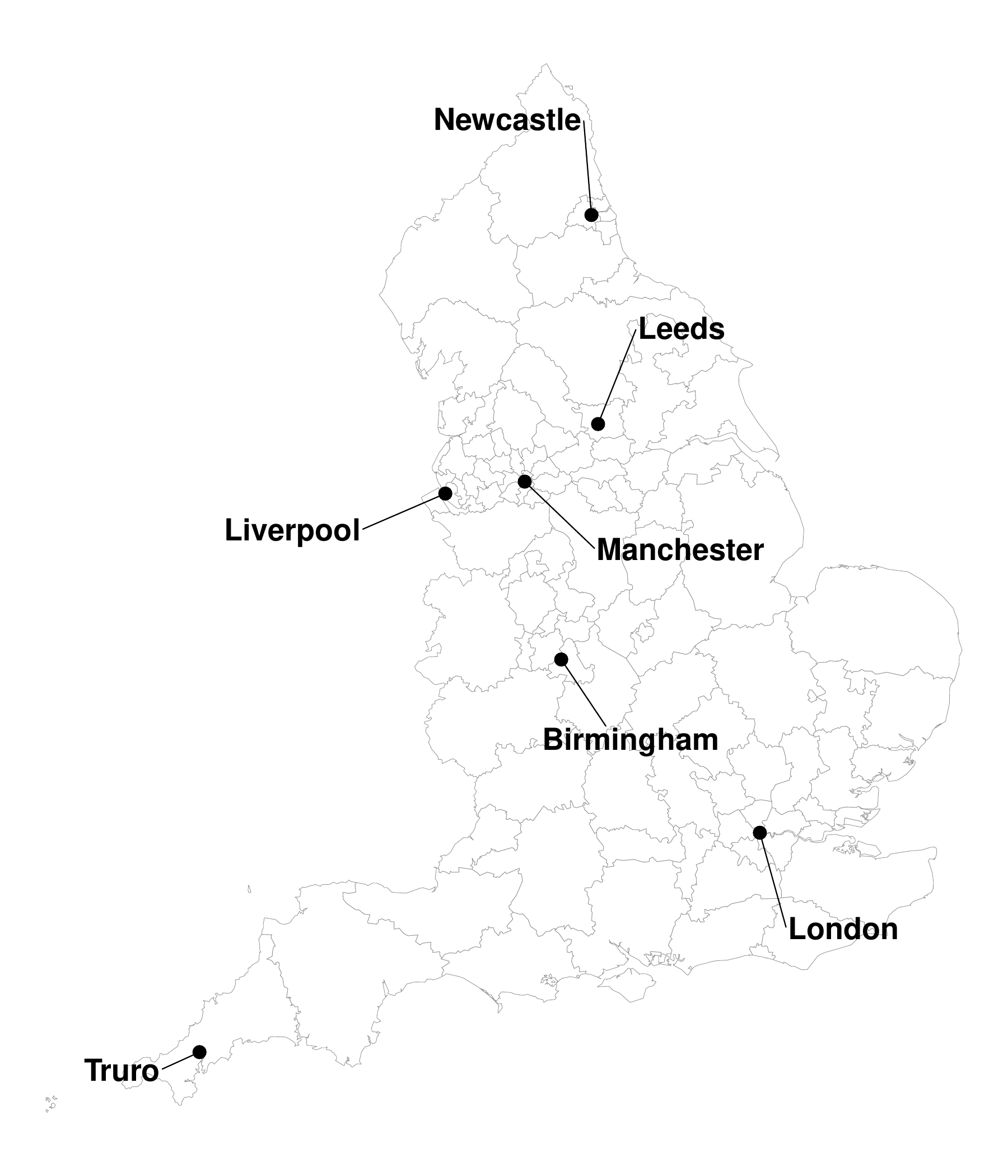}
		\end{center} \vspace{-0.9cm}
		\caption{\label{FigA13} Administrative division of England into 106 clinical commissioning groups.}
	\end{figure}
	
\begin{figure}[t!]
		\begin{center}
			\includegraphics[width = 14.5cm]{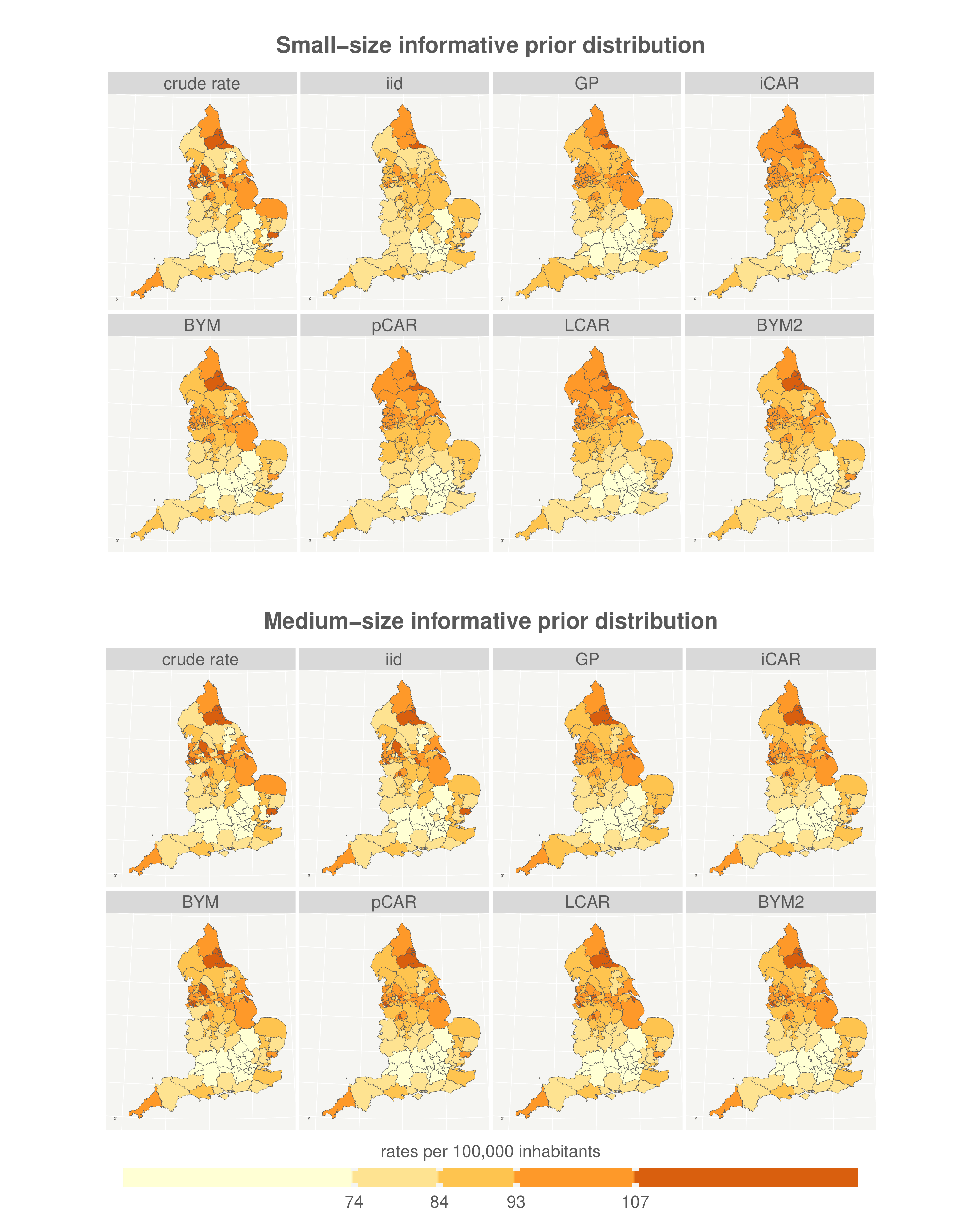}
		\end{center}
		\caption{\label{FigA14} Crude rates and posterior mean rates per 100,000 inhabitants for each spatial prior, assuming small-size (top) and medium-size (bottom) informative prior distributions for $\sigma^2$.}
	\end{figure}

	A total of 15,228 deaths from lung cancer in England were recorded in 2017. The administrative division used for this study ranges from 37,957 to about 720,010 inhabitants per unit in England. The spatial distribution of England, along with the localization of some cities mentioned in this section is provided in \autoref{FigA13}.
	As an initial overview of the diseases, crude mortality rates per 100,000 population by areas are provided in the first grid of \autoref{FigA14}. The crude rates vary between 40 and 140 deaths per 100,000 inhabitants. Generally, the highest crude rates are observed in the north-eastern areas and those near Manchester. Conversely, areas in the central-southern parts of England, particularly near London, exhibit the lowest mortality rates.

	
	\begin{table}[b!]
		\centering
		\caption{\label{tabA8} Empirical and theoretical expected smoothing for lung cancer dataset in England.} 
		\resizebox{\textwidth}{!}{
			\begin{tabular}{l|rrrrrrrrrr}
				\hline
				& MSS & RMSS & $maxMSS$ & $maxRMSS$ & TCV & $\sigma^2$ & $\tau^2$ & $\lambda$ & $\psi$ & SP \\ 
				\hline
				\multicolumn{10}{l}{\textbf{Small-size informative prior distribution}}\\
				iid & 138.014 & 1.580 & 965.573 & 10.530 & 1.044 & 0.010 & && &0.386\\ 
				GP & 117.241 & 1.295 & 1237.231 & 13.349 & 0.004 & 0.010 &&& 0.600 & 0.327\\ 
				iCAR & 134.106 & 1.526 & 1239.800 & 13.390 & 0.275 & 0.010 &  &&&0.375\\ 
				BYM & 89.806 & 1.006 & 611.916 & 7.301 & 0.012 & 0.009 & 0.009&& &0.251\\ 
				pCAR & 134.303 & 1.529 & 1225.952 & 13.267 & 0.275 & 0.010 & && &0.375\\ 
				LCAR & 133.323 & 1.516 & 1206.765 & 13.102 & 0.279 & 0.010 & &0.980 &&0.372\\ 
				BYM2 & 103.750 & 1.166 & 963.378 & 10.811 & 0.007 & 0.010 & &0.816& &0.290\\ 
				\multicolumn{10}{l}{ }\\
				\multicolumn{10}{l}{\textbf{Medium-size informative prior distribution}}\\
				iid & 83.292 & 0.931 & 374.059 & 3.763 & 4.841 & 0.046 & &&& 0.233\\ 
				GP & 96.038 & 1.054 & 1070.971 & 11.810 & 0.083 & 0.092 &&& 2.658 &0.268\\ 
				iCAR & 83.764 & 0.926 & 658.146 & 7.744 & 1.967 & 0.070 &  &&&0.234\\ 
				BYM & 78.903 & 0.871 & 417.743 & 5.164 & 0.070 & 0.043 & 0.013&&&0.220 \\ 
				pCAR & 82.899 & 0.917 & 603.473 & 7.179 & 2.148 & 0.076 &  &&&0.232\\ 
				LCAR & 81.331 & 0.899 & 619.751 & 7.360 & 2.059 & 0.068 & &0.879& &0.227\\ 
				BYM2 & 82.327 & 0.910 & 573.168 & 6.858 & 0.054 & 0.029 & &0.853&&0.230 \\
				\multicolumn{10}{l}{ }\\
				\multicolumn{10}{l}{\textbf{Large-size informative prior distribution}}\\
				iid & 83.814 & 0.910 & 278.301 & 2.502 & 17.469 & 0.165 & &&&0.234 \\ 
				GP & 91.342 & 1.002 & 988.660 & 11.041 & 0.233 & 0.205 &&& 4.731&0.255 \\ 
				iCAR & 78.945 & 0.860 & 342.019 & 4.295 & 4.727 & 0.168 & &&&0.221 \\ 
				BYM & 85.417 & 0.915 & 273.736 & 2.437 & 4.021 & 0.177 & 0.165&& &0.239\\ 
				pCAR & 78.726 & 0.861 & 320.585 & 4.053 & 4.770 & 0.169 &  &&&0.220\\ 
				LCAR & 76.812 & 0.836 & 309.285 & 3.932 & 5.018 & 0.168 & &0.902 &&0.215\\ 
				BYM2 & 81.040 & 0.872 & 262.879 & 2.424 & 1.296 & 0.165 & &0.970&&0.226\\  
				\multicolumn{10}{l}{ }\\
				\multicolumn{10}{l}{\textbf{Uniform prior distribution}}\\
				iid & 83.300 & 0.931 & 368.163 & 3.702 & 4.833 & 0.046 & && &0.233 \\ 
				GP & 94.081 & 1.032 & 1044.851 & 11.558 & 0.078 & 0.078 & &&2.017 &0.263\\ 
				iCAR & 84.012 & 0.929 & 653.367 & 7.701 & 1.965 & 0.070 &  &&&0.235\\ 
				BYM & 81.683 & 0.903 & 561.653 & 6.734 & 0.058 & 0.058 & 0.004&&&0.228 \\ 
				pCAR & 82.835 & 0.916 & 608.204 & 7.227 & 2.151 & 0.076 &  &&&0.231\\ 
				LCAR & 80.601 & 0.892 & 599.436 & 7.143 & 2.078 & 0.068 && 0.875& &0.225\\ 
				BYM2 & 82.152 & 0.909 & 572.507 & 6.849 & 0.054 & 0.029 && 0.855& &0.229\\ 
				\hline
			\end{tabular}
		}
	\end{table}
	
	Using the four priors for $\sigma^2$ above, \autoref{tabA8} displays the empirical smoothing criteria, namely MSS and RMSS, along with their respective maximum values and smoothing proportion (SP). It also includes the TCV measure and posterior mean of the estimated hyperparameters for each spatial prior and prior distribution for the variance of the proposed spatial priors for the England dataset. We begin by analyzing the smoothing induced by the spatial priors under informative prior distributions. In general, as the values specified for the informative prior increase, the variance increases, which leads to an increase in the theoretical measure and a decrease in the empirical smoothing criteria. This behavior aligns with what we have observed for the TCV metric in the simulation study. Furthermore, it is evident that for some spatial priors we achieve nearly the minimum average smoothing when a medium-size informative prior is assumed, as the MSS and RMSS values remain nearly constant while the theoretical metric continues to increase at the same pace. However, the maximum MSS and RMSS values still decrease for most of the spatial priors when we transition from a medium-size to a large-size informative prior distribution. Similar to Spain, the GP spatial prior tends to induce more smoothing than the CAR (neighbor-based) priors as the prior on $\sigma^2$ becomes weaker. Moreover, the uniform prior yields values nearly identical to those of medium-size informative prior.

	\begin{figure}[t!]
		\begin{center}
			\includegraphics[width = 14.5cm]{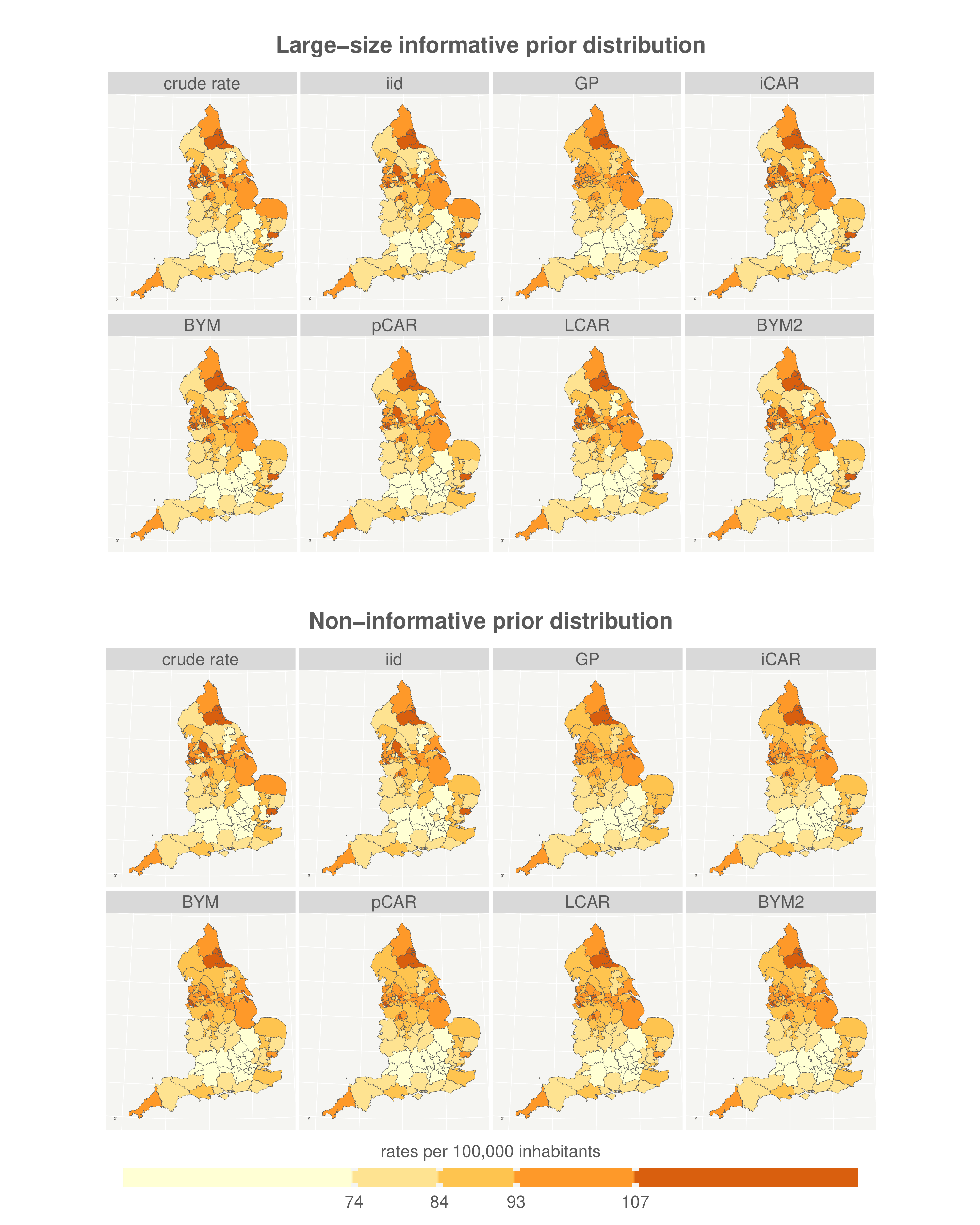}
		\end{center}
		\caption{\label{FigA15} Crude rates and posterior mean rates per 100,000 inhabitants for each spatial prior, assuming small-size (top) and medium-size (bottom) informative prior distributions for $\sigma^2$.}
	\end{figure}

	\autoref{FigA14} and \autoref{FigA15} show the crude rates and the posterior mean rates obtained for each spatial prior per 100,000 inhabitants. When a small-size informative prior distribution is used, more smoothing is observed compared to the medium-size distribution. For the small-size prior distribution, the smoothing effect is more pronounced with the CAR priors, except for the BYM and BYM2 priors, which show less smoothing. The remaining CAR priors exhibit higher rates in northern England and lower rates in London and its surrounding areas, revealing a south-east to north-west pattern that is not present in the crude rates. This observed pattern could be a result of the substantial smoothing effect introduced by these spatial priors rather than a reflection of the actual crude rates. 
	In contrast, when assuming medium-size informative prior distributions, all spatial priors yield similar results, with slight disparities observed in areas near Liverpool and Manchester. In these regions, the iid and BYM priors produce estimates more closely aligned with the crude rates. Results obtained with large-size informative are comparable to those with medium-size informative prior distribution, but larger discrepancies are noticeable with the CAR priors. Notably, with the medium-size informative prior distribution, smoother results are observed in the north-western side of England compared to the large-size informative prior. Additionally, the GP prior exhibits more smoothing in this region when using a large-size informative prior distribution compared to the other spatial priors. Results with the uniform prior distribution are identical to those obtained with the small-size informative prior distribution.

	\end{appendices}
	
\end{document}